\documentclass[journal]{IEEEtran}

\usepackage[T1]{fontenc}
\usepackage[utf8]{inputenc}
\usepackage{amsmath, amssymb, amsfonts, bm}
\usepackage{tikz}
\usetikzlibrary{arrows.meta, patterns, decorations.pathmorphing}
\usepackage{graphicx}
\usepackage{tcolorbox}
\usepackage{xcolor}

\usepackage{booktabs}
\usepackage{multirow}
\usepackage{float}
\usepackage{longtable}
\usepackage{array}
\usepackage{tabularx}

\usepackage{microtype}
\usepackage[caption=false,font=footnotesize]{subfig}
\usepackage[hidelinks]{hyperref}
\usepackage{placeins}

\newcommand{\xv}{\mathbf{x}}
\newcommand{\xs}{\mathbf{x}_s}

\newcommand{\xA}{\mathbf{x}_A}
\newcommand{\xH}{\mathbf{x}_H}
\newcommand{\xB}{\mathbf{x}_B}
\newcommand{\dD}{\partial\mathbb{D}}
\newcommand{\conv}{\ast}
\newcommand{\kH}{\mathbf{k}_H}

\ifCLASSINFOpdf
\else
\fi

\begin{document}

\title{Marchenko Theory, Algorithms, and Applications: A Review}

\author{
Hammed~A.~Oyekan
\thanks{Hammed A. Oyekan is with the College of Petroleum and Geoscience, King Fahd University of Petroleum and Minerals (email: oyekan.geophysics@gmail.com)}
}

\maketitle

\begin{abstract}
Seismic reflection data collected at the Earth's surface carry information from every depth in the subsurface, including the internal multiples that conventional migration treats as noise or artefacts. The Marchenko method retrieves subsurface Green's functions directly from these single-sided measurements. The only model information it requires is a smooth macro-velocity model, used to estimate the direct-wave traveltime from the surface to a virtual point at depth. The method originates in one-dimensional quantum mechanical inverse scattering and has been generalised over the past two decades to three-dimensional acoustic media using acoustic reciprocity theorems, and to elastic and electromagnetic media using the corresponding elastodynamic and Maxwell reciprocity theorems. This review covers the classical Gelfand--Levitan--Marchenko equation and its derivation through the Schr\"{o}dinger scattering formalism, the seismic interferometry results that motivated the multidimensional extension, and the coupled Marchenko equations for wave focusing in higher dimensions. On the application side it treats Marchenko redatuming, single- and double-sided imaging conditions, internal multiple elimination, target-oriented processing, and full-waveform inversion. Extensions to dissipative media, elastodynamics, electromagnetic fields, and plane-wave acquisition are also covered, along with practical topics such as least-squares implementations, GPU acceleration, compressive-sensing acquisition, and recent machine-learning approaches to predicting the focusing functions. Field-data results from marine, land, sub-salt, and time-lapse monitoring settings are summarised. The review closes with an assessment of current limitations and open problems, among them elastic multi-component extensions, joint inversion, and neural-network accelerators.
\end{abstract}

\begin{IEEEkeywords}
Green's functions, inverse scattering, Marchenko imaging, Marchenko theory, redatuming, Schr\"{o}dinger equation, seismic imaging.
\end{IEEEkeywords}

\IEEEpeerreviewmaketitle

\section{Introduction}
\label{sec:intro}

\IEEEPARstart{I}{n} seismic exploration, seismic energy is injected into the subsurface and the wavefields scattered from impedance contrasts are recorded at the surface. These recorded data contain a superposition of reflections from all interfaces at all depths, intermingled with free-surface reverberations, inter-bed and other multiples, mode-converted energy, and coherent noise. Getting an interpretable image of the subsurface out of that mixture is the central problem of reflection seismology.
The classical migration methods (Kirchhoff, reverse-time migration (RTM), and one-way wave-equation migration) address this by back-propagating the recorded wavefield through a smooth background velocity model and applying an imaging condition at each subsurface point. Two assumptions drive the approach: that the subsurface can be divided into a smooth background that determines wave kinematics, and that primary reflections dominate the data. Both fail in the presence of strong overburden velocity contrasts such as salt bodies or carbonates, and where internal multiples carry significant energy. Conventional migration then produces false events and distorted amplitudes.
The Marchenko method drops both assumptions. It retrieves the full subsurface Green's function, including contributions from all multiple reflections, directly from the single-sided reflection response, using only a smooth macro-velocity model to estimate the direct-wave traveltime. No structural model of the overburden is required, and the resulting image is free of artefacts caused by internal multiples.

The Green's function between any two points fully characterises wave propagation in the medium and is therefore what imaging needs. Classically it is computed by solving the forward wave equation in an assumed velocity model. The Marchenko method instead retrieves it directly from the data, without requiring a complete velocity model. This is possible because the single-sided reflection response encodes sufficient information about internal wave propagation to reconstruct the full two-sided response through a system of integral equations. Accurate Green's functions in turn allow virtual sources and receivers to be constructed at any depth without physically placing sensors there, a procedure known as \emph{redatuming}.

\subsection{Inverse Scattering and the Marchenko Framework}

Inverse scattering is the fundamental basis of the Marchenko framework, which aims to recover the properties of a medium from observations of scattered waves \cite{chadan1997introduction}. In quantum mechanics, the one-dimensional version was solved exactly by Gelfand and Levitan \cite{gelfand1951determination} and independently by Marchenko \cite{marchenko1955reconstruction} in the 1950s. Their approach converts the spectral data of a Schr\"{o}dinger operator \cite{berezin2012schrodinger}, equivalently the reflection coefficient as a function of frequency, into a linear integral equation (the Gelfand--Levitan--Marchenko, or GLM, equation) whose solution yields the scattering potential and hence the medium profile.

Extending this formalism to multi-dimensional seismology proved difficult, largely because the traveltime transformation that converts the wave equation into a Schr\"{o}dinger-type equation does not generalise beyond one dimension. Broggini and Snieder \cite{broggini2012connection} showed that the one-dimensional focusing problem, that of finding a wavefield which collapses to a spatial delta function at a target depth from a single-sided injection, is directly related to the GLM equation. Combined with the reciprocity-based representation theorems of Wapenaar and several other collaborators \cite{wapenaar2014marchenko, slob2014seismic}, that result produced the multidimensional Marchenko equations in their present form.

\subsection{Historical Overview}

The development of the Marchenko method in geophysics can be categorised into three broad periods.

\textbf{Classical foundations (1950s to 1990s).} The GLM equation was derived for one-dimensional quantum mechanical inverse scattering \cite{marchenko1955reconstruction, gelfand1951determination}. It relates the scattering potential within a medium to an internal focusing field defined by the reflection response measured on one side of the medium \cite{yue2020diffraction}. In 1980, Burridge \cite{burridge1980gelfand} recast the classical inverse scattering formulations, namely the Gelfand--Levitan, Marchenko, and Gopinath--Sondhi integral equations, as inverse impulse-response problems of the kind reflection seismology actually poses. The move that mattered was deriving them strictly in the time domain. These equations had originally been obtained as solutions to an inverse \emph{spectral} problem, naturally posed in the frequency domain; Burridge showed that a time-dependent formulation exists and that the derivation is simpler there than in its original setting, and he regarded the Gopinath--Sondhi equation as the one directly applicable to seismic reflection data. Related ideas appeared in Bremmer's \cite{bremmer1951wkb} treatment of wave propagation in layered media and, later, in the inverse scattering series of Weglein et al.\ \cite{weglein1997inverse, weglein2003inverse}. Work in this period exploited the connection between reflection data and the medium profile through WKBJ and layer-stripping approaches. Newton \cite{newton1982inverse} extended classical inverse scattering theory to non-spherically-symmetric 3D media as early as 1982, but his construction required data unavailable in seismic acquisition, so a practical multi-dimensional inversion remained out of reach.

\textbf{The seismic interferometry connection (2000s to early 2010s).} This period produced the ideas that made the multidimensional generalisation possible. Seismic interferometry showed that the Green's function between two receivers can be retrieved by cross-correlating their ambient noise or coda records \cite{lobkis2001emergence, wapenaar2004retrieving, wapenaar2006green, curtis2006seismic}, but the method requires sources on a closed surface surrounding the medium. The single-sided case was addressed by Broggini and Snieder \cite{broggini2012connection, broggini2012focusing}, who, building on Rose's single-sided autofocusing \cite{rose2002single}, clarified the relationship between interferometry and the Marchenko equation and showed that a focusing wavefield converging to a delta function at a target depth can be constructed from the reflection response alone. Combining this 1D result with the reciprocity theorems for wavefields \cite{wapenaar1996reciprocity, wapenaar1996reciprocityb} led Wapenaar and co-workers to the three-dimensional single-sided Marchenko equations \cite{wapenaar2013three}. By the early 2010s the multidimensional Marchenko framework had made data-driven Green's function retrieval and internal multiple elimination possible without a detailed velocity model of the overburden.

\textbf{The multidimensional and AI era (early 2010s to present).} Wapenaar et al.\ \cite{wapenaar2014marchenko} derived the coupled multidimensional Marchenko equations from acoustic reciprocity theorems. Extensions followed rapidly: three dimensions \cite{wapenaar2013three, brackenhoff2022three}, dissipative media \cite{slob2016green, slob2016electromagnetic, marchenko2016dissipative}, elastodynamics \cite{da2016elastic, da2014elastodynamic}, evanescent waves \cite{brackenhoff2023evanescent, wapenaar2020marchenko}, electromagnetic media \cite{slob2016electromagnetic, slob2013coupled, zhang2017electromagnetic}, and field data applications in marine \cite{jia2017subsalt, da2018marchenko}, land \cite{cheng2024marchenko}, subsalt \cite{jia2017subsalt, jia2018practical}, and time-lapse \cite{brackenhoff2020timelapse} settings.

The iterative calculation of focusing functions across large 3D data volumes remains the main computational bottleneck of the Marchenko method and its extensions. Machine learning is now being used to attack this bottleneck; the first papers on learned prediction of focusing functions appeared only recently \cite{wang2025accelerating, wang2025acceleratingdeep}.

\subsection{Scope and Organisation of This Study}

This review is self-contained. Sections~\ref{sec:math} through~\ref{sec:multidim} develop the mathematical foundation from first principles. Section~\ref{sec:math} establishes notation and reviews the essential tools: the acoustic wave equation, Green's functions, wavefield decomposition, and reciprocity theorems. Section~\ref{sec:1D} develops the one-dimensional Marchenko theory in full, including the transformation to the Schr\"{o}dinger equation, the ray expansion solution, the GLM equation, and medium reconstruction. Section~\ref{sec:bridge} describes the conceptual bridge from one dimension to many, focusing on the seismic interferometry connection and the focusing problem. Section~\ref{sec:multidim} presents the full derivation of the multidimensional Marchenko equations and their iterative solution. Sections~\ref{sec:greens} through~\ref{sec:targeting} cover the principal applications: Green's function retrieval, focusing functions, imaging, redatuming, multiple elimination, and target-oriented methods. Section~\ref{sec:extensions} reviews extensions to elastic, electromagnetic, dissipative, and other settings. Section~\ref{sec:implementation} addresses computational aspects. Sections~\ref{sec:fielddata} and~\ref{sec:timelapse} survey field data results. Section~\ref{sec:connections} discusses connections to other methods. Sections~\ref{sec:challenges} and~\ref{sec:future} address challenges and future directions. Appendices provide detailed derivations of the GLM equation (Appendix~\ref{app:1D}), the multidimensional Marchenko equations (Appendix~\ref{app:2D3D}), wavefield decomposition (Appendix~\ref{app:decomp}), and the iterative algorithm (Appendix~\ref{app:algorithm}).

\section{Mathematical Foundations}
\label{sec:math}

Throughout this review, three-dimensional Cartesian coordinates are denoted $\xv = (x_1, x_2, x_3)$, with $x_3$ positive in the downward direction. The horizontal coordinate vector is $\xH = (x_1, x_2)$. The acquisition surface lies at $x_3 = 0$ and is denoted $\dD_0$. A virtual depth level at depth $x_{3,i}$ is denoted $\dD_i$. The symbol $\ast$ denotes temporal convolution. The Fourier transform of a time-domain quantity $f(t)$ is
\[
\hat{f}(\omega) = \int_{-\infty}^{\infty} f(t)\, e^{i\omega t}\, dt
\]

\noindent So the time derivative corresponds to multiplication by $-i\omega$ in the frequency domain. Bold symbols denote vectors; $\nabla = (\partial_{x_1}, \partial_{x_2}, \partial_{x_3})$ is the spatial gradient operator.

\subsection{The Acoustic Wave Equation}

For a compressible, heterogeneous acoustic medium with spatially varying bulk modulus $\kappa(\xv)$ and density $\rho(\xv)$, the pressure wavefield $P(\xv, t)$ and particle velocity $\mathbf{v}(\xv, t)$ satisfy the coupled first-order system:
\begin{align}
\rho(\xv)\frac{\partial \mathbf{v}}{\partial t} + \nabla P &= \rho(\xv) \mathbf{f}(\xv, t) \label{eq:euler}\\
\frac{1}{\kappa(\xv)}\frac{\partial P}{\partial t} + \nabla \cdot \mathbf{v} &= q(\xv, t) \label{eq:continuity}
\end{align}

\noindent where $\mathbf{f}$ is an external body force density and $q$ is a volume injection rate. Eliminating $\mathbf{v}$ from equations \eqref{eq:euler}--\eqref{eq:continuity}, and setting $\mathbf{f} = \mathbf{0}$, yields the second-order wave equation for pressure:

\begin{equation}
\nabla \cdot \left[ \frac{1}{\rho(\xv)} \nabla P(\xv, t) \right] - \frac{1}{\kappa(\xv)} \frac{\partial^2 P}{\partial t^2} = -\frac{\partial q}{\partial t} \label{eq:waveP}
\end{equation}

\noindent In a homogeneous medium, $\rho(\xv) = \rho_0$ and $\kappa(\xv) = \kappa_0 = \rho_0 c_0^2$, and multiplying equation \eqref{eq:waveP} by $\rho_0$ reduces it to the standard wave equation

\[
\nabla^2 P - \frac{1}{c_0^2}\frac{\partial^2 P}{\partial t^2} = -\rho_0\frac{\partial q}{\partial t}
\]

\noindent For a unit-volume injection rate at $\xv = \xv'$, i.e.\ $q(\xv, t) = \delta(\xv - \xv')\delta(t)$, equation \eqref{eq:waveP} defines the Green's function $G(\xv, \xv', t)$:

\begin{equation}
\nabla \cdot \left[\frac{1}{\rho(\xv)}\nabla G(\xv, \xv', t)\right] - \frac{1}{\kappa(\xv)}\frac{\partial^2 G}{\partial t^2} = -\delta(\xv - \xv')\frac{d \delta(t)}{d t} \label{eq:greendef}
\end{equation}

\noindent The Green's function satisfies source-receiver reciprocity: $G(\xv, \xv', t) = G(\xv', \xv, t)$.

\subsection{Wavefield Decomposition into One-Way Components}

The Marchenko framework requires decomposing the pressure wavefield into downgoing ($+$) and upgoing ($-$) components. In a reference medium, the pressure field is written as
\begin{equation}
P(\xv, t) = P^+(\xv, t) + P^-(\xv, t)
\end{equation}
and the vertical component of particle velocity decomposes as
\begin{equation}
V_3(\xv, t) = V_3^+(\xv, t) + V_3^-(\xv, t)
\end{equation}

\noindent In the frequency-wavenumber domain $(\omega, \kH)$, where $\kH = (k_1, k_2)$ are horizontal wavenumbers and the medium is locally laterally invariant, the decomposed fields are related by
\begin{align}
\hat{P}^{\pm}(\mathbf{k}_H, x_3, \omega)
&= \frac{1}{2}\hat{P}(\mathbf{k}_H, x_3, \omega) \nonumber \\
&\quad \pm \frac{\omega\,\rho(x_3)}{2 k_{x_3}(\mathbf{k}_H, x_3, \omega)} \, \hat{V}_3(\mathbf{k}_H, x_3, \omega)
\end{align}

\noindent Where $k_{x_3} = \sqrt{\omega^2/c^2(x_3) - |\kH|^2}$ is the vertical wavenumber. This is the pressure-normalised decomposition; rescaling $\hat{P}^{\pm}$ by $\sqrt{k_{x_3}/(\omega\rho)}$ gives the \emph{flux-normalised} one-way fields, for which $|P^+|^2 - |P^-|^2$ is proportional to the vertical component of the acoustic power flux density \cite{wapenaar1990elastic}.

\noindent In the spatial domain, the decomposition takes the form of a convolutional relation involving a pseudo-differential operator $L$ \cite{wapenaar2014marchenko}:
\begin{align}
P^{\pm}(\mathbf{x}_H, x_3, t)
= \frac{1}{2} \Big[ & P(\xH, x_3, t) \nonumber \\
& \pm (L \ast V_3)(\xH, x_3, t) \Big]
\end{align}

\noindent For a plane wave at incidence angle $\theta$ in a constant medium, $L$ reduces to multiplication by $\rho c / \cos\theta$; at normal incidence, $L V_3 = \rho c V_3$, the familiar impedance relation.

The one-way Green's functions $G^+$ and $G^-$ denote the downgoing and upgoing components at a receiver location $\xv_r$ due to a monopole source at $\xv_s$:

\begin{align}
G(\xv_r, \xv_s, t) &= G^+(\xv_r, \xv_s, t) + G^-(\xv_r, \xv_s, t) \\
G^-(\xv_r, \xv_s, t) &\approx R(\xv_r, \xv_s, t) \; \text{when both points are on } \dD_0
\end{align}

where $R(\xv_r, \xv_s, t)$ is the reflection response. The transmission response, written $T(\xv_r, \xv_s, t)$ with $\xv_r$ on a depth level below $\xv_s$, is used from Section~\ref{sec:multidim} onward.

\subsection{Reciprocity Theorems}

Reciprocity theorems relate wavefields in two different states, state A and state B, within a domain $\mathbb{D}$ bounded by $\partial\mathbb{D}$. There are two types \cite{wapenaar1996reciprocity, fokkema1993seismic}.

\subsubsection*{Convolution-Type Reciprocity (Rayleigh's Reciprocity Theorem)}

For two frequency-domain pressure fields $\hat{P}_A$ and $\hat{P}_B$ satisfying equation \eqref{eq:waveP} with sources $q_A$ and $q_B$, respectively:
\begin{align}
\oint_{\partial\mathbb{D}} \frac{1}{\rho} & \left( \hat{P}_A \frac{\partial \hat{P}_B}{\partial n} - \hat{P}_B \frac{\partial \hat{P}_A}{\partial n} \right) d^2\xv \nonumber \\ &= i\omega\int_{\mathbb{D}} \left( \hat{P}_A\, \hat{q}_B - \hat{P}_B\, \hat{q}_A \right) d^3\xv \label{eq:recipconv}
\end{align}

where $\partial/\partial n$ is the outward normal derivative. The sign of the right-hand side follows from the transform convention of Section~\ref{sec:math}, in which $\partial_t \rightarrow -i\omega$; the opposite convention flips it. When both states are impulsive point sources inside $\mathbb{D}$ and the boundary integral vanishes under the radiation condition, equation \eqref{eq:recipconv} reduces to source-receiver reciprocity $G(\xv_A, \xv_B, t) = G(\xv_B, \xv_A, t)$.

\subsubsection*{Correlation-Type Reciprocity}

For state B and the time reversal of state A (denoted $\hat{P}_A^*$ in the frequency domain, corresponding to $P_A(-t)$ in the time domain):
\begin{align}
\oint_{\partial\mathbb{D}} \frac{1}{\rho} & \left( \hat{P}_A^* \frac{\partial \hat{P}_B}{\partial n} - \hat{P}_B \frac{\partial \hat{P}_A^*}{\partial n} \right) d^2\xv \nonumber \\ &= i\omega\int_{\mathbb{D}} \left( \hat{P}_A^*\, \hat{q}_B + \hat{P}_B\, \hat{q}_A^* \right) d^3\xv \nonumber \\ &\quad - \omega^2 \int_{\mathbb{D}} \left(\frac{1}{\kappa} - \frac{1}{\kappa^*}\right) \hat{P}_A^* \hat{P}_B\, d^3\xv \label{eq:recipcorr}
\end{align}

\noindent The final term is the dissipation term: $\kappa$ is real for a lossless medium, so the bracket vanishes and only the source terms survive. For a dissipative medium $\kappa$ is complex and frequency dependent, and the bracket equals $2i\,\mathrm{Im}(1/\kappa)$. Equations \eqref{eq:recipconv} and \eqref{eq:recipcorr} assume real $\rho$; complex $\rho$ adds a further volume term. For heterogeneous density the correlation-type integrand carries the same $1/\rho(\xv)$ weighting as the convolution-type one \cite{wapenaar2006green, schuster2009interferometry}.

The two theorems differ in one respect that matters for what follows. In the convolution-type equation the contribution from a boundary at infinity vanishes by the Sommerfeld radiation condition, because both Green's functions in the integrand are causal. That argument is unavailable in the correlation-type equation, whose integrand pairs a causal with an anti-causal field. What is invoked instead is the \emph{anti-radiation condition} \cite{schuster2009interferometry}: in a sufficiently heterogeneous medium, internal scattering weakens the arrivals that reach the distant boundary quickly enough for the monopole--dipole products to decay faster than the surface area grows, so the contribution from infinity can be neglected. The single-sided Marchenko derivation inherits this assumption through equation \eqref{eq:recipcorr}, and it is a physical assumption about the medium rather than a purely mathematical one.

These two theorems are the foundation on which the multidimensional Marchenko equations are derived (Section~\ref{sec:multidim} and Appendix~\ref{app:2D3D}).

\subsection{Seismic Interferometry: A Brief Overview}

Seismic interferometry shows that, for sufficiently uniform source distributions surrounding a medium, the cross-correlation of wavefields at two receiver positions retrieves the Green's function between those receivers \cite{lobkis2001emergence, wapenaar2004retrieving}:
\begin{align}
G(\xv_A, \xv_B, t) + G(\xv_A, \xv_B, -t) \propto \nonumber \\ \oint_{\partial\mathbb{D}} C(\xv_A, \xv_s, t) \, \ast \, C^*(\xv_B, \xv_s, t)\, d^2\xv_s \label{eq:interferometry}
\end{align}

\noindent where $C(\xv, \xv_s, t)$ is the wavefield at $\xv$ due to a source at $\xv_s$ on $\partial\mathbb{D}$, and the superscript asterisk denotes time reversal, so that $C \conv C^*$ is a cross-correlation. This requires sources on a closed surface, that is, double-sided illumination, which is not achievable in exploration surveys confined to the Earth's surface.

The Marchenko method retrieves Green's functions using only single-sided illumination. It does so not by cross-correlation but by solving a system of integral equations that encodes the causality constraints of wave propagation in the medium.

\section{One-Dimensional Marchenko Theory}
\label{sec:1D}

\subsection{The 1D Inverse Scattering Problem}

When an incident wave travelling through a medium encounters a heterogeneity, waves are scattered, and the scattered waves carry information about the heterogeneity to the receivers where they are recorded. Reconstructing the properties of an unknown medium from recordings of the scattered waves is the inverse scattering problem. In one spatial dimension, waves propagate in only two directions (forward and backward), and the reflection data recorded at a single receiver contain sufficient information to reconstruct the medium entirely, provided the medium satisfies certain regularity conditions. The mathematical tool for this reconstruction is the Gelfand--Levitan--Marchenko equation \cite{gelfand1951determination, marchenko1955reconstruction}.

\noindent The 1D wave equation for a vertically varying medium is:
\begin{equation}
\rho(x)\frac{\partial^2}{\partial t^2}u(x,t) - \frac{\partial}{\partial x}\left[\rho(x)c^2(x)\frac{\partial u(x,t)}{\partial x}\right] = 0 \,
\label{eq:1Dwave}
\end{equation}

\noindent where $u(x,t)$ is the particle displacement, $\rho(x)$ the density, and $c(x)$ the wave speed. The objective is to recover $\rho(x)$ and $c(x)$ from the reflection response $R(t)$ at a fixed surface location $x_0$. So, we try to transform \eqref{eq:1Dwave} into a form of Schr\"{o}dinger Equation.

\subsection{Step 1: Transformation to the Traveltime Domain and the Schr\"{o}dinger Equation}

To reduce equation \eqref{eq:1Dwave} to a canonical form amenable to inverse scattering, we introduce the one-way traveltime coordinate
\[
\tau(x) = \int_0^x \frac{dx'}{c(x')}
\]

which satisfies $d\tau = dx/c(x)$. By the chain rule, spatial derivatives transform as $d/dx = (1/c(x))\,d/d\tau$.

\textbf{Transforming the spatial term:} Applying the chain rule twice to the spatial term of \eqref{eq:1Dwave}:
\begin{align*}
\frac{\partial}{\partial x} \left( \rho(x) c^2(x) \frac{\partial u}{\partial x} \right) &= \frac{\partial}{\partial x} \left( \rho(x) c^2(x) \frac{d\tau}{dx} \frac{\partial u}{\partial \tau} \right) \\ &= \frac{\partial}{\partial x} \left( \rho(x) c(x) \frac{\partial u}{\partial \tau} \right)
\end{align*}
\begin{align*}
\frac{\partial}{\partial x} \left( \rho(x) c(x) \frac{\partial u}{\partial \tau} \right) &= \frac{d\tau}{dx} \frac{\partial}{\partial \tau} \left( \rho(x) c(x) \frac{\partial u}{\partial \tau} \right) \\ &= \frac{1}{c(x)} \frac{\partial}{\partial \tau} \left( \rho(x) c(x) \frac{\partial u}{\partial \tau} \right)
\end{align*}

\noindent Substitute the transformed term into the original equation \eqref{eq:1Dwave}:
\[
\rho(x) \frac{\partial^2 u}{\partial t^2} - \frac{1}{c(x)} \frac{\partial}{\partial \tau} \left( \rho(x) c(x) \frac{\partial u}{\partial \tau} \right) = 0
\]

\noindent Multiply through by \( c(x) \):

\begin{equation}
\rho c \frac{\partial^2 u}{\partial t^2} - \frac{\partial}{\partial \tau}\left[\rho c \frac{\partial u}{\partial \tau}\right] = 0
\label{eq:impedancewave}
\end{equation}

\noindent where the product $\rho c$ is the characteristic \textit{impedance} of the medium. In this form the equation is Webster's horn equation \cite{burridge1980gelfand}, and it already settles what the 1D problem can deliver: with data recorded only at $\tau = 0$, the density and the wave speed cannot be recovered separately, but only through the combination $\rho c$.

\textbf{Introduction of the scaled wavefield:} Define $\eta(\tau) = \sqrt{\rho(\tau)c(\tau)}$, the square root of impedance, so that $\rho c = \eta^2$ and equation \eqref{eq:impedancewave} reads

\[
\eta^2 \frac{\partial^2 u}{\partial t^2} - \frac{\partial}{\partial \tau} \left( \eta^2 \frac{\partial u}{\partial \tau} \right) = 0
\]

Now substitute $u = \psi/\eta$ and expand the spatial derivative:
\begin{align*}
\frac{\partial}{\partial \tau}\!\left(\eta^2 \frac{\partial u}{\partial \tau}\right)
= \frac{\partial}{\partial \tau}\!\left(\eta^2 \frac{\partial}{\partial\tau}\!\left(\frac{\psi}{\eta}\right)\right)
= \frac{\partial}{\partial\tau}\!\left(\eta\frac{\partial\psi}{\partial\tau} - \psi\frac{d\eta}{d\tau}\right)
\end{align*}
\begin{align*}
= \frac{d\eta}{d\tau}\frac{\partial\psi}{\partial\tau}
+ \eta\frac{\partial^2\psi}{\partial\tau^2}
- \frac{\partial\psi}{\partial\tau}\frac{d\eta}{d\tau}
- \psi\frac{d^2\eta}{d\tau^2}
= \eta\frac{\partial^2\psi}{\partial\tau^2} - \psi\frac{d^2\eta}{d\tau^2}
\end{align*}

Since $\eta^2\,\partial_t^2 u = \eta\,\partial_t^2\psi$, the equation for $\psi$ becomes:

\[
\eta \frac{\partial^2 \psi}{\partial t^2} - \left( \eta \frac{\partial^2 \psi}{\partial \tau^2} - \psi \frac{d^2 \eta}{d\tau^2} \right) = 0
\]

\[
\Rightarrow \eta \left( \frac{\partial^2 \psi}{\partial t^2} - \frac{\partial^2 \psi}{\partial \tau^2} \right) + \psi \frac{d^2 \eta}{d\tau^2} = 0
\]

Divide through by \( \eta \):

\begin{equation}
\frac{\partial^2 \psi}{\partial t^2} - \frac{\partial^2 \psi}{\partial \tau^2} + \frac{1}{\eta} \frac{d^2 \eta}{d\tau^2} \psi = 0
\end{equation}

Writing the coefficient of $\psi$ as the scattering potential $q$ gives
\begin{equation}
\frac{\partial^2\psi}{\partial t^2} - \frac{\partial^2\psi}{\partial \tau^2} + q(\tau)\,\psi = 0
\label{eq:schrodinger}
\end{equation}

where the \emph{scattering potential} is defined as

\begin{equation}
q(\tau) = \frac{1}{\eta(\tau)}\frac{d^2\eta}{d\tau^2}
\end{equation}

Equation \eqref{eq:schrodinger} is the 1D acoustic Schr\"{o}dinger equation \cite{Schrodinger1926undulatory}, also recognisable as the equation of motion of an elastically braced string \cite{burridge1980gelfand}. The potential $q(\tau)$ vanishes in homogeneous regions; non-zero values mark impedance variations. The chain from equation \eqref{eq:1Dwave} to this point, that is, the traveltime coordinate, Webster's horn equation, the impedance scaling, and the resulting potential $q = \eta''/\eta$, follows Burridge \cite{burridge1980gelfand} step for step.

Burridge's reason for working in the time domain rather than the frequency domain carries directly into the multidimensional method. Recovering the medium over a finite interval requires the reflection record only over the corresponding finite segment, whose length is the two-way traveltime across that interval. A frequency-domain treatment, by contrast, needs long records to transform. That finite-record property is the ancestor of the time-window operator in Section~\ref{sec:multidim}, and it is also why the time-domain formulation extends to media that are merely bounded and piecewise continuous, rather than requiring continuity of the unknown coefficient \cite{burridge1980gelfand}.

Two distinct and inequivalent 1D formulations have been identified in the literature under the same name, making their conflation a recurring source of misunderstanding. The route taken above is the Liouville transformation: a traveltime coordinate plus the impedance scaling $\psi = \eta u$, which produces a potential $q(\tau) = \eta''/\eta$ that is \emph{independent of frequency}. This is the setting in which the classical Gelfand--Levitan--Marchenko theory applies, because that theory requires a frequency-independent potential. The alternative writes a perturbed Helmholtz equation about a constant background, $\hat u'' + (\omega^2/c_0^2)[1+\alpha(x)]\hat u = 0$, in which the effective potential is proportional to $\omega^2\alpha(x)$ and therefore frequency dependent \cite{magalhaes2019comparison}. Both are used, and both are called the 1D Marchenko problem, but only the first reconstructs an impedance profile through the kernel machinery developed below.

\subsection{Step 2: Analytical Solution via Ray Expansion}

\noindent Solutions to equation \eqref{eq:schrodinger} can be written as propagating pulses. The ray expansion (progressing wave expansion) \cite{fichtner2012full} takes the form:
\begin{equation}
\psi(\tau, t) = \sum_{n=0}^{\infty} a_n(\tau)\, f_n\!\left[t - \phi(\tau)\right],
\label{eq:rayexpansion}
\end{equation}

\noindent where $\phi(\tau)$ is the phase function (traveltime), $a_n(\tau)$ are amplitude coefficients of the superposition of wave shapes or wavelets, and the wavelets $f_n$ satisfy the recursion

\[
\frac{df_n}{dz} = f_{n-1}(z).
\]

\noindent So for an impulsive ($\delta$-pulse) initial condition $f_0(z) = \delta(z)$, the higher-order wavelets are:

\begin{itemize}
    \item $f_0(z) = \delta(z)$ (Dirac delta, impulse)
    \item $f_1(z) = H(z)$ (Heaviside step function)
    \item \(f_2(z) = z H(z)\)
    \item \(f_3(z) = \frac{z^2}{2}H(z)\)
    \item \(f_4(z) = \frac{z^3}{6}H(z)\)
    \item $f_n(z) = \dfrac{z^{n-1}}{(n-1)!}H(z)$ for $n \geq 1$
\end{itemize}

\noindent Substituting the assumed solution, \eqref{eq:rayexpansion}, into the Schr\"{o}dinger equation, \eqref{eq:schrodinger} and collecting powers of $f_{n-2}$, $f_{n-1}$, and $f_n$:

\[
\frac{\partial^2 \psi}{\partial t^2} - \frac{\partial^2 \psi}{\partial \tau^2} + q \psi = 0
\]

\[
\sum_{n=0}^{\infty} \left[ \frac{\partial^2}{\partial t^2} - \frac{\partial^2}{\partial \tau^2} + q(\tau) \right] (a_n(\tau) f_n[t - \phi(\tau)]) = 0.
\]

\noindent Compute each term:

\begin{enumerate}
    \item \textbf{Time derivative} \((\partial_t^2)\):
    \[
    \frac{\partial \psi}{\partial t} = \sum_{n=0}^\infty a_n (\tau) \frac{\partial}{\partial t} f_n [t - \phi (\tau)] = \sum_{n=0}^\infty a_n (\tau) f_n' [t - \phi (\tau)]
    \]

     \[
    \frac{\partial^2 \psi}{\partial t^2} = \sum_{n=0}^\infty a_n (\tau) f_n'' [t - \phi (\tau)]
    \]
     
    From the recursion \(f_n' = f_{n-1}\), we have \(f_n'' = f_{n-2}\). We can compute the other wavelets either by differentiation (to get lower indices) or integration (to get higher indices).

    \item \textbf{Spatial derivative} \((\partial_\tau^2)\):
    \[
    \frac{\partial \psi}{\partial \tau} = \sum_{n=0}^\infty \left( \frac{da_n}{d\tau} f_n[t - \phi(\tau)] + a_n(\tau) \frac{\partial}{\partial \tau} f_n[t - \phi(\tau)] \right)
    \]
    
    Since $\frac{\partial}{\partial \tau} f_n[t - \phi(\tau)] = -f'_n[t - \phi(\tau)] \frac{d\phi}{d\tau}$, this becomes
    
    \[
    \frac{\partial \psi}{\partial \tau} = \sum_{n=0}^\infty \left( \frac{da_n}{d\tau} f_n[t - \phi (\tau)] - a_n \frac{d\phi}{d\tau} f'_n[t - \phi (\tau)] \right)
    \]

    \begin{align*}
    \frac{\partial^2 \psi}{\partial \tau^2} = \sum_{n=0}^\infty \bigg( 
    a_n'' f_n[t - \phi]
    - 2 a_n' \phi'\, f'_n[t - \phi] \\
    + \; a_n (\phi')^2 f''_n[t - \phi] - a_n \phi''\, f'_n[t - \phi]
    \bigg)
    \end{align*}

    \noindent where primes denote $d/d\tau$ for $a_n$ and $\phi$, and $d/dz$ for $f_n$.

    \item \textbf{Scattering potential term}:
    \[
    \sum_{n=0}^{\infty} \left[ q(\tau) a_n(\tau) f_n[t - \phi(\tau)]\right]
    \]
\end{enumerate}

\noindent Combine all terms into the Schr\"{o}dinger equation:
\begin{align*}
\sum_{n=0}^{\infty} \Bigg[ a_n f_n'' &- \Bigg( a_n'' f_n - 2 a_n' \phi'\, f_n' \\
&\quad + a_n (\phi')^2 f_n'' - a_n \phi''\, f_n' \Bigg) + q\,a_n f_n \Bigg] = 0
\end{align*}

\noindent Using $f_n' = f_{n-1}$ and $f_n'' = f_{n-2}$,
\begin{align*}
    \sum_{n=0}^{\infty} \Bigg[ a_n f_{n-2} &- a_n'' f_n + 2a_n' \phi' f_{n-1} - a_n (\phi')^2 f_{n-2} \\
    &\quad + a_n \phi'' f_{n-1} + q a_n f_n \Bigg] = 0
\end{align*}

\noindent Grouping terms with the same wavelet index,
\begin{align*}
    \sum_{n=0}^{\infty} \Bigg[ a_n\!\left[1 - (\phi')^2\right] f_{n-2} &+ \left( 2 a_n' \phi' + a_n \phi'' \right) f_{n-1} \\
    &\quad + \left( q\,a_n - a_n'' \right) f_n \Bigg] = 0
\end{align*}

\noindent The wavelets $f_m$ are linearly independent, so for the equation to hold for all \( t \) the total coefficient of each $f_m$ must vanish. Collecting the three contributions that multiply a given $f_m$ (from $n = m+2$, $n = m+1$, and $n = m$) gives

\[
a_{m+2}\!\left[1 - (\phi')^2\right] + \left[2a_{m+1}'\phi' + a_{m+1}\phi''\right] - a_m'' + q\,a_m = 0
\]

\noindent which unwinds into the following hierarchy, starting from the lowest order:

\subsubsection*{(a) The Eikonal Equation (lowest order)}
\[
a_0\!\left[1 - \left(\frac{d\phi}{d\tau}\right)^2\right] = 0
\quad \Longrightarrow \quad \frac{d\phi}{d\tau} = \pm 1
\]
\[
\Longrightarrow \quad \phi(\tau) = \pm\tau + C
\]

\noindent This describes waves propagating with unit speed in the traveltime coordinate: rightward ($+\tau$) and leftward ($-\tau$).

\subsubsection*{(b) The Leading Transport Equation (coefficient of $f_{n-1}$)}
\[
2\frac{da_0}{d\tau}\frac{d\phi}{d\tau} + a_0\frac{d^2\phi}{d\tau^2} = 0
\]

\noindent Since $\phi'' = 0$, this gives $da_0/d\tau = 0$, i.e.\ $a_0 = \text{const} = 1$ (normalisation).

\subsubsection*{(c) The Higher-Order Transport Equation (coefficient of $f_n$)}
\[
2\frac{da_n}{d\tau}\phi' + a_n\phi'' - \frac{d^2 a_{n-1}}{d\tau^2} + q(\tau)a_{n-1} = 0.
\]

For $n=1$, with $a_0 = 1$ and $\phi' = +1$:

\begin{equation}
\frac{da_1}{d\tau} = -\frac{q(\tau)}{2},
\quad\Longrightarrow\quad
a_1(\tau) = -\frac{1}{2}\int_0^{\tau} q(\tau')\,d\tau' \label{eq:a1integral}
\end{equation}

\noindent \textbf{Summary of the wavefield for a rightward-propagating $\delta$-pulse:}
\begin{align}
\psi(\tau, t) = \delta(t - \tau) + a_1(\tau)\,H(t - \tau) \, + \, \nonumber \\  \sum_{n=2}^{\infty} a_n(\tau)\frac{(t - \tau)^{n-1}}{(n-1)!}\,H(t-\tau)
\label{eq:wavefieldsol}
\end{align}

\noindent where the $\delta$-pulse propagates undistorted, and the Heaviside and polynomial tails encode the accumulated interaction with the scattering potential. The expansion, the eikonal and transport equations, and the coefficient $a_1$ follow Burridge \cite{burridge1980gelfand} term for term.

One condition on this construction deserves emphasis, because the media of interest in reflection seismology routinely violate it. The expansion and the recovery relation \eqref{eq:qrecovery} both assume the impedance $\eta$ is smooth. Where $\eta$ has jump discontinuities, which is exactly the piecewise-constant layered model, the leading coefficient $a_0$ is no longer constant, the leading $\delta$-pulse acquires a discontinuous multiplier that is in general not unity, and the kernel is no longer bounded but carries singularities of its own. Even the weaker case of continuous $\eta$ with a discontinuous derivative is enough to invalidate equation \eqref{eq:qrecovery}, and interpreting the derivative distributionally does not rescue it \cite{burridge1980gelfand}. This is the technical reason the layer-stripping schemes below, rather than differentiation along the diagonal, are the appropriate tool for piecewise-constant media.

\subsection{Step 3: The Annihilator Wavefield and the 1D Marchenko Equation}
\label{sec:step3_1D}

The goal is to recover $q(\tau)$ from the reflection response $R(t)$ measured at $\tau = 0$. To achieve this, we construct a special wavefield called the \emph{\textbf{annihilator}} that eliminates the scattered response, leading to a linear integral equation, the 1D Marchenko equation, whose solution yields $q(\tau)$.

\subsubsection*{The Fundamental Solution and Scattered Wavefield}

For an incident $\delta$-pulse from the left, the full wavefield is:
\begin{equation}
\psi_f(\tau, t) = \delta(\tau - t) + R(\tau, t)
\label{eq:fundsoln}
\end{equation}

where $R(\tau, t)$ is the scattered (reflected) component. For $\tau < 0$ (the region to the left of the medium), this simplifies to

\begin{equation}
\psi_f(\tau, t) = \delta(\tau - t) + R(t + \tau)
\label{eq:fundsoln2}
\end{equation}

and the reflection data $R(t)$ is recorded at the surface $\tau_0 = x_0 =  0$.

\subsubsection*{The Annihilator Wavefield}

We introduce a second solution to the Schr\"{o}dinger equation, the \emph{annihilator wavefield} \cite{fichtner2012full}:

\begin{equation}
\psi_a(\tau, t) = \delta(t - \tau) + A(\tau, t), \label{eq:annihilator}
\end{equation}

where $A(\tau, t)$ is the non-delta component, a smooth kernel whose support is confined to $|t| < \tau$. The structure is the one Wapenaar et al.\ later named the focusing function: a leading delta followed by a coda, the coda being shaped so that as it passes through the scattering region it cancels every forward-going pulse the leading delta generates, leaving a bare delta beyond \cite{magalhaes2019comparison}. The annihilator is designed so that $\psi_a$ focuses to a delta function at the target depth $\tau$ and has no energy scattered back to the surface for $t > \tau$. Because $\psi_a$ is a transformation-operator representation of the causal fundamental solution (Appendix~\ref{app:1D}), the kernel evaluated on the diagonal $t = \tau$, the trailing edge of its support, equals the leading amplitude coefficient of the ray expansion \eqref{eq:a1integral}:
\begin{equation}
A(\tau, \tau) = a_1(\tau) = -\frac{1}{2}\int_0^{\tau} q(\tau')\,d\tau'
\end{equation}

Differentiating with respect to $\tau$:

\begin{equation}
q(\tau) = -2\frac{d}{d\tau}A(\tau, \tau) \label{eq:qrecovery}
\end{equation}

Once $A(\tau, t)$ is known, the scattering potential is recovered by differentiation along the diagonal.

\subsubsection*{Derivation of the GLM Equation}

The Wronskian of two independent solutions to the Schr\"{o}dinger equation is constant. Using $\psi_f$ and $\psi_a$:
\[
W[\psi_f, \psi_a] = \psi_f \frac{\partial \psi_a}{\partial \tau} - \psi_a \frac{\partial \psi_f}{\partial \tau} = \text{const}
\]

The inner product between $\psi_f(\tau, s)$ and the kernel $A(\tau, t)$, combined with the boundary conditions at $\tau = 0$, yields (see Appendix~\ref{app:1D} for the full derivation):
\begin{equation}
A(\tau, t) + R(t + \tau) + \int_{-\tau}^{\tau} A(\tau, s)\, R(s + t)\, ds = 0, \quad |t| < \tau
\label{eq:GLM}
\end{equation}

This is the \textbf{Gelfand--Levitan--Marchenko (GLM) equation}: a Fredholm integral equation of the second kind for the unknown kernel $A(\tau, t)$, with the measured reflection response $R(t)$ as input. Once $A$ is found, the medium is reconstructed via equation \eqref{eq:qrecovery}. Three families of solution method are in common use.

\textbf{Layer stripping:} For piecewise-constant media, the GLM equation can be solved exactly by a sequence of layer-stripping operations, equivalent to the Schur recursions for a discrete transmission line. At each traveltime step $\Delta\tau$, the reflector amplitude is extracted and the medium is updated. The recursion costs $O(N^2)$ in the number of layers \cite{bruckstein1987inverse}.

\textbf{Iterative Neumann series:} Equation \eqref{eq:GLM} can be solved by successive substitution,
\[
A^{(0)}(\tau, t) = -R(t + \tau)
\]
\[
A^{(k+1)}(\tau, t) = -R(t + \tau) - \int_{-\tau}^{\tau} A^{(k)}(\tau, s)\, R(s + t)\, ds
\]

with the $k$-th iterate accounting for internal multiples up to order $k$. Convergence requires the truncated operator to be a contraction, which holds when scattering in the interval $[0,\tau]$ is not too strong; Appendix~\ref{app:1D} states the condition.

\textbf{Direct inversion method:} Discretising $A$ and $R$ on $N$ time samples turns equation \eqref{eq:GLM} into a dense linear system for each $\tau$, solved directly at $O(N^3)$ cost \cite{burridge1976some}. This is the method of choice when the Neumann series converges slowly or not at all, and it is the 1D counterpart of the least-squares Marchenko schemes of Section~\ref{sec:lsmi}.

\subsection{Reconstruction of Medium Properties}

Once $A(\tau, t)$ has been determined, the scattering potential follows from equation \eqref{eq:qrecovery}, and the impedance profile is recovered by solving the second-order ordinary differential equation implied by the definition of $q$,
\begin{equation}
\frac{d^2\eta}{d\tau^2} = q(\tau)\,\eta(\tau),
\end{equation}
subject to the known surface values $\eta(0)$ and $\eta'(0)$.

The physical wave speed $c(x)$ and density $\rho(x)$ are not individually recoverable from pressure data alone. Only their product, the impedance $\eta^2 = \rho c$, is uniquely determined as a function of traveltime; separating the two requires particle velocity measurements. As an example, for a single interface at one-way traveltime $\tau_0$ separating two homogeneous half-spaces, with reflection coefficient $r$, the reflection response is a single spike at the two-way traveltime,
\begin{equation}
R(t) = r\,\delta(t - 2\tau_0),
\end{equation}
and solving the GLM equation recovers the impedance contrast at $\tau_0$. The sign convention here needs care, because $R$ is the reflection response of the scaled field $\psi$, not of the displacement or the pressure. Imposing continuity of displacement and of traction across the interface on $\psi = \eta u$ gives
\begin{equation}
r = \frac{\eta_1^2 - \eta_2^2}{\eta_1^2 + \eta_2^2} = \frac{Z_1 - Z_2}{Z_1 + Z_2},
\qquad \frac{Z_2}{Z_1} = \frac{1-r}{1+r}
\end{equation}

\noindent with $Z = \rho c = \eta^2$. This is minus the familiar normal-incidence pressure reflection coefficient, the sign difference being the usual one between displacement- and pressure-normalised conventions. Adopting $\eta = (\rho c)^{-1/2}$, which is the natural scaling if the derivation is run on pressure rather than displacement, returns the more familiar $r = (Z_2 - Z_1)/(Z_2 + Z_1)$. Either convention is workable; mixing them inverts every recovered contrast.

\subsection{Relationship to Classical Inversion Methods}

The GLM approach is related to, but distinct from, other classical 1D inversion methods.

\noindent \textbf{Inverse scattering series (ISS):} The ISS \cite{weglein1997inverse} expresses the potential as a series in the data, each term involving increasing powers of the reflection response. The GLM equation resums this series into a single integral equation, so that solving it accounts for the whole series at once rather than term by term. In practice the ISS is truncated for efficiency, which costs it the exactness the GLM solution retains.

\noindent \textbf{Recursive impedance inversion:} The conventional approach uses normal-incidence reflectivity $r(t)$ obtained by deconvolution to compute $\ln\eta$ by integration, since $r(t) = \tfrac{1}{2}\,d(\ln\eta^2)/dt$ for weak contrasts. This is the Born approximation of the GLM solution and it ignores all multiple reflections. The GLM equation accounts for all orders of internal multiples automatically.

\noindent \textbf{WKBJ (Wentzel--Kramers--Brillouin--Jeffreys)  approximation:} For smooth media, the WKBJ solution replaces the full Green's function with an amplitude-corrected plane wave. The GLM equation reduces to the WKBJ result in the high-frequency, smooth-medium limit \cite{bremmer1951wkb}.

\section{Road to Multidimensions: The Conceptual Bridge}
\label{sec:bridge}

\subsection{The Limitations of Strict 1D Traveltime Transformation}

The transformation from the wave equation to the Schr\"{o}dinger equation in Section~\ref{sec:1D} relied on the existence of a unique travel-time coordinate $\tau(x)$ parameterising all wave paths. In one spatial dimension this is trivially satisfied: waves can only travel left or right, and $\tau(x) = \int_0^x dx'/c(x')$ is unique. In two or three dimensions the situation differs fundamentally:

\begin{enumerate}
    \item \textbf{Non-uniqueness of traveltime paths:} Multiple ray paths connect a surface source to a subsurface point, each with a different traveltime. No single $\tau(\xv)$ can describe all of them simultaneously.
    \item \textbf{Directional dependence:} The travel time from a surface source to a subsurface point depends on the source position; there is no analogue of the single depth-variable $\tau$.
    \item \textbf{Wavefront curvature (Ray bending):} Wavefronts are surfaces, not points; their curvature introduces geometric spreading without a 1D counterpart.
    \item \textbf{Multi-pathing and caustics:} Strong lateral velocity contrasts cause ray paths to cross, generating multiple arrivals that a single-valued traveltime function cannot capture.
\end{enumerate}

Taken together, these four obstructions leave no single-valued $\tau(\xv)$ to play the role that the traveltime coordinate plays in one dimension, so the reduction to a Schr\"{o}dinger equation has no multidimensional counterpart. This is why the multidimensional Marchenko equations are derived instead from wave reciprocity theorems, which require no such coordinate and never pass through a Schr\"{o}dinger form \cite{wapenaar2014marchenko, slob2014seismic}. Table~\ref{tab:comp} compares the 1D and multidimensional formulations.


\begin{table*}[!t]
\centering
\caption{Comparison between 1D and 2D/3D Marchenko methods}
\label{tab:comp}
\begin{tabular}{|l|p{4.5cm}|p{6.5cm}|}
\hline
\textbf{Aspect} & \textbf{1D Marchenko} & \textbf{2D/3D Marchenko} \\ \hline
Route to the equations & Reduction to a Schr\"{o}dinger equation via the traveltime coordinate $\tau(x) = \int dx/c(x)$ & No such reduction exists; the equations follow from convolution- and correlation-type reciprocity theorems \\ \hline
Focal point specified by & One-way traveltime $\tau$, not depth & Position $\xA$ in space, given a background model for $t_d$ \\ \hline
Input data & Single reflection trace & Reflection response over the full acquisition surface, deconvolved and deghosted \\ \hline
Unknowns & One kernel $A(\tau,t)$ & Two coupled focusing functions $f_1^{\pm}$ \\ \hline
Time-separation condition & Exact & Approximate: holds for layered media with moderately curved interfaces at finite offset; may fail otherwise \cite{wapenaar2014multicomponent} \\ \hline
Solution method & Layer stripping, Neumann series, or direct matrix inversion & Neumann iteration, or least-squares inversion where the series does not contract \\ \hline
Internal multiples & Accounted for exactly by the converged solution & Accounted for to the order reached by the iteration \\ \hline
Cost & Negligible & Dominated by the multidimensional convolution, $O(N_s^4 N_t)$ in 3D \\ \hline
\end{tabular}
\end{table*}

The conceptual step toward the multidimensional Marchenko method came through seismic interferometry. Equation~\eqref{eq:interferometry} shows that the Green's function between any two points can be retrieved by cross-correlating records at those points over a sufficiently complete source distribution. In exploration seismology, sources are at the surface; no natural or artificial sources exist at depth. The virtual-source method of Bakulin and Calvert \cite{bakulin2006virtual} addressed this with surface shots and downhole receivers placed beneath the most troublesome part of the overburden. Time-reversal logic applied to that geometry synthesises a downward-continued dataset whose sources sit at the geophone locations, and it does so with \emph{no knowledge of the velocity model between the shots and the receivers} \cite{bakulin2006virtual}. That last property is the one the Marchenko method later inherited and extended.

Two limitations kept the approach from being general. It requires a physical receiver at the position where the virtual source is wanted, so it needs a borehole. And because the illumination is one-sided, correlation leaves spurious arrivals in the result. Wapenaar et al.\ \cite{wapenaar2011virtual} showed that replacing the correlation by multidimensional deconvolution, which solves a least-squares problem for the virtual-source response instead of simply correlating, suppresses those artefacts.

\subsection{Broggini and Snieder (2012): Focusing in a 1D Medium}

Broggini et al.\ \cite{broggini2012focusing}, building on Rose's autofocusing \cite{rose2002single}, asked whether a focusing wavefield could be constructed, one that collapses to a spatial delta function at a chosen depth and time, using only the reflection response at the surface and no receiver at depth. For a 1D medium the answer is yes, and it holds even when velocity and density vary independently, so the construction does not presuppose a known impedance relationship. The companion tutorial of Broggini and Snieder \cite{broggini2012connection} places this focusing result alongside inverse scattering, Green's function reconstruction, and imaging, and shows that the equations governing all four share a functional form. The focusing condition requires the annihilator wavefield $\psi_a$ defined in equation \eqref{eq:annihilator}: injected from $\tau = 0$, it focuses exactly at $\tau = \tau_0$, with no energy propagating downward past the focal point. This focusing wavefield cannot be generated by a physical source, but it can be computed from the reflection data alone via the GLM equation.

Broggini and Snieder further showed that this focusing property is equivalent to a connection between time-reversal focusing (which requires both transmitted and reflected waves) and single-sided interferometry. That single-sided reflection data contain enough information to construct a focusing wavefield at any depth is the conceptual foundation of the multidimensional Marchenko method.

\subsection{Extension to Multiple Dimensions: The Role of Reciprocity}

The extension to multiple dimensions by Wapenaar et al.\ \cite{wapenaar2014marchenko} and Slob et al.\ \cite{slob2014seismic} proceeds not by generalising the Schr\"{o}dinger equation but by using wave reciprocity to derive representations for one-way Green's functions. The central observation is that two coupled integral equations, analogous to the two needed for the 1D GLM system, can be derived from the convolution- and correlation-type reciprocity theorems by choosing one state to be the physical Green's function and the other to be the focusing wavefield. The resulting system, the multidimensional Marchenko equations, reduces exactly to the GLM equation in one spatial dimension \cite{wapenaar2014marchenko}, and gives a practical algorithm for Green's function retrieval in 2D and 3D.

\section{The Multidimensional Marchenko Method}
\label{sec:multidim}
The standard configuration for the multidimensional Marchenko method (Figure~\ref{fig:config}) consists of:

\begin{figure}[htpb]
    \centering
    \includegraphics[width=1.0\linewidth]{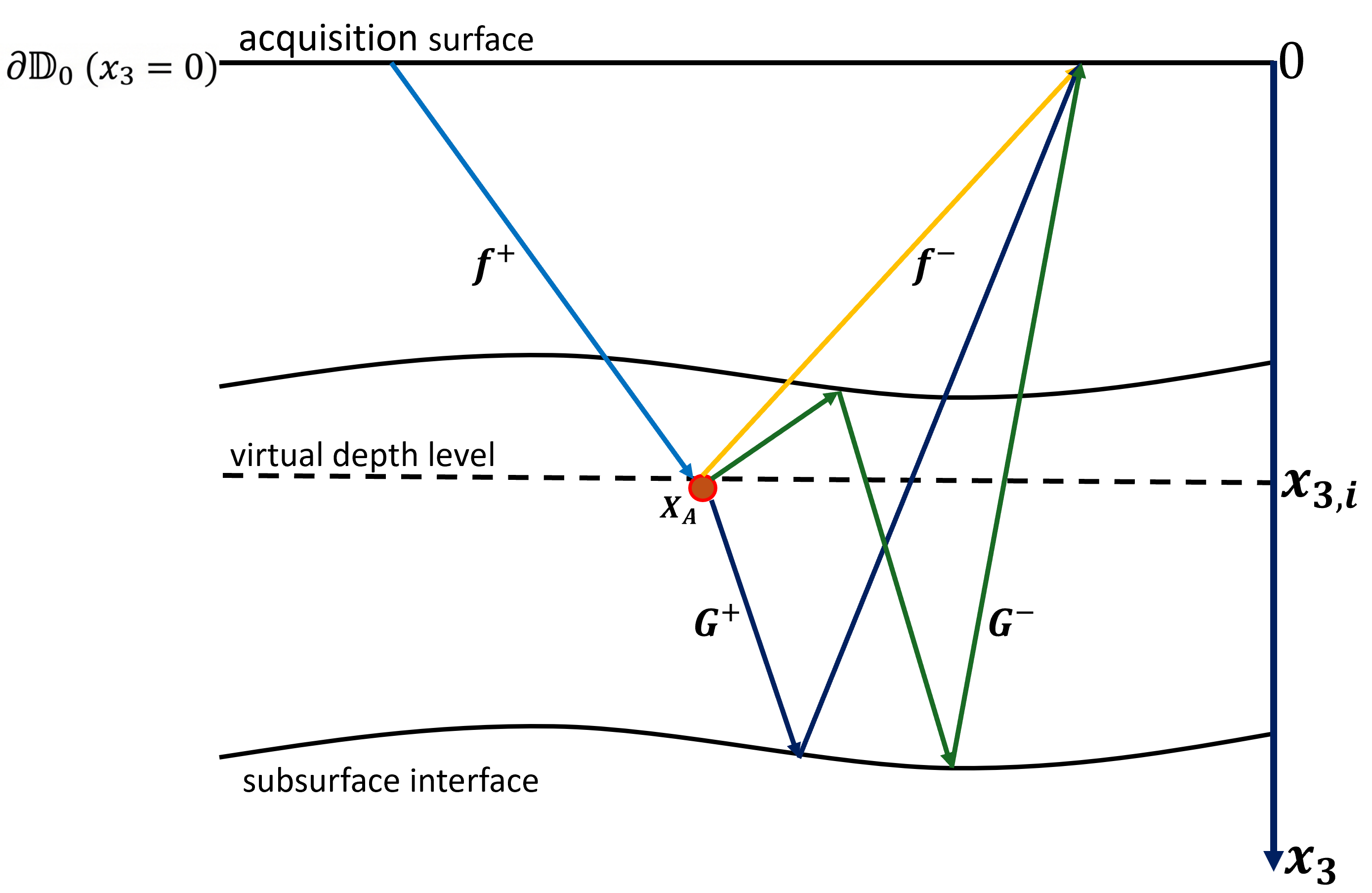}
        
    \caption{Standard Marchenko configuration. Modified from \cite{lomas2019introduction}. Sources and receivers are distributed along the acquisition surface $\partial\mathbb{D}_0$ at $x_3 = 0$; the focal point $\mathbf{x}_A$, marked by the red dot, lies on the virtual depth level $\partial\mathbb{D}_i$ at $x_3 = x_{3,i}$ (dashed).
    Two families of wavefield are superimposed, and they belong to \emph{different} media. The focusing functions are defined in the truncated medium, which matches the actual medium above $\partial\mathbb{D}_i$ and is reflection-free below it: the downgoing $f^+$ (light blue) is injected from the surface and focuses to a band-limited spatial delta function at $\mathbf{x}_A$, and the upgoing $f^-$ (gold) is its reflected response, carrying only the scattering from above $\partial\mathbb{D}_i$.
    The Green's functions are defined in the actual medium and describe the response to a virtual source at $\mathbf{x}_A$, which is why they interact with the deeper interface as well. Their superscript records the direction in which energy \emph{leaves} that source: $G^+$ (dark blue) radiates downward and reflects once from the interface below, while $G^-$ (green) radiates upward, reflects from the shallower interface, and reaches the deeper one before returning. Both arrive upgoing at $\partial\mathbb{D}_0$, since the acquisition surface is treated as transparent, and both are retrieved from the converged focusing functions with no sensor at depth.
    Ray paths are schematic. The figure omits the subscript~1 carried by $f_1^{\pm}$ in the text, and writes the focal point as $X_A$.}
    
    \label{fig:config}
\end{figure}

\begin{itemize}
    \item An \emph{acquisition surface} $\dD_0$ at $x_3 = 0$, where sources and receivers are located.
    \item A \emph{virtual depth level} $\dD_i$ at $x_3 = x_{3,i}$. The \emph{truncated medium}, also called the reference medium, is identical to the physical medium between $\dD_0$ and $\dD_i$ and homogeneous on both sides of that slab: reflection-free below $\dD_i$, and reflection-free above $\dD_0$ so that the acquisition surface is transparent \cite{wapenaar2016homogeneous}. It is an auxiliary construction used only to define the focusing condition; it is not a medium with a reflecting boundary at $\dD_i$.
    \item A \emph{focal point} $\xA = (\mathbf{x}_{A,H}, x_{3,i})$ on $\dD_i$.
    \item The slab $\mathbb{D}^+$ bounded above by $\dD_0$ and below by $\dD_i$.
\end{itemize}

The reflection response $R(\xH, \xH', t)$ (where $\xH$ and $\xH'$ are horizontal coordinates on $\dD_0$) is the upgoing wavefield at $\xH$ due to a downgoing point source at $\xH'$. Together with an estimate of the direct arrival, this is the entire input. It is worth being precise about the division of labour, because it is what distinguishes the method from model-driven alternatives: the macro model is used only to estimate the first arrival, and the information needed to deal with the internal multiples comes entirely from the measured reflection response \cite{wapenaar2021marchenko}. The input requirement is therefore no greater than that of standard redatuming and imaging of primaries. Before use, $R$ must be preprocessed to:

\begin{enumerate}
    \item Deconvolve for the source wavelet, so that $R$ approximates the impulse response. This step is not optional: the derivation below assumes an impulsive source, and residual wavelet effects propagate directly into the focusing functions \cite{thorbecke2013green, thorbecke2017implementation}.
    \item Remove the direct wave (the source wavelet in a homogeneous half-space).
    \item Apply source and receiver deghosting to remove free-surface effects. In the formulation that incorporates free-surface multiples, this step is omitted \cite{davydenko2017full, wapenaar2021marchenko}.
    \item Interpolate to a regular, gap-free source--receiver grid and compensate for 3D-to-2D amplitude effects where a 2D scheme is applied to 3D data.
\end{enumerate}

\subsection{Focusing Functions}

The one-way \emph{downgoing focusing function} $f_1^+(\xH, \xA, t)$ is the downgoing wavefield at $\dD_0$ that, when injected into the truncated medium, focuses to a band-limited spatial delta function at $\dD_i$ at $t = 0$. Writing $T(\xB, \xH, t)$ for the downgoing transmission response of the truncated medium between $\dD_0$ and $\dD_i$, the focusing condition is
\begin{align}
    \int_{\dD_0} T(\xB, \xH', t) \ast f_1^+(\xH', \xA, t)\, d\mathbf{x}_{H}' \nonumber \\
    = \delta(\mathbf{x}_{B,H} - \mathbf{x}_{A,H})\,\delta(t), \quad \xB \in \dD_i \label{eq:focusdef}
\end{align}

\noindent The delta is two-dimensional because both $\xA$ and $\xB$ lie on the depth level $\dD_i$.

The \emph{upgoing focusing function} $f_1^-(\xH, \xA, t)$ is the reflected response of the truncated medium to $f_1^+$:

\begin{equation}
f_1^-(\xH, \xA, t) = \int_{\dD_0} R_t(\xH, \xH', t) \ast f_1^+(\xH', \xA, t)\, d\mathbf{x}_{H}' \label{eq:f1minus}
\end{equation}

\noindent where $R_t$ is the reflection response \emph{of the truncated medium}, not the measured response $R$ of the actual medium. The two differ by everything reflected from below $\dD_i$, and keeping them distinct is what makes equation \eqref{eq:rep1} below a non-trivial statement rather than a restatement of \eqref{eq:f1minus}. Note also that $R_t$ is not available as data; it never has to be, because the Marchenko equations are driven by $R$ alone.

Both $f_1^+$ and $f_1^-$ are non-causal wavefields: they exist for $t < 0$ as well as $t > 0$, unlike the causal Green's functions. Their support in time is bounded by the direct-wave traveltime $t_d(\xH, \xA)$ from $\xA$ to $\xH$ through the background model:
\begin{align}
f_1^+(\xH, \xA, t) &= 0 \quad \text{for } |t| > t_d(\xH, \xA), \label{eq:f1plus_support}\\
f_1^-(\xH, \xA, t) &= 0 \quad \text{for } |t| > t_d(\xH, \xA) \label{eq:f1minus_support}
\end{align}

\noindent The support of $f_1^+$ is bounded on both sides: its direct arrival sits at $t = -t_d$, at the left edge of the window, and its coda occupies $-t_d < t < t_d$. The two-sided bound, not merely $t > t_d$, is what the time-window argument below relies on, and it is worth seeing where the left-hand bound comes from, because it is not obvious.

Since $f_1^+$ is the inverse of the transmission response of the truncated medium, its support follows from the structure of that response. The transmission coda is causal and minimum-phase, a classical result of O'Doherty and Anstey \cite{odoherty1971reflections}, and the inverse of a minimum-phase signal is itself causal and minimum-phase \cite{wapenaar2015causality}. The inverse therefore consists of the time-reversed direct arrival at $t = -t_d$ followed by a causal coda, so $f_1^+$ vanishes for $t < -t_d$. Nothing but minimum-phase gives this; the bound is a property of layered transmission, not a general wave-theoretic fact.

The Green's functions are bounded on the other side, and the two are not bounded identically. Both vanish before the direct arrival, but $G^{+}$ contains that arrival while $G^{-}$ does not, so the intervals differ at the single instant $t = t_d$: $G^{+}$ vanishes for $t < t_d$ and $G^{-}$ for $t \leq t_d$ \cite{wapenaar2015causality}. That one point is the only overlap between the Green's functions and the focusing functions, which is precisely what makes a time window sufficient to separate them.

These causality constraints, which follow from the finiteness of wave speed in the medium, are what allow the coupled Marchenko equations to separate the focusing functions from the Green's functions.

\subsection{Derivation of the Green's Function Representations}

The coupled relations between focusing functions and Green's functions are derived using the two reciprocity theorems of Section~\ref{sec:math} (full derivation in Appendix~\ref{app:2D3D}). Setting state A to the focusing function and state B to the physical Green's function, and integrating over the domain $\mathbb{D}^+$ bounded by $\dD_0$ above and $\dD_i$ below:

A note on the Green's function superscripts is needed first, because they are a common source of confusion. The acquisition surface $\dD_0$ is treated as transparent, so the field there is purely upgoing and there is no downgoing Green's function to speak of at the receiver. The superscript instead records the direction in which the wave \emph{leaves the virtual source} at $\xA$: $G^-$ for energy radiated upward and $G^+$ for energy radiated downward, both observed as upgoing at $\dD_0$. In the fuller notation of Wapenaar et al.\ \cite{wapenaar2014marchenko} and Brackenhoff et al.\ \cite{brackenhoff2022three} these are $G^{-,+}$ and $G^{-,-}$, the first superscript being the direction at the receiver. Both are causal.

Two integral operators appear. The first convolves the reflection response with a wavefield and integrates over the surface, an operation the implementation literature calls \emph{synthesis} \cite{thorbecke2017implementation}:
\begin{align}
(\mathcal{R}g)(\xH, \xA, t) = \int_{\dD_0}\!\int_{0}^{\infty} & R(\xH, \xH', t') \nonumber \\ &\times g(\xH', \xA, t - t')\, dt'\, d\xH' \label{eq:synthesis}
\end{align}

\noindent The second correlates instead of convolving, by time-reversing the reflection response:
\begin{align}
(\mathcal{R}^{\star}g)(\xH, \xA, t) = \int_{\dD_0}\!\int_{-\infty}^{0} & R(\xH, \xH', -t') \nonumber \\ &\times g(\xH', \xA, t - t')\, dt'\, d\xH' \label{eq:synthesiscorr}
\end{align}

From the \textbf{convolution-type} theorem:
\begin{align}
(\mathcal{R}f_1^+)(\xH, \xA, t) = f_1^-(\xH, \xA, t) + G^{+}(\xH, \xA, t) \label{eq:rep1}
\end{align}

From the \textbf{correlation-type} theorem:
\begin{align}
(\mathcal{R}^{\star}f_1^-)(\xH, \xA, t) = f_1^+(\xH, \xA, t) + G^{-}(\xH, \xA, -t) \label{eq:rep2}
\end{align}

Equations \eqref{eq:rep1}--\eqref{eq:rep2} each contain two unknowns. To extract the focusing functions alone, we apply the \emph{time-window operator}
\begin{align}
\Theta(\xH, \xA, t) = \theta\!\left(t + t_d - \varepsilon\right) \nonumber \\ - \; \theta\!\left(t - t_d + \varepsilon\right) \label{eq:window}
\end{align}

\noindent with $\theta$ the Heaviside step and $t_d = t_d(\xH, \xA)$, so that $\Theta$ passes $|t| < t_d - \varepsilon$ and mutes everything outside. Because $f_1^{\pm}$ are time-limited to that window (equations \eqref{eq:f1plus_support}--\eqref{eq:f1minus_support}) while $G^{\pm}$ arrive only at and after $t_d$, applying $\Theta$ removes the Green's function contributions. The window is symmetric in time, so it makes no difference whether it is applied to a wavefield or to its time reversal \cite{brackenhoff2022three}. Applying $\Theta$ to $f_1^+$ removes its direct arrival and leaves the coda, which is why $f_{1,d}^+$ reappears outside the window below.

The Coupled Multidimensional Marchenko Equations
\begin{align}
f_1^-(\xH, \xA, t) &= \Theta\!\left\{ (\mathcal{R}f_1^+)(\xH, \xA, t) \right\} \label{eq:marchenko1}
\end{align}
\begin{align}
f_1^+(\xH, \xA, t) &= f_{1,d}^+(\xH, \xA, t) \nonumber \\ &\quad + \Theta\!\left\{ (\mathcal{R}^{\star}f_1^-)(\xH, \xA, t) \right\} \label{eq:marchenko2}
\end{align}

where $f_{1,d}^+(\xH, \xA, t)$ is the initial estimate of the focusing function (Section~\ref{sec:initestimate}). Both equations carry a positive $\Theta$ term; the minus signs enter instead in equations \eqref{eq:Gm}--\eqref{eq:Gp}.

These two equations, together with the support conditions \eqref{eq:f1plus_support}--\eqref{eq:f1minus_support}, are the multidimensional Marchenko equations \cite{wapenaar2014marchenko, slob2014seismic}. They involve only the reflection response $R$ (measured data) and the focusing functions $f_1^{\pm}$ (unknowns). The Green's functions follow by substituting the converged focusing functions back into equations \eqref{eq:rep1}--\eqref{eq:rep2}.

\subsection{The Initial Estimate of the Focusing Function}
\label{sec:initestimate}

Exactly, the downgoing focusing function is the inverse of the transmission response of the truncated medium, $f_1^+ = T^{\,\mathrm{inv}}$, and splits into a direct part and a scattering coda, $f_1^+ = T_d^{\,\mathrm{inv}} + M^+$ \cite{wapenaar2014marchenko, thorbecke2017implementation}. Iterating equations \eqref{eq:marchenko1}--\eqref{eq:marchenko2} requires the direct part $f_{1,d}^+ = T_d^{\,\mathrm{inv}}$, which is a scaled delta at the negated direct traveltime:
\begin{equation}
f_{1,d}^+(\xH, \xA, t) = \frac{\delta\!\left(t + t_d(\xH, \xA)\right)}{A_d(\xH, \xA)} \label{eq:initestimate}
\end{equation}

where $T_d(\xH, \xA, t) = A_d(\xH, \xA)\,\delta(t - t_d(\xH, \xA))$ is the direct component of the transmission response, $t_d$ is the direct-wave traveltime from $\xA$ to $\xH$, and $A_d$ is the geometrical spreading factor.

In practice $T_d^{\,\mathrm{inv}}$ is not available, because inverting the transmission response requires the very medium the method is trying to avoid modelling. The standard substitution is the time-reversed direct arrival of the Green's function computed in the smooth background model \cite{thorbecke2017implementation}:
\begin{align}
f_{1,d}^+(\xH, \xA, t) \approx G_d(\xH, \xA, -t) \nonumber \\ = A_d(\xH, \xA)\,\delta\!\left(t + t_d(\xH, \xA)\right)
\end{align}

\noindent This replaces $1/A_d$ by $A_d$, so it introduces an overall scaling error and an offset-dependent amplitude error proportional to the transmission losses through the overburden \cite{thorbecke2017implementation}. The kinematics are unaffected, which is why the approximation is acceptable for structural imaging and why true-amplitude work (Section~\ref{sec:imaging}) needs a separate illumination correction rather than a better initial estimate.

The traveltime $t_d$ is computed from the smooth background velocity model by solving the eikonal equation with fast-marching methods \cite{sethian1996fast, sethian1999fast}, fast sweeping methods \cite{zhao2005fast, wong2016fast}, or other iterative methods \cite{mokry2016iterative}, or by picking the first arrival from a finite-difference simulation. Setting $A_d = 1$ gives a purely kinematic initialisation, which is common.

The initial estimate is the most practically sensitive input to the scheme, and errors in it propagate into everything downstream. Chen et al.\ \cite{chen2020marchenko} address this with a self-adaptive traveltime update built on the principle of equal traveltime: the time shift a velocity error imposes on the direct wave is measured, the direct wave is corrected by that shift, and the Green's functions are reconstructed from the corrected estimate. On synthetic tests the scheme restored mispositioned structures to their true depths, which is the specific damage an erroneous $t_d$ inflicts.

Van der Neut et al.\ \cite{van2015practical} catalogued what else the iteration is sensitive to, and the list is a useful corrective to the impression that the traveltime estimate is the only fragile ingredient: accurate knowledge of the source signature, which is not always available, along with source and receiver ghosts, coupling effects, attenuation, and noise. They also examined adding the series terms adaptively rather than exactly, using only the first two terms, and found it useful for internal multiple suppression on synthetic data with severe event interference and on field data. Their stated condition on that procedure is the one that recurs throughout the field-data literature: it succeeds when the internal multiples do not interfere with the primary reflections, and it has limitations in more complex media.

\subsection{Iterative Solution of the Marchenko Equations}
\label{sec:iterative}

The coupled equations \eqref{eq:marchenko1}--\eqref{eq:marchenko2} are solved by a simple alternating Neumann iteration \cite{wapenaar2014marchenko}:

\begin{enumerate}
    \item \textbf{Initialisation:} $f_1^{+,0}(\xH, \xA, t) = f_{1,d}^+(\xH, \xA, t)$
    
    \item \textbf{Upgoing update:}
    \begin{align}
        f_1^{-,k+1}(\xH, \xA, t) = \Theta\!\left\{ (\mathcal{R}f_1^{+,k})(\xH, \xA, t) \right\} \label{eq:iter1}
    \end{align}

    \item \textbf{Downgoing update:}
    \begin{align}
        f_1^{+,k+1}(\xH, \xA, t) = f_{1,d}^+(\xH, \xA, t) \nonumber \\ + \; \Theta\!\left\{ (\mathcal{R}^{\star}f_1^{-,k+1})(\xH, \xA, t) \right\} \label{eq:iter2}
    \end{align}
    
    \item \textbf{Convergence check:} If $\|f_1^{+,k+1} - f_1^{+,k}\|_2 / \|f_1^{+,k}\|_2 < \epsilon$, stop; otherwise, return to step 2.
\end{enumerate}

The operator $\Theta$ in steps \eqref{eq:iter1}--\eqref{eq:iter2} is applied using the traveltime $t_d(\xH, \xA)$ from the background model: it passes the window $|t| < t_d(\xH, \xA) - \varepsilon$ and mutes everything outside it. The margin $\varepsilon$ is a small positive constant chosen to exclude the wavelet in the direct arrival, and in implementations the window edge is tapered rather than sharp, to suppress high-frequency truncation artefacts \cite{thorbecke2017implementation}. It is distinct from the convergence tolerance $\epsilon$. Applying this mute is not an optimisation but a requirement: without it the scheme does not separate the focusing functions from the Green's functions and the method is simply incorrect \cite{thorbecke2017implementation}.

Physically, each iteration accounts for one additional order of internal multiple scattering. At iteration $k$, the focusing function $f_1^{+,k}$ accounts for all multiples involving up to $k$ reflection legs in the overburden above the focal depth. In the synthetic examples of Thorbecke et al.\ \cite{thorbecke2013green}, 10 to 15 iterations were sufficient; Koehne et al.\ \cite{koehne2021multi} used 8 in their benchmark models. Strong reflectors such as salt bodies may need more iterations or damping \cite{jia2017subsalt}, and where the Neumann series does not contract at all the least-squares formulation of Section~\ref{sec:lsmi} replaces it.

Once the iteration has converged, the one-way Green's functions are retrieved from equations \eqref{eq:rep1}--\eqref{eq:rep2}:
\begin{align}
G^{+}(\xH, \xA, t) &= (\mathcal{R}f_1^+)(\xH, \xA, t) - f_1^-(\xH, \xA, t) \label{eq:Gm}
\end{align}
\begin{align}
G^{-}(\xH, \xA, -t) &= (\mathcal{R}^{\star}f_1^-)(\xH, \xA, t) - f_1^+(\xH, \xA, t) \label{eq:Gp}
\end{align}

\noindent Each Green's function is exactly what the time window rejected. Comparing \eqref{eq:Gm} with \eqref{eq:marchenko1} gives $G^{+} = (1 - \Theta)\{\mathcal{R}f_1^+\}$, and \eqref{eq:Gp} with \eqref{eq:marchenko2} gives $G^{-}(-t) = (1 - \Theta)\{\mathcal{R}^{\star}f_1^-\} - f_{1,d}^+$. No further synthesis is needed once the iteration has converged, which is also the cheapest internal check that the signs have been implemented consistently \cite{thorbecke2017implementation, brackenhoff2022three}.

The two-way Green's function at the surface is:
\begin{equation}
G = G^+ + G^-
\end{equation}

\noindent all three evaluated at $(\xH, \xA, t)$.

These Green's functions include the full internal multiple content. They describe the response of a virtual source at $\xA$ as measured at the surface $\dD_0$, including all orders of internal reverberations above $\dD_i$. All equations above are written in 2D notation (single horizontal coordinate $x_H$) for clarity. In three dimensions, $\xH = (x_1, x_2)$ and all surface integrals extend over two dimensions. The coupled 3D Marchenko equations take the same form as equations \eqref{eq:marchenko1}--\eqref{eq:marchenko2}, with $dx_H \rightarrow d^2\mathbf{x}_H$ and coordinates in 3D \cite{wapenaar2014marchenko, brackenhoff2022three}. The practical difference is computational: for $N$ surface positions per horizontal dimension, the 3D problem involves $O(N^4)$ data compared with $O(N^2)$ in 2D.

The three-dimensional Marchenko equations require full-azimuth 3D reflection data, now increasingly available from ocean-bottom-node (OBN) and wide-azimuth towed-streamer surveys \cite{brackenhoff2022three}. Brackenhoff et al.\ \cite{brackenhoff2022three} implemented the 3D case using plane-wave injection to reduce the computational load, demonstrating 3D Marchenko redatuming, imaging, and homogeneous Green's function retrieval at field scale.

\section{Green's Function Retrieval}
\label{sec:greens}


Equations \eqref{eq:Gm}--\eqref{eq:Gp} give single-sided representations of the one-way Green's functions: both $G^+$ and $G^-$ at $\dD_0$ are expressed entirely in terms of the reflection response $R$ (measured at $\dD_0$) and the focusing functions $f_1^{\pm}$ (computed from $R$ via the Marchenko equations). No information about the medium below $\dD_i$ or about the transmission response is needed \cite{wapenaar2014marchenko}. The retrieved $G^-(\xH, \xA, t)$ represents the pressure response at the surface due to an impulsive source at $\xA$: a virtual shot gather with a virtual source located at subsurface depth. Together with $G^+(\xH, \xA, -t)$, the time-reversed downgoing response, interpretable as a virtual receiver gather, these two Green's functions provide everything needed for subsurface imaging without any subsurface sensor \cite{wapenaar2014marchenko}.

\subsection{Homogeneous Green's Function Retrieval}

The \emph{homogeneous Green's function} $G_h(\xA, \xB, t)$ between two subsurface points is the Green's function superposed with its own time reversal,
\begin{equation}
G_h(\xA, \xB, t) = G(\xA, \xB, t) + G(\xA, \xB, -t) \label{eq:homgreendef}
\end{equation}

\noindent equivalently $2\Re\{\hat G\}$ in the frequency domain. The superposition cancels the source term, so $G_h$ solves the wave equation without a right-hand side, and in doing so removes the singularity at the source position \cite{wapenaar2016homogeneous}.

Classically, $G_h(\xA, \xB, \omega)$ is expressed as a \emph{closed} boundary integral of $G$ and $G^*$ over a surface enclosing both points. That representation is exact and underlies seismic interferometry and holographic imaging, but measurements are rarely available on a closed boundary, which is what limits its use. Wapenaar et al.\ \cite{wapenaar2016homogeneous} replaced the closed boundary with an open one at the acquisition surface by introducing the focusing function, obtaining
\begin{align}
G_h(\xA, \xB, \omega) = \int_{\dD_0} \frac{2}{\omega\rho(\xv)} \Big( & \Im\{\hat f_1\}\, \partial_3 \hat G_h \nonumber \\ - \; & \Im\{\partial_3 \hat f_1\}\, \hat G_h \Big)\, d^2\xv \label{eq:homgreen}
\end{align}

\noindent where the focusing function $\hat f_1 = \hat f_1(\xv, \xA, \omega)$ and the homogeneous Green's function $\hat G_h = \hat G_h(\xv, \xB, \omega)$ are both evaluated at $\xv$ on $\dD_0$, $\Im$ denotes the imaginary part, and $\partial_3$ the derivative normal to the acquisition surface. Here $f_1 = f_1^+ + f_1^-$ is the two-way focusing function for focal point $\xA$. Only its imaginary part enters, which is what makes the contributions from the lower boundary and the cylindrical closing surface drop out.

One approximation is built into equation \eqref{eq:homgreen} and is easy to miss. The convolution-type half of the derivation is exact, but the correlation-type half neglects evanescent components at the focal depth level \cite{wapenaar2016homogeneous}. The reason is not convenience: a focusing function that satisfies the focusing condition exactly is \emph{unstable} in the evanescent field, so the spatial delta at the focus has to be treated as band limited. This is the same restriction that motivates the dedicated evanescent-wave formulations discussed in Section~\ref{sec:extensions}.

Applied once, equation \eqref{eq:homgreen} moves one endpoint into the subsurface. Applied twice, first to redatum the sources and then the receivers, it places both endpoints at depth and yields $G_h(\xA, \xB, t)$ between two arbitrary subsurface points, from surface reflection data and an estimate of the direct arrivals alone. No information about the positions or shapes of the scattering interfaces is used, yet the retrieved response shows the scattering at those interfaces explicitly. Setting $\xA = \xB$ and $t = 0$ recovers a multiple-free image, but $G_h(\xA, \xB, t)$ for $\xA \neq \xB$ carries more, including local amplitude-versus-angle information \cite{wapenaar2016homogeneous}.

One practical caveat is specific to this construction. The derivation assumes a lossless medium, and unlike the imaging conditions, the representation carries no term that absorbs a departure from that assumption. Retrieval of $G_h$ is correspondingly sensitive to attenuation, which is the main reason it has been demonstrated far more often on synthetic than on field data \cite{brackenhoff2020virtual}. Correcting the reflection data for energy loss before applying the representation is what makes field application viable, and it is consistent with the behaviour Thorbecke et al.\ \cite{thorbecke2013green} observed for the underlying scheme, where a medium with $Q = 50$ produced ghost events that $Q$ compensation could suppress. The quality of the retrieved Green's functions depends further on:

\begin{enumerate}
    \item \textbf{Accuracy of the initial estimate:} Errors in $t_d(\xH, \xA)$ shift the time window $\Theta$ and introduce traveltime shifts in the retrieved Green's functions, which appear as mispositioned events in the image \cite{thorbecke2013green, van2015practical}.
    \item \textbf{Acquisition aperture:} Limited horizontal aperture truncates the surface integrals in equations \eqref{eq:marchenko1}--\eqref{eq:marchenko2}. The consequence is more specific than a loss of resolution. Sripanich and Vasconcelos \cite{effects2019aperture} showed that the focused response is not isotropic: it radiates in the direction perpendicular to the line joining the centre of the surface array to the focal position in time-migration coordinates, with a further rotation on conversion to depth. The virtual source therefore has a radiation pattern set by the acquisition geometry rather than by the medium, which matters directly for amplitude-preserving redatuming and for any subsequent amplitude-versus-angle analysis.
    \item \textbf{Data bandwidth:} The Marchenko equations assume a broadband reflection response. Missing near offsets or limited low-frequency content degrade focusing quality and introduce Gibbs-like oscillations.
    \item \textbf{Noise:} The iterative scheme amplifies coherent noise that falls within the time window $\Theta$. Pre-processing noise attenuation is therefore critical for field data \cite{da2018marchenko}.
    \item \textbf{Elasticity and mode conversion:} Ignoring mode conversion introduces errors at large incidence angles \cite{da2016elastic}.
\end{enumerate}

Thorbecke et al.\ \cite{thorbecke2013green} tested this sensitivity directly on a strongly multiple-generating layered model. With a $+15\%$ velocity error in the top layer, the focusing operator no longer collapsed to a band-limited point but to a blurred focal area, and the retrieved wavefields acquired a traveltime shift together with a small number of extra artefacts. The main reflections and internal multiples survived, and the scheme neither diverged nor introduced spurious ghost events. Their conclusion was that the iteration is stable against phase and amplitude errors in the direct arrival, with the errors showing up as low-amplitude artefacts rather than as a breakdown of the retrieval. The same study found that when the reflection data come from a medium with losses ($Q = 50$) while the scheme assumes a lossless medium, ghost events do appear, but only in the upgoing Green's function, and can be reduced by $Q$ compensation applied before the iteration.

\subsection{Green's Functions in Dissipative Media}

In attenuative media, the frequency-domain wavenumber acquires an imaginary part; for a constant-$Q$ model,
\begin{equation}
k(\omega) = \frac{\omega}{c}\left(1 + \frac{i}{2Q}\right),
\end{equation}

where $Q$ is the quality factor, so wave amplitudes decay with propagation distance and energy is no longer conserved. This is the weak-attenuation form, retaining the loss term and omitting the accompanying velocity dispersion that a strictly constant $Q$ implies through causality; the omission is harmless for the argument below, which turns on the loss of energy conservation rather than on the phase behaviour. The standard Marchenko equations fail in dissipative media because the correlation-type reciprocity theorem \eqref{eq:recipcorr} acquires the volume term identified in Section~\ref{sec:math}, which no longer vanishes.

Slob \cite{slob2016green} restored the structure by pairing the physical medium with an auxiliary one. The \emph{effectual medium} is identical to the physical medium except that its dissipation is negative, so that the two together conserve energy in the sense the correlation theorem requires. Writing $q$ for quantities in the dissipative medium and $e$ for the effectual one, two coupled sets of Marchenko equations follow, the first relating the dissipative focusing functions to the reflection responses of both media,
\begin{align}
R_q f_{1q}^+ &= f_{1q}^- + G_q^- \\
R_e f_{1q}^{-*} &= f_{1q}^{+*} - G_e^+
\end{align}

\noindent and a second, similar pair for the effectual focusing functions $f_{1e}^{\pm}$. Each set on its own is single-sided and is solved by the same kind of iteration as the lossless scheme. Solving the first set retrieves the upgoing Green's function in the dissipative medium and the downgoing one in the effectual medium; solving the second exchanges them \cite{slob2016green, marchenko2016dissipative}.

There is a cost, and it is not incidental. The effectual reflection response $R_e$ is not measurable, and computing it from $R_q$ requires reflection \emph{and} transmission measurements on both sides of the medium \cite{slob2016green}. The dissipative scheme is therefore single-sided only in its solution step; its input is double-sided. That restricts it to configurations where the target is accessible from two sides, which describes laboratory and non-destructive-testing geometries and borehole-to-borehole surveys, but not surface seismic acquisition.

A second requirement falls on the initial estimate. In a lossless medium $f_{1,d}^+$ needs only a traveltime; here it also needs the energy lost along the direct path, which has to be compensated before the iteration starts \cite{slob2016green}. Omitting that compensation does not simply degrade the result, it biases the two media in opposite directions: the focusing field in the dissipative medium comes out too weak and the one in the effectual medium too strong, by the same factor, and the error carries through to the Green's functions and to any reflection response built from them.

That symmetry is also what makes the loss recoverable. At the focal point the two media must give the same frequency-integrated reflection response,
\begin{equation}
\int_{-\infty}^{\infty} \hat{R}_q(\xv_i, \xv_i, \omega)\, d\omega = \int_{-\infty}^{\infty} \hat{R}_e(\xv_i, \xv_i, \omega)\, d\omega
\end{equation}

\noindent and the discrepancy between the two sides equals twice the two-way path loss from the surface to that point \cite{slob2016green}. Smooth loss and velocity models can therefore be built by analysing the two reflection responses together, which is what allows the scheme to run without an attenuation profile known in advance.

Wapenaar et al.\ \cite{wapenaar2022marchenko} expressed the propagator and transfer matrix of a dissipative (energy-losing) acoustic medium in terms of Marchenko focusing functions, an alternative route within the same framework.

\section{Focusing Functions: Properties and Mutual Relations}
\label{sec:focusing}
The downgoing focusing function $f_1^+(\xH, \xA, t)$ represents an input wavefield at the surface that, when injected into the truncated medium, focuses to a delta function at $\xA$ at $t = 0$. In time reversal, achieving this would require injecting the time-reversal of the medium's response to a source at $\xA$ which is the time-reversed Green's function $G^-(\xH, \xA, -t)$ from all surface positions \cite{fink2000time}. But time reversal requires the full medium response, including transmitted waves. The focusing function achieves the same focus from the reflection response alone, by constructing a wavefield that pre-cancels all internal multiple reflections above $\xA$. The upgoing component $f_1^-(\xH, \xA, t)$ is the reflected response of the truncated medium to the injection of $f_1^+$. It represents the wavefield that leaks back to the surface before the focus is achieved, and carries information about all reflectors above $\dD_i$ \cite{wapenaar2021marchenko}.


A fundamental property of the downgoing focusing function is that it acts as the \emph{inverse of the transmission response} of the truncated medium \cite{wapenaar2021marchenko}: injecting $f_1^+$ from the surface undoes, arrival by arrival, all the scattering that transmission through the overburden imposes, which is precisely what the focusing condition \eqref{eq:focusdef} demands. Symbolically,
\begin{align}
\int_{\dD_0} T(\xB, \xH, t) \ast f_1^+(\xH, \xA, t)\, d\xH \nonumber \\
= \delta(\mathbf{x}_{B,H} - \mathbf{x}_{A,H})\,\delta(t)
\end{align}

for $\xB$ on $\dD_i$, where $T$ is the downgoing transmission response of the truncated medium. This restates the focusing condition \eqref{eq:focusdef} and makes the interpretation explicit: the focusing function unravels the scattering accumulated on the way from $\dD_0$ to $\dD_i$.

For two depth levels $\dD_i$ and $\dD_j$ with $x_{3,i} < x_{3,j}$, the focusing function for $\dD_j$ can be expressed in terms of that for $\dD_i$ and the reflection response between the two levels \cite{wapenaar2021marchenko}. This \emph{cascading property} enables layer-by-layer processing of deep targets without repeating the full Marchenko iteration for each depth.

\subsection{Marchenko Focusing Functions and Free-Surface Multiples}

The standard Marchenko derivation assumes the reflection response $R$ has been deghosted. It is also possible to formulate the Marchenko equations using the \emph{total} reflection response, including free-surface multiples \cite{davydenko2017full, wapenaar2021marchenko}. In this formulation, the focusing functions focus at $\xA$ through both internal and free-surface multiple reverberations, and the retrieved Green's functions include contributions from all orders of free-surface multiples. Subsequent processing then separates primaries, internal multiples, and free-surface multiples as needed. Combining the Marchenko scheme with free-surface multiple elimination via seismic interferometry \cite{he2022elimination} or the SRME algorithm \cite{verschuur2013seismic} allows a sequential workflow that handles all categories of multiples systematically.

\section{Marchenko Imaging}
\label{sec:imaging}

\subsection{Imaging Conditions}

Once the one-way Green's functions have been retrieved for a range of focal points $\xA$, an image is formed by applying an imaging condition. van der Neut et al.\ \cite{van2017single} distinguish three, and the terms \emph{single-sided} and \emph{double-sided} describe the illumination each one needs, not the mathematics of the condition itself.

\subsubsection*{Imaging by Deconvolution}

The first condition deconvolves the retrieved upgoing field by the downgoing one, by multidimensional deconvolution, and evaluates the resulting redatumed reflection response $R'$ at zero time and zero offset \cite{wapenaar2014marchenko, van2017single}:
\begin{equation}
I(\xA) = R'(\xA, \xA, t = 0) \label{eq:IC1}
\end{equation}

Deconvolution carries an implicit normalisation, so the amplitudes come out correct without the wavefields having to be normalised beforehand. The cost is that a large, ill-conditioned inverse problem has to be solved and stabilised at every image point.

\subsubsection*{Imaging by Double Focusing}

The second condition avoids that inversion by focusing a second time, exactly as in the redatuming scheme of Section~\ref{sec:srredatum}, and then taking the ratio of the reflected and incident fields at the image point. Imaging the two fields independently and dividing is dramatically cheaper than deconvolution \cite{van2017single}. Here the normalisation is not automatic. van der Neut et al.\ distinguish \emph{focal} from \emph{physical} normalisation of the Marchenko solution, and accurate amplitudes with this condition require the physical variant \cite{van2017single}.

Both of these conditions need illumination from one side only, which is what makes them applicable to surface seismic acquisition. Both also retrieve the internal reflectivity from \emph{primary reflections alone}. The multiples have been accounted for, in that they no longer generate false events, but they do not contribute to the image.

That last point is counterintuitive enough to be worth isolating, since the appeal of the method is usually stated as its ability to use multiples. Wapenaar et al.\ \cite{wapenaar2017whymultiples} showed why deconvolution behaves this way. Given the full downgoing and upgoing fields at a depth level, both containing all multiples, deconvolving the upgoing by the downgoing field yields an image to which the multiples make no contribution at all, a result they explain through minimum-phase arguments. What the deconvolution buys in exchange is that the image is true-amplitude and free of the cross-talk artefacts that correlation produces.

\subsubsection*{Imaging by Cross Correlation}

The third condition correlates the up- and downgoing fields at zero time lag \cite{wapenaar2014marchenko, wapenaar2021marchenko, van2017single}:
\begin{equation}
I(\xA) = \int_{-\infty}^{\infty} \int_{\dD_0} G^+(\xH, \xA, t)\, G^-(\xH, \xA, t)\, d\xH\, dt \label{eq:IC2}
\end{equation}

\noindent which is formally the standard RTM condition applied to Marchenko-retrieved wavefields. Correlating the full up- and downgoing fields does let both primaries and multiples contribute to the image, but at a cost: the image is not true-amplitude and carries cross-talk artefacts between events that the deconvolution route avoids \cite{wapenaar2017whymultiples}. Its requirement is also stricter, since those artefacts cancel only under \emph{double-sided} illumination, with sources and receivers on two boundaries enclosing the target volume. The reflection response at the lower boundary can be physically recorded where a suitable geometry exists, or retrieved from the response at the upper boundary, but it cannot simply be dispensed with.

What is bought at that price is unusual and worth stating plainly. Under double-sided illumination the multiple reflections focus at the image points and contribute physically to the retrieved reflectivity values. In the single-sided conditions they are only prevented from doing harm; here they carry signal. Weakly illuminated parts of a strongly heterogeneous medium can therefore be imaged using energy that the primary-only conditions discard \cite{van2017single}.

\subsubsection*{Imaging with Free-Surface Multiples Included}

The standard scheme retrieves a Green's function containing primaries and internal multiples, which is why the reflection response has to be stripped of surface-related multiples beforehand. Singh et al.\ \cite{singh2015marchenko} extended the Marchenko equation so that the retrieved Green's function includes free-surface multiples as well, that is, so that it is the Green's function of the medium \emph{with} its free surface present.

The practical consequence is that surface-related multiple elimination becomes unnecessary: the free-surface multiples are carried into the imaging operator and used there instead of being removed first. The information required is the same as for the standard scheme, the only difference being that the input reflection response now retains its free-surface multiples \cite{singh2015marchenko, singh2015incorporating, singh2017accounting}. Since SRME is itself an imperfect and sometimes unreliable step, particularly in shallow water (Section~\ref{sec:challenges}), removing it from the workflow is worth more than the extra bookkeeping costs.

\subsection{True-Amplitude Marchenko Imaging}

Getting the amplitude right is a question of normalisation, and where that normalisation has to be imposed depends on the imaging condition. Deconvolution supplies it implicitly. Double focusing and cross correlation do not, and for those the retrieved wavefields must be physically normalised with respect to power flux if the image is to carry true reflectivity amplitudes rather than something merely proportional to them \cite{van2017single}.

Where a full normalisation is impractical, the standard approximation is to divide out the source-side illumination, in the manner of illumination-compensated migration \cite{vasconcelos2010nonlinear}:
\begin{equation}
I_{\text{TA}}(\xA) = \frac{\displaystyle\int_{\dD_0}\int dt\, G^+(\xH, \xA, t) G^-(\xH, \xA, t)}{\displaystyle\int_{\dD_0}\int dt\, \left|G^+(\xH, \xA, t)\right|^2}
\end{equation}

This is the least-squares estimate of the local reflectivity treated as a scalar at each image point, and it is an approximation to the full multidimensional deconvolution rather than an alternative formulation of it: the matrix inverse is replaced by a division by the illumination energy. That is what makes it cheap and also what limits it, since a scalar cannot represent the angle dependence a matrix inverse retains. Regularisation of the denominator is required wherever illumination is weak, or the division amplifies noise instead of correcting amplitudes \cite{vasconcelos2010nonlinear}.

\subsection{Least-Squares Marchenko Imaging}
\label{sec:lsmi}

The iterative scheme of Section~\ref{sec:iterative} solves the coupled equations by expanding the inverse of the Marchenko operator as a Neumann series, which presumes that the series contracts. There are cases where it does not, the free-surface-inclusive formulation being the standard example \cite{de2021marchenko, dukalski2017marchenko}. Recasting the coupled equations as a single linear system and solving it by least squares removes that dependence entirely, and at the same time compensates for incomplete illumination and the amplitude inconsistencies left by irregular acquisition geometry \cite{de2021marchenko, ravasi2020implementation}:
\begin{equation}
\min_m \left\| \mathbf{d} - \mathbf{L}\mathbf{m} \right\|_2^2 + \lambda \|\mathbf{m}\|_2^2
\end{equation}

where $\mathbf{m}$ is the subsurface reflectivity model, $\mathbf{d}$ is the observed reflection data, $\mathbf{L}$ is the Marchenko-based demigration operator, and $\lambda$ is a regularisation parameter. The least-squares solution iteratively applies the Marchenko migration operator and its adjoint, producing a high-resolution, balanced-amplitude image \cite{de2021marchenko}.

de Paula et al.\ \cite{de2021marchenko} implemented least-squares Marchenko imaging in Julia, using the same matrix-free linear-operator style as PyLops \cite{ravasi2020implementation, ravasi2020pylops}, and validated it against RTM on a synthetic constant-velocity, variable-density model with a damping factor of $10^{-2}$. The cost relative to single-pass imaging scales with the number of LSQR iterations required.

\subsection{Plane-Wave Marchenko Imaging}

Forming an image over a dense grid of focal points requires solving the Marchenko equations at every point independently. The cost scales as $n_x n_z$ in 2D and $n_x n_y n_z$ in 3D, which is the practical barrier to routine application even though the solves are embarrassingly parallel.

Meles et al.\ \cite{meles2018virtual} removed most of that cost by changing what the focusing function focuses \emph{on}. The standard function $f_1$ focuses in space and time at a point, returning a point-source Green's function. Their function $F_1$ focuses in time only, across an entire depth plane, and returns an \emph{areal-source} response: the field for a source distributed over that whole plane and fired simultaneously. Substituting this condition into the representations gives a coupled system with the same structure as the point-source one,
\begin{align}
\mathbf{F}_1^- + \mathbf{G}^- &= \mathbf{R}\,\mathbf{F}_1^+, \qquad
\mathbf{F}_1^+ - \mathbf{G}^{+*} = \mathbf{R}^*\mathbf{F}_1^-
\end{align}

\noindent which, given a suitable separation operator $\Theta_F$ built from the first arrival of the areal response, is solved by the same Neumann iteration,
\begin{equation}
\mathbf{F}_1^+ = \sum_{k=0}^{\infty}\left(\Theta_F\mathbf{R}^*\Theta_F\mathbf{R}\right)^k \mathbf{F}_{1d}^+
\end{equation}

The saving is that one solve now serves an entire depth level, so the number of Marchenko solutions drops to $n_z$, and it drops to $n_z$ in three dimensions as well as in two. That is the property that makes the approach interesting for 3D: the cost stops growing with the lateral extent of the target.

The saving is not free. Because the retrieved response is an integral along the focal plane, the imaging condition inherits that integration, and the resulting image has poorer angle illumination and reduced lateral resolution than the point-source construction \cite{meles2018virtual}. Recovering the lost angular information is what motivates imaging with several plane waves at different horizontal slownesses rather than the single horizontal one, and Almobarak et al.\ \cite{almobarak2021plane} applied that multi-slowness version to field data. Meles et al.\ demonstrated the original scheme on 2D synthetics, noting that focusing degrades where a focal plane crosses an interface, since the separation operator is not strictly valid there.

\subsection{Marchenko Imaging from Rugged Topography}

Standard Marchenko imaging assumes sources and receivers on a flat surface at $x_3 = 0$. In mountainous terrain that assumption is not merely inconvenient but hard to recover by the usual route, since elevation-static correction restores it only where surface consistency holds, and where it does not the correction itself distorts the image. Chen et al.\ \cite{chen2023marchenko} avoid the flat-datum requirement rather than correcting to it. Their topography-Marchenko scheme takes data corrected to a \emph{floating} datum that follows the topography, and estimates the initial downgoing focusing function between that floating datum and the subsurface focal point. The coupled equations themselves are unchanged; what changes is the surface on which they are posed. Tests on a synthetic model with rugged topography and on a land dataset from northwest Sichuan, China, showed the distortions introduced by elevation-static correction to be avoided \cite{chen2023marchenko}.

\subsection{Diffraction Imaging via Marchenko}

Diffraction imaging exploits the distinct moveout of diffracted energy to image point scatterers, faults, and discontinuities. Combining the Marchenko method with diffraction separation \cite{yue2020diffraction} allows diffraction imaging from Marchenko-redatumed data, which is free of multiple contamination. The result is an image of subsurface diffractors less affected by overburden artefacts than conventional diffraction imaging.

\section{Marchenko Redatuming}
\label{sec:redatuming}


Redatuming transforms seismic data recorded at one datum, usually the Earth's surface, to what would have been recorded at a different datum, such as a virtual level just above a target formation. Classically, redatuming was done by wave-equation extrapolation in a known velocity model. The Marchenko method achieves it from the data alone, without detailed knowledge of the medium between the acquisition surface and the new datum. This data-driven process produces a new dataset $R'(\mathbf{x}_{A,H}, \mathbf{x}_{B,H}, t)$ representing the reflection response of the subsurface below $\dD_i$, as if both sources and receivers were located on $\dD_i$. This redatumed dataset contains only reflections from below $\dD_i$; the overburden response is removed \cite{wapenaar2014marchenko, van2014interferometric}.

Marchenko redatuming has three advantages over model-driven datuming. A detailed velocity model of the overburden is not needed, since a smooth background suffices for the initial traveltime estimate. Overburden-related multiple reflections, internal multiples included, are removed without explicit multiple modelling. And the redatumed response satisfies the wave equation with correct amplitude scaling, so it can be fed to amplitude-sensitive downstream processing \cite{wapenaar2021marchenko}.

The limitations are equally specific. The method as formulated is acoustic, and mode conversion in the overburden introduces errors that the acoustic approximation cannot absorb \cite{da2016elastic}. The accuracy of the initial traveltime estimate propagates directly into the focusing functions, and does so most severely for the complex overburdens where redatuming is most wanted. The time-window argument requires the direct wave and the coda to be separable, which fails for very shallow targets where they overlap. Limited acquisition aperture truncates the surface integrals and introduces artefacts that are worst for steep dips and for focal points near the array edge \cite{effects2019aperture}.

\subsection{Receiver Redatuming}

Receiver redatuming transforms a dataset with sources and receivers at $\dD_0$ into a dataset with sources at $\dD_0$ and virtual receivers at $\dD_i$ \cite{wapenaar2021marchenko}. The virtual receiver gather at $\xA$ for a surface source at $\xs$ is exactly the Marchenko-retrieved two-way Green's function:
\begin{equation}
P(\xs, \xA, t) = G^+(\xA, \xs, t) + G^-(\xA, \xs, t)
\end{equation}

Summing over sources and applying an imaging condition gives the reflectivity below $\dD_i$ as seen from the surface.

\subsection{Source-Receiver Marchenko Redatuming}
\label{sec:srredatum}

\emph{Source-receiver Marchenko redatuming} \cite{staring2018source} goes further and places both the virtual source and the virtual receiver at depth. Solving the coupled Marchenko equations is itself the first focusing step: it brings the receivers down, giving the Green's functions $G^{\pm}(\xA, \xs, t)$ for a receiver at the focal point $\xA$ and a source still at the surface. Two routes then bring the sources down.

The conventional route is multidimensional deconvolution. The receiver-redatumed Green's functions satisfy
\begin{align}
    G^-(\xA, \xs, t) = \int_{\dD_i} R'(\xA, \xB, t) \conv G^+(\xB, \xs, t)\, d\xB \label{eq:mddredatum}
\end{align}

\noindent and inverting for $R'$ inside the integral yields the redatumed reflection response \cite{staring2018source}. That response lives in the \emph{truncated} medium, so all overburden effects are removed. The price is that the inversion is a large, fundamentally ill-posed problem that has to be stabilised, and incomplete illumination, finite aperture, and noise all make a correct solution harder to reach on field data \cite{staring2018source}.

The alternative replaces that inversion with a second focusing step, which is what gives \emph{double focusing} its name and echoes Berkhout's double dynamic focusing \cite{berkhout2008pushing}. Convolving the retrieved upgoing Green's function with the downgoing focusing function creates downward-radiating virtual sources at the redatuming level \cite{staring2018source}:
\begin{align}
    R'(\xA, \xB, t) = \int_{\dD_0} G^-(\xA, \xs, t) \conv f_1^+(\xs, \xB, t)\, d\xs \label{eq:srredatum}
\end{align}

\noindent Two conditions attach to equation \eqref{eq:srredatum}. The virtual sources must sit slightly above the virtual receivers for the relation to hold, and the result carries a double band limitation, because both wavefields entering the convolution are already dressed with the wavelet; removing it requires deconvolving the redatumed response by the autoconvolution of the wavelet on the direct arrival of $f_1^+$ \cite{staring2018source}.

The two routes are not equivalent. Multidimensional deconvolution redatums into the truncated medium and removes the overburden entirely; double focusing redatums into the physical medium and removes only the strongest overburden interactions, leaving a residue that Staring et al.\ \cite{staring2018source} quantify. What is bought with that residue is robustness. Double focusing needs no inversion, so it is cheaper, simpler to implement, and parallel over pairs of focal points, since the integral runs over the acquisition surface rather than over the redatuming level. On field data with sparse geometry and noise, that trade favours double focusing.

Staring et al.\ \cite{staring2018source} make the scheme adaptive by writing the retrieval of $G^-$ and $f^+$ as a series in the operator $\Omega = \theta\mathcal{R}^{\star}\theta\mathcal{R}$ and matching the predicted overburden interactions to the data by least-squares subtraction, which absorbs the amplitude and phase mismatches that velocity errors and imperfect acquisition introduce. Field applications include the Santos Basin \cite{staring2018marchenko, staring2020three} and a North Sea field \cite{ravasi2016target}; Marchenko-based isolation of a target zone has also been applied to land data \cite{shirmohammadi2025application, shirmohammadi2024application}.

\section{Multiple Elimination and Attenuation}
\label{sec:multiples}

Multiple reflections are among the most persistent coherent noise sources in marine seismic data. These reflections exist as free-surface or internal multiples. Free-surface multiples are well handled by surface-related multiple elimination (SRME) \cite{verschuur2013seismic} or similar surface-related methods. Internal multiples are harder: they require knowledge of the subsurface to model, and model-based approaches such as the inverse scattering series (ISS) \cite{weglein1997inverse} are computationally expensive and sensitive to velocity errors. Table~\ref{tab:multiples} compares the Marchenko multiple elimination scheme with the two other main approaches. The structural difference from the ISS is that the Marchenko method poses the problem as a Fredholm integral equation of the second kind over a finite time window, which can be solved directly when an iterative expansion fails to converge, whereas the ISS exists only as a series and offers no such fallback. Compared with SRME, the Marchenko method addresses the harder problem of internal multiples without the specific free-surface source--receiver pairing that SRME requires.

\subsection{Marchenko Internal Multiple Elimination}

Three distinct routes go under the heading of Marchenko internal multiple removal. Two of them differ in where the sources and receivers end up \cite{wapenaar2021marchenko}; the third avoids the subtraction step altogether.

The first is a by-product of redatuming. The retrieved focusing functions encode the scattering the overburden imposes, so the redatumed response $R'$ at $\dD_i$ contains only reflections generated below $\dD_i$. Which multiples this removes depends on how the redatuming was done: multidimensional deconvolution eliminates the internal multiples related to the overburden entirely, whereas double focusing, though more stable and better suited to an adaptive implementation, does not eliminate the multiples between the target and the overburden \cite{wapenaar2021marchenko}. That difference is the same one described in Section~\ref{sec:srredatum}, and it is the reason the cheaper route is not simply the better one.

The second route, \emph{Marchenko multiple elimination}, keeps the sources and receivers where they were. Rather than redatuming, it predicts the internal multiples in the surface reflection response and subtracts them there, producing a primary-only dataset at $\dD_0$. Two properties set it apart. It predicts and subtracts all orders of internal multiples \emph{with the correct amplitude}, and it does so \emph{without needing a macro subsurface model} at all \cite{wapenaar2021marchenko}. The macro model, which the redatuming route needs to estimate the first arrival, is dispensed with because nothing has to be positioned in depth. Like redatuming, this class admits several implementations: as a multidimensional deconvolution, as a double dereverberation process, or in an efficient plane-wave form.

The third route sidesteps subtraction entirely, which matters because subtracting predicted multiples is the step that most often fails on field data, even when the prediction itself is accurate. Rather than removing multiples from the recorded data, Meles et al.\ \cite{meles2016reconstructing} \emph{construct} a parallel dataset containing only primaries. Marchenko redatuming supplies the upgoing wavefields at virtual receivers, and convolutional interferometry then combines the first arrivals of those upgoing fields with the direct-wave downgoing Green's functions to synthesise the primary reflections directly. Nothing is predicted and nothing is subtracted, so no adaptive matching filter is needed. The input requirement is no greater than for migration, namely surface reflection data and estimates of the direct arrivals, and on a stratified synclinal model the construction tolerated errors in the reference velocity model while still improving the migrated image.

Formally, the elimination scheme produces a \emph{primary-only} reflection response $P_{\text{prim}}(\xH, \xH', t)$ \cite{wapenaar2021marchenko}:
\begin{equation}
P_{\text{prim}}(\xH, \xH', t) = R(\xH, \xH', t) - M(\xH, \xH', t) \label{eq:meelim}
\end{equation}

where $M(\xH, \xH', t)$ is the contribution of all orders of internal multiples, computed from the Marchenko focusing functions without explicit multiple modelling. The method has been demonstrated on synthetic data for complex models \cite{meles2016reconstructing} and on field data from the North Sea \cite{da2018marchenko}, the Santos Basin \cite{staring2018marchenko, staring2020three}, and land surveys \cite{shirmohammadi2025application, shirmohammadi2024application}.

van der Neut and Wapenaar \cite{van2016adaptive} reworked the scheme to remove its dependence on a macro velocity model altogether. In place of the model they require only an estimate of the two-way traveltime surface of a chosen horizon, which can be picked from the data, and they introduce an approximation that makes the prediction amenable to adaptive subtraction. The output is a dataset in which every interaction, primary and multiple alike, with the medium above the picked horizon has been eliminated. Two properties distinguish it from layer-stripping approaches to internal multiple elimination: it can be applied at any chosen target horizon without first resolving the multiples generated by shallower ones, and it sits early in a processing flow rather than after velocity model building.

Their results also mark the boundary honestly. On synthetics the method worked, with limitations traced to thin layers, diffraction-like discontinuities, and finite acquisition aperture. On field data the kinematics of the predicted updates matched the internal multiples in the recorded data, but subtracting them proved difficult, which is the amplitude-matching problem discussed in Section~\ref{sec:challenges}.

\subsection{Adaptive Marchenko Internal Multiple Attenuation}
\label{sec:adaptive}

The standard scheme \eqref{eq:meelim} performs an exact subtraction, which requires the predicted multiples to match those in the data in both amplitude and phase. Errors in the initial estimate and irregular acquisition make that match imperfect, and the result is over- or under-subtraction. Replacing the exact step with an adaptive one, in which the prediction is matched to the data by a least-squares filter before subtraction, absorbs the mismatch and is what makes the scheme survive field conditions \cite{staring2018source, van2016adaptive}.

Staring and Wapenaar \cite{staring2020three} asked the question that matters for whether any of this reaches production: does adaptive double focusing survive a real three-dimensional acquisition geometry? Marchenko methods had been tested extensively on 2D synthetics and applied to 2D field data, but their behaviour on sparse 3D geometries was largely unknown. Starting from a wide-azimuth dense grid of sources and receivers, they decimated in stages until a narrow-azimuth streamer geometry remained, removing sail lines, near offsets, far offsets, and outer cables in turn, and recorded which of those the method could tolerate.

The answer is specific and useful. The scheme proved most sensitive to two things: the sail line spacing, and the limited aperture in the crossline direction. The distinction between them matters, because sail line spacing can be recovered by interpolation whereas crossline aperture cannot; it is fixed by the acquisition and no processing recovers it. Applying the resulting interpolation strategy to narrow-azimuth streamer field data from the Santos Basin, they predicted and adaptively subtracted the internal multiples and improved the geological interpretation of the target. Their own qualification is worth carrying over: the method is robust enough for 3D field data, but which acquisition parameters bind will differ with the geology and the survey type.

\begin{table*}[!t]
\centering
\caption{Comparison of multiple elimination methods}
\label{tab:multiples}
\resizebox{\textwidth}{!}{%
\begin{tabular}{lccc}
\toprule
\textbf{Criterion} & \textbf{SRME} \cite{verschuur2013seismic} & \textbf{ISS} \cite{weglein1997inverse} & \textbf{Marchenko} \cite{wapenaar2021marchenko} \\
\midrule
Multiples addressed & Free-surface & Free-surface and internal, & Internal, all orders; free-surface \\
 & & via separate subseries & in the extended formulation \\
Model information required & None & None & Smooth background traveltime \\
Absolute amplitude calibration & Not required & Not required & Required \cite{da2018using, jia2017subsalt} \\
Adaptive subtraction in practice & Standard & Standard \cite{reinicke2020monotonicity} & Optional \cite{reinicke2020monotonicity, staring2018source} \\
Behaviour when scattering is strong & Unaffected & Series may diverge & Iteration may fail; direct \\
 & & & inversion remains available \\
Computational cost & Low & High & Moderate \\
Field-data maturity & Industry standard & Established & Growing; mostly marine \\
\bottomrule
\end{tabular}%
}
\end{table*}

The table compresses a comparison that has since been made precisely. Reinicke et al.\ \cite{reinicke2020monotonicity} observed that the requirements of the two methods had only ever been stated verbally, which made them hard to compare, and recast both as \emph{separability conditions} depending on the medium, the angle of incidence, and the redatuming depth.

Series-based prediction, whether by the Jakubowicz method or the ISS, rests on two \emph{monotonicity} assumptions: that the temporal ordering of primaries matches the ordering of reflectors in depth, and that internal multiples are recorded after the primaries that generate them. Both hold for acoustic waves outside special cases. The conventional Marchenko method needs only a weaker form of the same condition, which is the source of its advantage, and it buys that advantage by requiring more prior information in the form of the initial estimate. The two are therefore not ranked but traded: less stringent event ordering in exchange for a model-derived input. Remixing the Marchenko scheme to remove the need for that prior information tightens the condition again, to nearly the strictness of the ISS.

Where the Marchenko method does differ in kind is what it treats as the object of removal. It handles the overburden as a single complex multiple generator rather than as a stack of independent generators \cite{reinicke2020monotonicity}, which is why it removes all internal multiples associated with a group of layers at once, and why it can do so without adaptive subtraction where the series methods predict and then subtract adaptively.

A second line of work uses free-surface multiples within imaging rather than removing them beforehand. Singh et al.\ \cite{singh2015incorporating, singh2017accounting} incorporated them directly into the Marchenko scheme. Davydenko and Verschuur \cite{davydenko2017full} developed full-wavefield migration, which images with all orders of multiples. Dukalski and de Vos \cite{dukalski2017marchenko} extended Marchenko inversion to strong-scattering regimes that include surface-related multiples, where the Neumann series no longer converges and an inversion is required. Wapenaar et al.\ \cite{wapenaar2021marchenko} reformulated the equations in terms of the total reflection response, free-surface reverberations included, and showed that the same iterative scheme applies with minimal modification.

\section{Target-Oriented Methods}
\label{sec:targeting}

\subsection{Target-Enclosed Seismic Imaging}

van der Neut et al.\ \cite{van2017target} proposed \emph{target-enclosed imaging}. Instead of imaging from the acquisition surface downward, one selects two redatuming boundaries, one above and one below a target volume, and redatums to both. The required input is the upgoing and downgoing wavefield at each boundary due to impulsive sources at the surface. Those wavefields can come from actual subsurface measurements or from numerical modelling, but they can equally be retrieved by solving the multidimensional Marchenko equation, in which case the whole construction runs on surface reflection data alone.

The output is a set of virtual reflection and transmission responses, as though sources and receivers sat on the two enclosing boundaries. What makes them useful is what they exclude: they contain all orders of reflection inside the target volume and no interaction whatsoever with the medium outside it. The target can then be imaged from above or from below, and van der Neut et al.\ found the two images to agree when the wavefields came from the Marchenko equation. Because the enclosed multiples contribute rather than being discarded, resolution improves. Because the extended images do not depend on the medium outside the volume, localised inversion can afterwards be run inside the target alone, without contamination from interactions with the rest of the model.

\subsection{Target-Oriented Full-Waveform Inversion}

Full-waveform inversion (FWI) is the most computationally expensive seismic inversion method; applied to the full data volume from the surface, it requires hundreds of forward simulations and is susceptible to cycle-skipping and local minima. Marchenko redatuming reduces the dimensionality of the inversion problem: instead of inverting for the entire medium from the surface, one inverts only for the target zone using the redatumed data $R'$ \cite{cui2020marchenko}.

The steps in \emph{target-oriented FWI via Marchenko redatuming} are:

\begin{enumerate}
    \item Apply Marchenko redatuming to produce the virtual experiment $R'(\xA, \xB, t)$ with virtual sources and receivers on $\dD_i$.
    \item Build a smooth initial velocity model for the target zone below $\dD_i$.
    \item Run FWI using $R'$ as the target data and a model that includes only the target zone, injecting sources at $\dD_i$.
\end{enumerate}

The inversion domain shrinks to the target, and the internal multiples generated above $\dD_i$ have already been removed, so a major source of artefacts in the FWI gradient is gone before the inversion starts.

Cui et al.\ \cite{cui2020marchenko} built the scheme around two matching pieces: Marchenko redatuming to supply subsurface wavefields accurate both inside and around the target, and a local forward-modelling operator, derived from the convolution-type representation theorem, that propagates only within the target while retaining full acoustic coupling to the surrounding medium. That coupling is the point. Simply truncating the modelling domain neglects second- and higher-order scattering between the target and its exterior, which shows up as inversion error. On a 2D Barrett Unconventional P-wave velocity model, their local FWI resolved the reservoir zone considerably more cheaply than full-domain FWI without a significant loss of accuracy, given a sufficiently accurate starting macromodel. The demonstration is numerical; a field-data application of target-oriented FWI on Marchenko-redatumed wavefields has not yet been published.

\subsection{Target-Oriented Waveform Inversion Based on Marchenko Redatumed Data}

Lin and Liu \cite{lin2023target} assembled the same two ingredients differently. Marchenko redatuming again supplies the virtual data at the datum, but the local forward operators are built in crosscorrelation and cross-convolution form, and the misfit is posed as a subsurface-domain interferometric objective function rather than a surface data misfit. The inversion resolves reservoir parameters in the target area. On the Chevron 2014 benchmark dataset it produced high-resolution results at reduced computational cost, and they closed with a time-lapse inversion to show the approach carries over to 4D. A point worth noting for target-oriented work generally is one they make explicitly: Marchenko redatuming needs only a smoothed macromodel to compute the traveltime at the datum level, which is what relieves the inversion of having to know the details outside the target.

\subsection{Marchenko-Based Target Replacement}

\emph{Target replacement} \cite{wapenaar2018target} replaces the reflection response of a target zone in the data with that of a modified version, while keeping all overburden effects intact. The procedure is:

\begin{enumerate}
    \item Use the Marchenko method to \emph{remove} the response of the original target zone from the measured surface reflection response, leaving the overburden and underburden responses.
    \item Insert the modelled response of a new target zone between them, and propagate the result back to the surface.
\end{enumerate}

The output is the reflection response that would have been recorded at the surface had the target zone been the modelled one, with all orders of multiple scattering between target, overburden, and underburden accounted for, and, in the elastodynamic case, with wave conversion included \cite{wapenaar2018target}.

The efficiency argument is what makes this more than a curiosity. Only the second step depends on the target model, so a monitoring study that evaluates many target-zone scenarios repeats only that step, while the expensive Marchenko step is performed once. Since the target occupies a small fraction of the medium, this is far cheaper than remodelling the full reflection response for each scenario, which is the conventional alternative. The applications Wapenaar and Staring name are those where a small region changes inside a static medium: fluid flow in an aquifer, subsurface storage of waste, and hydrocarbon production.

\subsection{Marchenko-Based Immersive Wave Simulation}

\emph{Immersive boundary conditions} let a wave simulation confined to a small subvolume behave as though the rest of the medium were present, by supplying time-dependent boundary conditions that couple the simulation to its surroundings at every time step. Earlier implementations needed an explicit model of the reflectors outside the simulated region, which is normally what one does not have.

Elison et al.\ \cite{elison2018marchenko} removed that requirement. The interaction with the surrounding medium is carried entirely by Marchenko-derived Green's functions, computed from surface reflection data, a smooth velocity model, and density information for the simulated area alone. The surrounding medium is then known only implicitly: the locations of the external scatterers never enter. Once those Green's functions exist, arbitrary perturbations can be introduced inside the target and a whole suite of models differing only there can be simulated efficiently, with the simulation fully coupled to its surroundings including all orders of multiple scattering.

Two conditions attach. The Marchenko step assumes preferential vertical wave propagation, and extending the result to predict reflection responses at the acquisition surface for a perturbed target requires additional amplitude information beyond what the scheme itself supplies \cite{elison2018marchenko}.

\section{Extensions and Generalisations of the Marchenko Method}
\label{sec:extensions}
\subsection{Elastodynamic Marchenko Method}

The real Earth is elastic, not acoustic. It supports both P-waves and S-waves, and mode conversion occurs at interfaces. Extending the Marchenko method to elastic media is substantially more involved than the acoustic case because the wavefield is a vector quantity and the stiffness tensor has up to 21 independent components in the general anisotropic case. da Costa Filho et al.\ \cite{da2014elastodynamic} derived the elastodynamic Marchenko equations using the two-way elastodynamic wave equation. The scalar pressure $P$ is replaced by the stress tensor $\sigma_{ij}$, and the particle velocity $V_i$ becomes the vector particle velocity. The elastodynamic reciprocity theorems have the same structure as the acoustic ones, but with matrix-valued Green's functions and focusing operators.

Wapenaar \cite{wapenaar2014elastic} showed what the focusing itself looks like once mode conversion is admitted. Complex focusing functions emitted from the surface produce \emph{separate and independent} foci for compressional and shear waves at the same subsurface point, and each focus acts as an independent virtual source for its own wave type. The physical reason the acoustic argument does not simply carry over is stated there directly: single-sided focusing works because backpropagating multiply scattered energy reconstructs the focus, and in the elastic case that reconstruction depends on multiply scattered \emph{and mode-converted} waves acting together, which breaks down when a significant part of the field is not recorded.

\noindent The coupled elastodynamic Marchenko equations are \cite{da2014elastodynamic, da2016elastic}:
\begin{align}
\mathbf{f}_1^-(\xH, \xA, t) &= \Theta \int_{\dD_0} \mathbf{R}(\xH, \xH', t) \conv \mathbf{f}_1^+(\xH', \xA, t)\, d\xH'
\end{align}
\begin{align}
\mathbf{f}_1^+(\xH, \xA, t) &= \mathbf{f}_{1,d}^+(\xH, \xA, t) + \Theta \int_{\dD_0} \mathbf{R}(\xH, \xH', -t) \nonumber \\ &\qquad \qquad \qquad \conv \quad \mathbf{f}_1^-(\xH', \xA, t)\, d\xH'
\end{align}

\noindent where $\mathbf{f}_1^{\pm}$ and $\mathbf{R}$ are $N \times N$ matrices acting on flux-normalised, mode-decomposed wavefields, a different normalisation from the pressure-normalised acoustic scheme of Section~\ref{sec:multidim}. The indices run over \emph{wave modes}, not Cartesian components: $N = 2$ in two dimensions, giving the $PP$, $PS$, $SP$, and $SS$ entries, and $N = 3$ in three dimensions, where one source excites downgoing compressional waves and two excite downgoing shear waves \cite{wapenaar2014multicomponent}. The first index is the mode at the observation point, the second the mode radiated at the source.

Because these are matrices, \emph{the order of multiplication matters} \cite{wapenaar2014multicomponent}. The reflection matrix stands to the left of the focusing matrix in both equations, and reversing that order does not give an equivalent system. The second equation uses the time-reversed reflection matrix, the matrix counterpart of the correlation operator \eqref{eq:synthesiscorr}.

Writing the equations in matrix form is the easy part. Wapenaar and Slob \cite{wapenaar2014multicomponent} identified what actually stands in the way, and it is not computational cost. Every scheme in this review rests on a causality condition: that the coda of the focusing function separates cleanly in time from the Green's function, which is what allows the window $\Theta$ to divide them. That condition is strictly obeyed for 1D scalar waves. For vectorial waves, in one dimension or three, the extent to which it holds has \emph{not been established}, and no general remedy exists for the cases where it fails. This, rather than data availability, is the reason a fully multicomponent Marchenko scheme remained out of reach for so long.

They also identified the additional requirement that an elastic implementation must satisfy: the multicomponent direct arrival used to initiate the iteration has to include the \emph{forward-scattered} field, not merely the direct wave computed in a smooth model. Under that condition, and others, they retrieved the multicomponent Green's function from single-sided reflection data and demonstrated it on a 1D numerical example. The contrast with earlier elastic inverse scattering is instructive: the Newton--Marchenko approach of Budreck and Rose requires illumination and measurement from all directions, involving transmission as well as reflection, whereas this scheme needs a one-sided reflection response and a specific initial estimate \cite{wapenaar2014multicomponent}.

da Costa Filho et al.\ \cite{da2016elastic} combined the Marchenko method with convolutional interferometry to predict prestack internal multiples in general elastic media, either to identify them in prestack data and migrated sections or to remove them by adaptive subtraction. Testing on synthetic elastic data with both horizontal and vertical density and velocity variation, they reported two things that together set the current practical position.

The full elastic method is expensive, and it needs data components that are not usually recorded. They therefore also ran an \emph{acoustic} approximation on the same elastic data. Spatial resolution of the resulting image dropped, but the multiples were still predicted accurately, with only minor artefacts, and their conclusion was that where cost matters and some resolution can be given up, the acoustic version of the demultiple method may well be sufficient.

Beneath the practical obstacles lies a structural one. The acoustic method works because the Green's functions and the focusing functions occupy disjoint time intervals, touching at the single instant of the direct arrival, which is what allows a time window to separate them. That separation does not survive the elastodynamic extension. Wapenaar and Slob \cite{wapenaar2014multicomponent} traced the cause directly to the modes travelling at different speeds: in their layered example, two events of the Green's function and the time-reversed focusing function already overlap, and in more complex media more events will. The separation also fails when the focal depth sits close to an interface, where the focusing functions extend into the region occupied by the Green's function.

Reinicke et al.\ \cite{reinicke2020monotonicity} separate the two effects. Even acoustically there is one unavoidable overlap between the Green's and focusing functions, which muting handles. Elasticity adds a \emph{second} overlap, caused by the speed difference between modes, which cannot be predicted without knowing the medium and vanishes only conditionally; it also makes the first overlap no longer straightforward to predict without extra prior information. So the overlap is not a single instant but an \emph{extended interval} \cite{wapenaar2015causality}. The parts of the focusing functions falling inside it cannot be retrieved by the Marchenko method at all and must be supplied separately, which is why the multicomponent scheme's central additional requirement is that the direct arrival used to initiate the iteration \emph{include the forward-scattered field} \cite{wapenaar2014multicomponent}. In the acoustic case a traveltime and an amplitude suffice. The elastic problem is therefore not the acoustic problem with more components: something the acoustic scheme gets for free has to be provided from outside the data.

Reinicke et al.\ \cite{reinicke2020elastic} pinned down why the elastic scheme is hard to apply in practice, and when the acoustic substitute is defensible. The elastic extension rests on two assumptions that marine acquisition violates. It needs \emph{all} components of the reflection response to predict the converted multiples, whereas streamer data supply only $R_{pp}$, leaving the shear components $R_{ps}$, $R_{sp}$, and $R_{ss}$ absent and the elastic inverse transmission undefined. It also requires the temporal ordering of primaries to match the ordering of reflectors in depth, which fails when a target-related P-primary overtakes an overburden-related S-primary. Both conditions are thought to bind other data-driven elastic de-multiple methods as well.

Their response was to treat elastic $R_{pp}$ as though it were an acoustic response and test what that costs. On a synthetic Middle East model built from well logs, with gentle regional dip, varying layer thicknesses, minor faults, and small relief structures, acoustic Marchenko de-multiple applied to elastic $R_{pp}$ data gave images almost identical to those from a genuinely acoustic response, attenuating near to all internal multiples and doing so without adaptive subtraction. Their conclusion sets the boundary the earlier work left open: the acoustic substitute is sufficient for \emph{structural} imaging in close-to-1.5D media, because structural imaging relies on stationary points where elastic effects vanish at normal incidence and grow only gradually with angle, but it falls short for amplitude-versus-offset analysis, where those same angle-dependent effects are the signal.

\subsection{Marchenko Without Up/Down Decomposition}
\label{sec:nodecomp}

The standard Marchenko equations require the reflection response to be \emph{decomposed} into upgoing and downgoing components. This decomposition requires knowledge of velocity and density at the acquisition surface and relies on a flat, horizontal surface. For land surveys with rough topography or uncertain near-surface properties, this decomposition can introduce errors.

The deeper objection to decomposition is not that it is awkward to apply but that it embeds an assumption about the wavefield: that at the focal point the field propagates purely up or down. At large offsets in layered media, and where refracted waves or near-surface inhomogeneity are present, that assumption fails and the decomposed algorithm degrades with it \cite{kiraz2021marchenko}.

Kiraz et al.\ \cite{kiraz2021marchenko} retrieved the Green's function on the Marmousi model without using the up/down components at all, and obtained a match with the numerically modelled Green's function. Their secondary observation is a caution worth carrying. Refracted waves were retrieved, and independently of whether the acquisition was single- or two-sided; but refracted arrivals that precede the first primary contradict the requirement that the Green's function vanish before the direct wave. Tracing the cause, they found these events come from injecting the direct wave into sufficiently detailed background velocity and density models, not from the Marchenko algorithm operating on the recorded wavefield. The lesson generalises: an event appearing in a retrieved Green's function is not thereby a product of the retrieval, and where it violates a causality condition the background model is the first place to look.

\subsection{Velocity-Independent Marchenko Method}

The initial estimate $f_{1,d}^+$ requires a background velocity to compute travel times, and Sripanich et al.\ \cite{sripanich2019velocity} showed where that requirement actually bites. Reformulating the method in the time-imaging domain, they observed that the direct-wave traveltime is the Cheop's pyramid familiar from time-domain processing, which can be read directly off the local slopes of common-midpoint gathers. The velocity-model-based focusing operator is then replaced by a data-driven slope estimate, and the focal point is specified by vertical time rather than by depth.

Their conclusion is a sharpening rather than an elimination: the prior velocity model is needed only to state the focal position in depth, and that need disappears if one is content to work in vertical time. Where the usual time-imaging assumption of mild lateral heterogeneity holds, the retrieved Green's functions are of comparable quality to those obtained with a known velocity model. Where it does not, the time-to-depth correspondence the method rests on fails, so this is not a route around the strongly heterogeneous near-surface of Section~\ref{sec:challenges}.

\subsection{Marchenko Equations for Electromagnetic Media}
\label{sec:EM}

The Marchenko framework extends to electromagnetic wave propagation \cite{slob2013coupled, slob2016electromagnetic}. Maxwell's equations and the reciprocity theorems for electromagnetic fields have the same convolution- and correlation-type structure as the acoustic ones \cite{wapenaar1996reciprocity, wapenaar1996reciprocityb}, so the derivation follows the same path once the field has been decomposed appropriately.

That decomposition is where the electromagnetic case differs. Slob and Wapenaar \cite{slob2013coupled} split the field twice: into up- and downgoing constituents, as in the acoustic case, and additionally into transverse electric and transverse magnetic modes. In a vertically transverse isotropic layered medium whose permittivity and permeability vary smoothly in the horizontal direction within each layer, the TE and TM modes decouple and each can be carried through the Marchenko scheme separately, then recombined. The result is a pair of coupled Marchenko equations per mode, yielding the up- and downgoing Green's functions for a virtual receiver at depth, which corresponds physically to a virtual vertical radar profile. Zhang and Slob \cite{zhang2016electromagnetic, zhang2017electromagnetic} built the corresponding internal multiple elimination scheme on top.

The application these results point to is ground-penetrating radar, where interbed reverberation obscures deeper reflectors much as it does in seismic data and where the retrieved Green's functions support true-amplitude imaging free of internal multiple effects \cite{slob2013coupled}. Demonstrations remain numerical. Whether the same construction reaches the diffusive regime of controlled-source electromagnetics is a separate question: the causality and time-windowing arguments that separate focusing functions from Green's functions presuppose propagating wavefronts with a well-defined arrival time, and that premise is what a diffusive field lacks.

\subsection{Marchenko Method in Dissipative and Attenuative Media}
\label{sec:dissipative}

As described in Section~\ref{sec:greens}, the standard Marchenko equations assume a lossless medium, and the dissipative extension of Slob and co-workers \cite{slob2016green, marchenko2016dissipative} buys exactness at the price of double-sided input. Wapenaar et al.\ \cite{wapenaar2022marchenko} reach the same setting through propagator and transfer-matrix representations.

Cui et al.\ \cite{cui2018marchenko} implemented the dissipative scheme numerically and experimentally in a one-dimensional acoustic wave tube, 3D-printed in sections so that impedance contrasts could be created by varying the tube diameter. Working from double-sided scattering measurements, they retrieved the focusing and Green's functions inside the tube \emph{without any prior knowledge of the medium properties}, and found offline processing and real-time focusing to be comparably accurate. True-amplitude Green's functions were also recovered, but required a scaling correction.

The honest practical position is narrower than the theory might suggest. Because the effectual reflection response cannot be measured from one side, the exact dissipative scheme does not transfer to surface seismic acquisition. What is done instead in exploration practice is to apply $Q$ compensation to the reflection data before running the lossless scheme, which is an approximation whose quality degrades as attenuation grows. Thorbecke et al.\ \cite{thorbecke2013green} quantified what happens when attenuation is ignored altogether: for a medium with $Q = 50$, ghost events appeared in the retrieved upgoing Green's function while the downgoing one stayed clean, and the artefacts were reducible by $Q$ compensation applied beforehand. Attenuating targets such as gas-bearing sands and fractured carbonates therefore remain accessible, but through a corrected lossless scheme rather than through the dissipative equations.

\subsection{Plane-Wave and Virtual Plane-Wave Marchenko}
\label{sec:planewave}

Section~\ref{sec:imaging} introduced the areal-source construction of Meles et al.\ \cite{meles2018virtual}, in which the focusing function focuses in time across a whole depth level and one solve serves that level. That is the zero-slowness case. Generalising it, a plane-wave focusing function focuses at all points of $\dD_i$ simultaneously with linear moveout corresponding to a horizontal slowness $p$, and the family over $p$ restores the angular illumination that the single areal source loses \cite{almobarak2021plane}. The plane-wave Marchenko equations are:
\begin{align}
f_1^{-,\text{pw}}(p, t) &= \Theta \int R(p, t')\, f_1^{+,\text{pw}}(p, t - t')\, dt'
\end{align}
\begin{align}
f_1^{+,\text{pw}}(p, t) &= f_{1,d}^{+,\text{pw}}(p, t) + \Theta \int R(p, t') \nonumber \\
&\qquad \times f_1^{-,\text{pw}}(p, t' - t)\, dt'
\end{align}

\noindent where $R(p, t)$ is the plane-wave reflection response obtained by linear moveout slant-stacking. Imaging then sums over slownesses rather than over focal points, so the number of Marchenko solves is set by the number of slownesses retained rather than by the lateral sampling of the target. Almobarak et al.\ \cite{almobarak2021plane} applied this to field data. How many slownesses are needed is the governing trade-off: too few and the angular illumination is as poor as the single areal source, too many and the saving over the point-source scheme disappears.

\subsection{Marchenko Equations for Imperfectly Sampled Data}

Real acquisition grids are non-uniform: missing shots, missing receivers, cable feathering, and near-offset gaps all produce an incomplete reflection dataset. The standard Marchenko equations assume a uniformly sampled, gap-free reflection response. Several strategies handle imperfect sampling \cite{van2021imperfect, wapenaar2020discrete}:

\begin{enumerate}
    \item \textbf{Data reconstruction:} Interpolate and regularise the data to a uniform grid before applying the Marchenko equations. Methods include minimum-weighted norm interpolation, anti-leakage Fourier methods, and deep-learning-based reconstruction.
    \item \textbf{Compressive-sensing acquisition:} Design the survey to be sparse and irregular from the outset, then reconstruct the dense response by exploiting the sparsity of seismic data in a suitable transform domain before the Marchenko equations are applied. Zhang \cite{zhang2020marchenko} combined a randomised survey design with post-acquisition compressive reconstruction on this principle.
    \item \textbf{Summation representations:} Attack the discretisation itself. The Marchenko representations are integrals over sources and receivers, replaced in practice by finite summations, and it is that replacement which degrades under imperfect sampling. Wapenaar and van IJsseldijk \cite{wapenaar2020discrete, van2021imperfect} reformulated the representations directly as summations that account for irregular sampling, introducing point-spread functions in the manner of multidimensional deconvolution for irregular source distributions. This yields representations valid for imperfectly sampled data; resolving them to recover the focusing functions is, by their own account, ongoing research rather than a finished method.
\end{enumerate}

\subsection{Rayleigh--Marchenko Redatuming}

The requirements the standard scheme places on its input are what have most restricted it in practice: an accurately deconvolved source wavelet including its absolute scaling factor, sources and receivers co-located at the same horizontal positions, and removal of the direct wave, the ghosts, and the surface-related multiples. Ravasi \cite{ravasi2017rayleigh} showed that most of these can be dropped by combining the coupled Marchenko equations with a one-way version of the Rayleigh integral representation.

The price of admission is dual-sensor data, that is, pressure together with vertical particle velocity. Given that, the resulting \emph{Rayleigh--Marchenko} scheme handles internal and free-surface multiples together, works with band-limited data whose source signature is unknown, and no longer requires sources and receivers to be co-located. What it requires instead is a line of regularly sampled receivers with arbitrarily positioned sources somewhere above them, a description that fits ocean-bottom acquisition, source-over-spread streamer geometries, and horizontal borehole surveys alike.

Removing the absolute-scaling requirement matters because it removes a step rather than improving one: the wavelet inversions of Section~\ref{sec:challenges} recover that factor, whereas Rayleigh--Marchenko never needs it. Ravasi validated the scheme on synthetic and field data and used the retrieved wavefields both for structural imaging and to compute true-amplitude angle gathers, which is what makes it a route to quantitative amplitude-versus-angle interpretation rather than structural imaging alone. Reciprocal and upside-down variants have since adapted it to the sparse, irregular geometries of real ocean-bottom surveys \cite{ravasi2023reciprocal, wang2023upside, wang2024upside}, at roughly double the input data and computational burden of the standard scheme \cite{wang2025accelerating}, which is what motivated the machine-learning acceleration of Section~\ref{sec:ML} \cite{wang2025efficient}.

\section{Practical Implementation}
\label{sec:implementation}
\subsection{The Standard Iterative Scheme}

Thorbecke et al.\ \cite{thorbecke2017implementation} described a complete 2D implementation of the Marchenko method for synthetic data using the Neumann iteration of Section~\ref{sec:iterative}. The main computational steps are:

\begin{enumerate}
    \item \textbf{Data preparation:} Load the reflection data $R(\xH, \xH', t)$, apply source and receiver deghosting, and compute the direct-wave traveltime $t_d(\xH, \xA)$ for the focal point $\xA$.
    \item \textbf{Initialise focusing function:} Set $f_1^{+,0}(\xH, \xA, t) = T_d^{\,\mathrm{inv}}(\xH, \xA, t)$, a scaled delta at $t = -t_d$, using the background model.
    \item \textbf{Compute upgoing response:} Convolve $f_1^{+,k}$ with $R$ (a multi-dimensional convolution), apply the time-window $\Theta$, and store as $f_1^{-,k+1}$.
    \item \textbf{Compute downgoing update:} Time-reverse $f_1^{-,k+1}$, convolve with $R$, apply $\Theta$, add to $f_{1,d}^+$, and store as $f_1^{+,k+1}$.
    \item \textbf{Convergence check and repeat until converged.}
    \item \textbf{Retrieve Green's functions:} Apply equations \eqref{eq:Gm}--\eqref{eq:Gp}.
    \item \textbf{Form image:} Apply imaging condition \eqref{eq:IC1} or \eqref{eq:IC2}.
\end{enumerate}

The dominant computational cost is the multi-dimensional convolution in steps 3-4: for a 2D survey with $N_s$ surface points and $N_t$ time samples, each iteration requires $O(N_s^2 N_t \log N_t)$ operations.

\subsection{Least-Squares Implementation in Julia}

de Paula et al.\ \cite{de2021marchenko} implemented least-squares Marchenko imaging in Julia, in the matrix-free linear-operator style popularised by \href{https://github.com/PyLops/pylops}{PyLops} \cite{ravasi2020implementation, ravasi2020pylops}. Julia's just-in-time compilation and native support for complex arithmetic suit large-scale linear-algebra-based seismic processing. The implementation solves the damped least-squares system with LSQR (from the \texttt{IterativeSolvers} package) and applies the Marchenko operators on the fly, so the system matrix is never assembled. It was validated on a 2D synthetic model against an RTM image of the same data.

\subsection{GPU-Accelerated and Multi-GPU Implementation}
\label{sec:gpu}

The multi-dimensional convolution and space integration of the Marchenko iteration, which Thorbecke et al.\ \cite{thorbecke2017implementation} call \emph{synthesis}, parallelises across source positions, receiver positions, and frequencies, and is therefore well suited to GPUs. Koehne et al.\ \cite{koehne2021multi} implemented it in CUDA C on a node with four Tesla V100 cards and benchmarked it against the same code on 36 CPU cores, over reflection datasets of 3.3, 30, and 249 GB. Formulating the kernel as a segmented dot product and reducing it with warp-shuffle instructions through the CUDA CUB library, they reached synthesis speedups of 3.7 to 6.0 over the 36-core version.

Those figures apply when many focal points are batched together. For a single focal point the GPU version was slower than the CPU version, with speedups of 0.34 to 0.51, because data loading and host-to-device transfer then dominate: reading the reflection response and first arrivals accounted for 37\% of GPU runtime on the smallest dataset and 87\% of CPU runtime. The break-even point fell from roughly 35 batched focal points on the 3.3 GB dataset to about 3 on the 249 GB dataset, where the data no longer fit in GPU memory and had to be pipelined. That last case is the informative one for 3D, where pipelining is unavoidable.

\subsection{Machine Learning Approaches}
\label{sec:ML}

Machine learning enters Marchenko imaging as a surrogate for the iterative solve. The cost that motivates it is structural: each focal point requires its own solution of the Marchenko equations, whether by truncated Neumann series or by an LSQR inversion, and the Rayleigh--Marchenko variants roughly double both the input data and the computational burden \cite{wang2025accelerating}.

\subsubsection*{Self-Supervised Prediction of Focusing Functions}

Wang et al.\ \cite{wang2025accelerating} replace the inversion with a trained U-Net. The network learns to map an initial estimate of the \emph{upgoing} focusing function, cheap to compute from the smooth velocity model and the surface reflection response, to its converged counterpart. Training labels come from a small subset of focal points in the target area for which the converged focusing functions have been computed with an LSQR solver. Once $f_1^-$ is predicted at the remaining focal points, the downgoing focusing function and the Green's functions follow from a direct forward evaluation of the Marchenko equations, with no iteration. Reported reduction in computation time is up to $50\times$.

The scheme is self-supervised in the sense that matters for field data: every training example comes from the survey being imaged, so no external training set and no ground-truth Green's function is required \cite{wang2025accelerating}. That is also its limitation, since the network is fitted to one imaging area and carries no guarantee of transfer to another. The method was validated first on synthetic data and then on Volve field data, where the images were comparable to those from the conventional iterative scheme and showed less signal leakage.

\subsubsection*{Physics-Informed Approaches}

A different route encodes the coupled equations \eqref{eq:marchenko1}--\eqref{eq:marchenko2} into the loss function itself, in the manner of physics-informed neural networks \cite{raissi2019physics}, penalising the network for violating the Marchenko relations rather than for departing from a precomputed reference. Wang and Alkhalifah \cite{wang2026deep} move in this direction by embedding the inverse solve in an optimisation-based network. A formulation that solves the Marchenko system for arbitrary focal points without any survey-specific training has not been demonstrated, and remains open.

\subsubsection*{Deep-Learning-Based Data Reconstruction}

A critical pre-processing step for Marchenko imaging is the reconstruction of missing offsets and interpolation to a regular grid. Deep learning methods, among them U-Nets, GANs, and transformer architectures, have been applied to seismic data reconstruction \cite{meng2021self, dodda2023simultaneous, liu2022seismic}, providing cleaner input for the Marchenko equations and reducing artefacts in the retrieved Green's functions.

\subsection{Compressive-Sensing Acquisition for Marchenko Imaging}

The Marchenko method needs densely and regularly sampled shots and receivers, which is precisely what makes it expensive to acquire for. Zhang \cite{zhang2020marchenko} attacked the cost at the acquisition stage rather than in the algorithm: a randomised survey with sparse, irregular receivers, followed by compressive-sensing reconstruction of the dense response, which is then fed to the standard Marchenko scheme. Because seismic data are sparse in a suitable transform domain, the dense response can be recovered from measurements below the Nyquist requirement.

Two aspects of the result matter. The Green's functions retrieved from the reconstructed data were comparable to those from a simulated dense acquisition, so the reconstruction step does not obviously degrade what the Marchenko equations then produce. And the reconstruction carries a natural denoising effect, which is useful given that the iteration amplifies coherent noise falling inside the time window. This is a different strategy from solving the Marchenko equations directly on incomplete data, which is the route taken by Wapenaar and van IJsseldijk \cite{wapenaar2020discrete, van2021imperfect}.

\section{Field Data Applications}
\label{sec:fielddata}

\subsection{Marine Seismic Applications: Santos Basin}

The Santos Basin, offshore Brazil, is where source-receiver Marchenko redatuming was first carried through on field data \cite{staring2018marchenko, staring2018source}. Its oil-bearing carbonate reservoirs sit beneath a highly reflective stratified salt layer, and that layer generates internal multiples strong enough that conventional migration interprets them as primaries from deeper reflectors, placing ghost reflectors in the target zone where none exist. Applying adaptive double focusing produced a redatumed dataset with virtual sources and receivers below the salt, and a sub-salt image with the multiple-related noise reduced. Staring and Wapenaar \cite{staring2020three} then took the same method to narrow-azimuth 3D streamer data from the basin, which is the step from a 2D demonstration to something resembling a production geometry.

\subsection{North Sea: Troll and Volve Fields}

Ravasi et al.\ \cite{ravasi2016target, ravasi2015volve} produced the first real target-oriented Marchenko images of a North Sea field, the Volve field, and the interesting part is what the focusing functions let them do rather than what they removed. Because the coda of the downgoing focusing function consists entirely of transmission-born multiple scattering, it can be used to \emph{synthesise underside illumination} from surface reflection data alone. They reconstructed underside reflections without first having to locate or image any reflector, which had not been done on field data before, and the resulting images showed features absent from a conventional migration of the same surface data. That reframes the multiples from contamination into a source of illumination that single-sided acquisition otherwise cannot provide.

da Costa Filho and Curtis \cite{da2018marchenko} later combined Marchenko-based and interferometric multiple attenuation on a 2D line from another North Sea field, reducing multiple energy and revealing structure the multiples had obscured. Wang et al.\ \cite{wang2025accelerating} used Volve field data to test learned prediction of the focusing functions against the iterative solution.

van IJsseldijk et al.\ \cite{brackenhoff2020timelapse} applied Marchenko-based isolation to a time-lapse marine dataset from the Troll field, off Norway. Isolating the reservoir removes both overburden and underburden interference from the target response, so the baseline and monitor can be compared without the static geology and its associated non-repeatable noise sitting in the difference. The target was the small traveltime difference in the reservoir, which is the quantity most easily lost in unredatumed 4D data.

\subsection{Subsalt Imaging: Gulf of Mexico}

Subsalt imaging tests a seismic imaging method about as hard as any setting does. Jia et al.\ \cite{jia2017subsalt} applied Marchenko redatuming and imaging to a marine towed-streamer line from the Gulf of Mexico, and their contribution is as much about what field data demands as about the images themselves. They singled out two obstacles. The first is the missing near offsets of a towed-streamer record, for which they give a criterion to decide whether reconstruction is actually necessary. The second is calibration: the reflection response and the first-arrival estimate enter the scheme as two separate inputs, and any mismatch between their amplitudes generates artefacts that survive into the image, so the relative scaling has to be set correctly rather than assumed.

Their conclusion is measured. The Marchenko images were consistent and, for the most part, comparable with conventional migration; the gains were in the continuity of geological structures and in the suppression of artefacts that they attribute to internal multiples. Jia et al.\ \cite{jia2018practical} followed with a practical implementation on a Gulf of Mexico dataset that works through the pre-processing and calibration in detail.

Taken together, the two papers make a point that recurs across the field-data literature: on a real subsalt line the Marchenko method does not so much outperform RTM as produce a cleaner version of it, and most of the work lies in pre-processing rather than in the inversion.

\subsection{Middle East Applications: Land Seismic}

Land seismic surveys present distinct difficulties for the Marchenko method: a strongly heterogeneous near-surface that causes statics problems, less regular sampling, and significant mode conversion. Marchenko results had accumulated offshore well before a convincing onshore application existed.

Cheng et al.\ \cite{cheng2024marchenko} supplied one, on a hydrocarbon field in Kuwait. The problem there is specific. Predominantly horizontal strata and low-relief structure generate strong short-period internal multiples whose character is nearly indistinguishable from the primaries reflected by the underlying Jurassic reservoirs. The industry remedy, multiple prediction followed by adaptive subtraction, is precisely the wrong tool when prediction and target look alike, because subtraction then carries a high risk of damaging primary amplitudes and with them any quantitative interpretation. This is the failure mode van der Neut et al.\ \cite{van2015practical} identified in the abstract: adaptive schemes work when the internal multiples do not interfere with the primaries, and Kuwait is a setting where they do. The attraction of the Marchenko method in that setting is that it retrieves Green's functions for target-oriented imaging without an adaptive subtraction step at all.

Their study integrated well logs, vertical seismic profiling, and surface seismic. Two of their findings are diagnostic rather than methodological: the poor image at the centre of the field is caused by destructive interference of internal multiples, and reverberation between evaporite formations in the overburden is the most likely generator of the multiples that degrade the Jurassic reservoir image. The VSP data served two purposes, and neither is what a reader might assume. The recorded downhole Green's functions were used to cross-check the Marchenko-retrieved ones, which they were found to match, and they were used to estimate the scaling factor of the Marchenko method, the quantity discussed in Section~\ref{sec:challenges}. Where a well is available this is the most direct route to that factor, though as noted there it is not the only one. Marchenko imaging improved the images of the Jurassic formations.

\subsection{Carbon Sequestration Monitoring}

Marchenko methods have also been applied to carbon capture and storage (CCS), where time-lapse monitoring must track injected CO$_2$ migration. Kiraz \cite{kiraz2018marchenko} worked with the vertical seismic profiling data from the Frio brine pilot in Texas, using the pre- and post-injection z-component records from a single shot. Marchenko redatuming placed a virtual source at depth and an inversion recovered the reflectivity, which was then compared against the corridor stack from conventional VSP processing of the same records. Differencing the pre- and post-injection results both ways allowed the injection-zone response from the Marchenko inversion to be set against the conventional 4D difference directly. Note that the configuration is borehole rather than surface acquisition, so the result speaks to VSP-based monitoring rather than to redatuming of a surface baseline and monitor pair.

\subsection{Vertical Seismic Profiling Applications}

Borehole recordings enter Marchenko workflows in two quite different ways, and the distinction matters.

The first is validation and calibration. Cheng et al.\ \cite{cheng2024marchenko} used downhole Green's functions to check the Marchenko-retrieved ones and to estimate the scaling factor, neither of which the surface data can supply on its own.

The second is illumination, and it addresses a limit that no amount of processing on surface data can lift. A vertical seismic profile provides a second acquisition boundary, and the extra illumination it brings is precisely what single-sided geometry lacks. Lomas et al.\ \cite{imaging2019vsp} exploited this to image what surface-only Marchenko imaging cannot: vertical interfaces and steeply dipping structures. On a variable-density synthetic and on a modified Marmousi~2 model, their VSP Marchenko scheme imaged horizontal and vertical structures alike, with the best results obtained by combining VSP-based images with standard Marchenko images rather than replacing one with the other. Applied to fault structures, the combination reduced internal multiple contamination and improved the imaging of the faults themselves. This is the practical counterpart of the shadow-zone limitation in Section~\ref{sec:challenges}: what single-sided acquisition cannot illuminate, a borehole boundary sometimes can.

\subsection{Laboratory Experiments}

Laboratory acquisition sits usefully between synthetic and field data: the medium is known, but the measurement is real. da Costa Filho et al.\ \cite{da2018using} used a submerged ultrasonic setup, a water column above steel, brass, and steel plates, recorded with a 128-element 2.25~MHz linear array at 0.7~mm element spacing. The water column was deep enough that several orders of internal multiple arrive before the water--air reflection, which isolates the internal multiple problem cleanly.

Their result is a caution rather than a confirmation, and it is worth stating plainly. Marchenko imaging improved on reverse-time migration, but by less than synthetic experiments had led them to expect, which they attribute to Marchenko assumptions being violated in the experiment and, by extension, in seismic data. The specific culprit is scaling. Marchenko theory requires the reflection response to be scaled to a particular absolute magnitude, and a real measurement supplies only $a R$ for some unknown factor $a$. Methods that depend on that absolute scaling proved ill suited to the data; methods that use adaptive subtraction, which absorbs the scaling error into the matching filter, performed better. Reliable estimation of $a$ without additional information such as VSP data remains unavailable \cite{da2018using}.

Cui et al.\ \cite{cui2018marchenko} took the dissipative case into the laboratory using a 1D acoustic wave tube built from 3D-printed cylindrical sections, in which impedance contrasts come from changes in tube diameter and the attenuation is that of air at room conditions, quantified beforehand with a nearly constant-$Q$ model. Retrieving the focusing and Green's functions from double-sided scattering measurements, without prior knowledge of the medium, validated the dissipative equations experimentally and reconciled them with the numerical implementation.

\section{Time-Lapse and Monitoring Applications}
\label{sec:timelapse}

\subsection{The 4D Seismic Challenge}

Time-lapse (4D) seismic monitoring aims to image changes in the subsurface such as fluid migration, saturation change, compaction, or pore-pressure change, by comparing baseline and monitor surveys acquired at different times. The primary challenge is separating genuine subsurface changes from non-repeatable noise: differences in acquisition geometry, weather, tides, and near-surface conditions between surveys. Careful normalisation, surface-consistent corrections, and noise suppression are required. The Marchenko method addresses one specific component of 4D noise which is the overburden contribution to the 4D difference. If the overburden does not change between surveys, a reasonable assumption for passive reservoirs overlain by stable shale, then any change in the overburden response are non-repeatable noise. Removing the overburden response by Marchenko redatuming suppresses this noise channel \cite{brackenhoff2020timelapse}.

\subsection{Marchenko-Based Virtual Seismology for Time-Lapse}

The single-sided homogeneous Green's function representation of Wapenaar et al.\ \cite{wapenaar2016homogeneous}, and the virtual-source and virtual-receiver constructions built on it \cite{brackenhoff2020virtual}, allow virtual experiments to be run with both sources and receivers at depth, free of overburden contamination. Applied to time-lapse monitoring, this produces a 4D signal that reflects only subsurface changes below the virtual datum, eliminating both overburden multiples and non-repeatable overburden noise in a single step.

van IJsseldijk et al.\ \cite{brackenhoff2020timelapse} demonstrated this on the Troll field, isolating the reservoir response from the overburden and underburden and retrieving time-lapse traveltime shifts that are difficult to detect reliably in the surface data. Their scheme turned out to do more than isolate the primaries: it also estimated internal multiples that fall outside the recording time, and both the primaries and those multiples could then be used to measure the traveltime difference. That is worth noting, because it inverts the usual framing. The multiples are not merely removed as contamination but carry an independent measurement of the same time-lapse change. The requirement is that the focusing functions stay stable between surveys, which holds when the overburden does not change appreciably between acquisitions.

\subsection{Carbon Sequestration Monitoring}

For CO$_2$ injection monitoring at a CCS site, the signal of interest, namely the change in reflection amplitude and traveltime from the growing CO$_2$ plume, is small compared with the total reflection amplitude, and non-repeatable overburden noise can mask it entirely. The Frio VSP study of Kiraz \cite{kiraz2018marchenko} and the Troll field study \cite{brackenhoff2020timelapse} point the same way: isolating the target zone before differencing makes time-lapse changes detectable that are ambiguous in the unredatumed data. Both remain single-site demonstrations, and neither has been repeated at the survey scale that operational CCS monitoring would require.

\section{Connections to Other Methods}
\label{sec:connections}

\subsection{Seismic Interferometry}

The Marchenko method and seismic interferometry both retrieve Green's functions from existing data. The difference is acquisition geometry. Seismic interferometry as traditionally formulated requires sources on a closed boundary surrounding the medium (double-sided illumination); the Marchenko method requires only single-sided illumination at the acquisition surface. The cost of this advantage is the need to solve the Marchenko integral equations, which requires an initial traveltime estimate, whereas interferometry requires only a cross-correlation and no model input.

van der Neut et al.\ \cite{van2014interferometric} read the iterative scheme as a sequence of linear filters acting on the initial focusing function, each combining a multidimensional crosscorrelation with the reflection response, a time reversal, and a truncation in time. Interpreting the crosscorrelation interferometrically, as the subtraction of traveltimes along stationary raypaths, makes the scheme's reach visible.

Their conclusion is a real constraint on what the method can deliver. An internal multiple arriving at the focal point from a particular angle is retrieved only if the initial focusing function, that is, the direct wave, reached the focal point from that same angle. A shadow zone that primary reflections cannot illuminate therefore cannot be illuminated by internal multiples either, at least not through the standard iterative scheme \cite{van2014interferometric}. This is worth setting against the imaging-condition results of Section~\ref{sec:imaging}: multiples do contribute physical illumination there, but only under double-sided acquisition. Under single-sided illumination they suppress artefacts rather than extend coverage. The same analysis notes that the scheme is designed for layered media with smooth interfaces and performs less well where point scatterers dominate, because the underlying ansatz is then imperfectly satisfied.

\subsection{Reverse-Time Migration}

Conventional RTM computes the source wavefield $S(\xv, t)$ by forward modelling in the velocity model, and the receiver wavefield $W(\xv, t)$ by back-propagating the data. The image is the zero-lag cross-correlation of $S$ and $W$:
\[
I_{\text{RTM}}(\xv) = \int_{-\infty}^{\infty} S(\xv, t)\, W(\xv, t)\, dt
\]

Marchenko imaging replaces $S$ and $W$ with the Marchenko-retrieved Green's functions $G^+$ and $G^-$, which carry the full multiple content. Within the target zone, and under the acoustic, well-sampled, single-sided assumptions of the derivation, this reproduces what RTM would give in a model that matched the true overburden, without that model being known \cite{wapenaar2014marchenko}. The equivalence is local rather than global: it holds for the reflectivity at and below the focal level, and it inherits whatever error the traveltime estimate $t_d$ carries. The Marchenko approach is a data-driven alternative to model-driven RTM, worth its extra cost when the overburden is complex and the velocity model is imprecise.

\subsection{Full-Waveform Inversion}

FWI iteratively updates a velocity model by minimising the misfit between observed and modelled data, over the full domain in which the waves propagate. Marchenko redatuming confines that domain to a target zone, as described in Section~\ref{sec:targeting}, which is where the cost saving comes from: a smaller model to update and a smaller volume to simulate, rather than faster convergence per iteration \cite{cui2020marchenko}. The retrieved wavefields also retain the multiple scattering that a Born-approximation method linearises away, so the target-zone objective function is sensitive to the target medium without further assumptions about the medium outside it.

\subsection{Inverse Scattering Series}

The ISS \cite{weglein1997inverse, weglein2003inverse} expresses the subsurface medium perturbation as a series in the data:
\[
m(\xv) = m^{(1)}[\mathbf{d}] + m^{(2)}[\mathbf{d}, \mathbf{d}] + m^{(3)}[\mathbf{d}, \mathbf{d}, \mathbf{d}] + \cdots
\]

where $m^{(k)}$ involves $k$-fold products of the data. The first term is the Born approximation and the higher terms correct for multiple scattering. Both methods therefore account for the same physics, and Broggini and Snieder \cite{broggini2012connection} set out the family resemblance between inverse scattering, focusing, Green's function reconstruction, and imaging: the governing equations share a functional form, and what links them is the interaction between the causal and anti-causal Green's functions.

The structural difference lies in how each is solved. The ISS exists only as a series, so where it fails to converge there is nothing to fall back on. The Marchenko equations are a Fredholm system over a finite time window, which admits direct solution when an iterative expansion does not contract. The Marchenko approach can be read as a resummation of the multiple-scattering content that the ISS accumulates term by term.

\subsection{Deconvolution-Based Imaging}

Multidimensional deconvolution (MDD) \cite{wapenaar2011virtual} retrieves the Green's function by least-squares deconvolution of an upgoing field by a downgoing one, and requires receivers at the redatuming level. The Marchenko method reaches the same result without them, replacing the physical measurement at depth with the temporal causality constraint on the focusing function. Their relation is close enough that Marchenko redatuming can be read as interferometry by deconvolution rather than by cross-correlation \cite{van2014interferometric}. The trade is a real one in both directions: MDD needs the downhole or seabed array but inherits no dependence on a traveltime estimate, while the Marchenko method needs only surface data but is sensitive to errors in $t_d$ and, in its iterative form, assumes regular sampling.

\section{Challenges, Limitations, and Open Research Problems}
\label{sec:challenges}

The constraints below are ordered roughly by how much they currently restrict practical application, beginning with those that admit workarounds and ending with those that do not.

\subsection{Free-Surface Effects and Acquisition Artefacts}

The standard Marchenko equations assume a free-surface-multiple-free reflection response. In practice, SRME removes free-surface multiples before the Marchenko equations are applied. SRME is effective at moderate water depths but degrades in shallow water, where multiple generators overlap in time with primaries (severe multiple crosstalk). In these cases, the Marchenko equations can be reformulated to include free-surface multiples explicitly \cite{davydenko2017full, wapenaar2021marchenko}, but at the cost of additional computation and more careful time-windowing.

\subsection{Elastic Effects and Mode Conversion}

The acoustic approximation ignores P-to-S mode conversion, which is significant on land and in marine surveys with hard seafloor geology. Elastic internal multiples convert between modes at each interface; the acoustic Marchenko equations cannot distinguish between P-P and P-S arrivals. For deep-water marine surveys with soft seabeds, mode-converted energy is usually a small fraction of the recorded wavefield; for land surveys it can dominate. The elastodynamic Marchenko equations address this in principle but require multi-component data and substantially higher computational cost \cite{da2016elastic}. Land applications of the method are, in consequence, harder than marine ones and are not yet routine.

\subsection{Accuracy of the Initial Estimate}

The initial estimate $f_{1,d}^+$ is the most sensitive input to the Marchenko iteration. Errors in the background velocity model propagate into errors in the traveltime $t_d(\xH, \xA)$, which shift the time-window operator $\Theta$ and contaminate the focusing functions. For smooth overburdens with small velocity errors, the retrieved Green's functions degrade gracefully \cite{thorbecke2013green}. For strong lateral velocity contrasts, such as salt bodies or carbonates, errors can be much larger, and model-updating strategies are needed \cite{chen2020marchenko}.

\subsection{Validity of the Time-Separation Condition}

Everything in this review depends on one assumption, and it is worth isolating because it is usually left implicit. The window $\Theta$ separates the focusing functions from the Green's functions only if their supports do not overlap, that is, if the coda of the focusing function is confined strictly inside $|t| < t_d$ while the Green's function arrives only at and after $t_d$.

That condition is exact for 1D scalar waves. In three dimensions it is not exact but approximate, and Wapenaar and Slob \cite{wapenaar2014multicomponent} state its domain of validity precisely: it holds in layered media with moderately curved interfaces at finite horizontal source--receiver offsets, and it may be violated for more complex media, at large offsets, or both. Since complex media at large offsets is a fair description of the settings where the method is most wanted, this is not a marginal caveat. It is also the cleanest explanation of why aperture limits and steep dips degrade the retrieval beyond what illumination arguments alone would predict. For vectorial waves the condition has not been investigated at all, which is the subject of Section~\ref{sec:extensions}.

\subsection{Absolute Scaling of the Reflection Response}

The derivation assumes $R$ is the impulse reflection response scaled to a particular absolute magnitude, namely twice the vertical velocity response to a negative pressure impulsive point source. A field or laboratory measurement delivers $aR$ for some unknown factor $a$ that absorbs source strength, receiver sensitivity, and the residual of wavelet deconvolution. The Marchenko equations are not invariant to $a$: because the iteration multiplies $R$ by itself, an error in the scaling compounds with the order of the multiple being predicted, so the retrieved coda is wrong by a factor that grows with iteration number even when the kinematics are right.

The consequences are documented across settings. On real ultrasonic data, da Costa Filho et al.\ \cite{da2018using} found that methods relying on the absolute scaling underperformed what synthetic experiments had promised, synthetics having fixed $a$ by construction. On a Gulf of Mexico line the same issue appears as calibration: Jia et al.\ \cite{jia2017subsalt, jia2018practical} traced image artefacts to the amplitude mismatch between the scheme's two inputs, the reflection response and the first-arrival estimate.

Several routes to estimating $a$ have been proposed, and they differ in what they require. Thomsen calibrated against VSP data, and Cheng et al.\ \cite{cheng2024marchenko} did the same on their Kuwait dataset, which works where a well exists and not otherwise. Ravasi et al.\ \cite{ravasi2016target} instead scaled the reflection data using the convergence behaviour of the Marchenko solution itself. Van der Neut et al.\ minimised a misfit built on the upgoing part of the redatumed Green's function, an approach later shown not always to recover the correct scale.

The most complete treatment comes from Mildner et al.\ \cite{mildner2019wavelet}, who moved the misfit from the Green's functions to the \emph{focusing functions}. An incorrect wavelet leaves artefact energy in the focusing functions, so minimising that energy inverts not merely for a scalar but for the frequency-dependent amplitude spectrum and the phase of the source wavelet, and in an extended form for its lateral variation across the survey. Applied to a deep-water sub-salt dataset from the Gulf of Mexico, the procedure iteratively and automatically updated the source function and produced Marchenko images with visibly better reflector continuity than either surface-based images or Marchenko images computed without the inverted wavelet \cite{mildner2019source}.

The problem is therefore solvable from surface data alone, which was not clear when the method was first applied to field data. Two qualifications remain. None of these procedures has become a standard step, so published Marchenko results differ in how, and whether, the scaling was determined. And the common fallback of adaptive subtraction \cite{van2016adaptive, staring2018source, staring2020three} sidesteps the estimation rather than performing it, absorbing the error into a matching filter; that works, but it converts an exact method into a partly heuristic one and the size of the correction is rarely reported.

\subsection{Missing Near Offsets and Bandwidth Limitations}

The Marchenko equations involve the reflection response at all offsets, including zero offset. Marine acquisition cannot supply that: the physical separation between the source and the nearest live channel leaves a near-offset gap of the order of a hundred metres, and the gap is a property of the geometry rather than of the processing. Near-offset reconstruction by interpolation or inversion is therefore a prerequisite, and imperfect reconstruction propagates into the focusing functions, most severely for shallow targets where the missing offsets carry a large share of the relevant illumination \cite{wapenaar2020discrete}.

Bandwidth constrains the method at the other end of the spectrum. The lowest few hertz are the hardest part of the band to record reliably, and they are the part that constrains long-wavelength impedance contrasts, so their absence degrades amplitude recovery in the focusing functions. The limitation is not specific to the Marchenko method: it restricts conventional full-waveform inversion in the same way and for the same reason.

\subsection{Aperture Limitations}

The horizontal aperture of the acquisition array limits the maximum dip of reflectors that can be imaged. The governing relation is the standard migration-aperture requirement: imaging a reflector of dip $\theta$ at depth $z$ displaces the specular point laterally by $z\tan\theta$, so an array half-aperture $X$ must satisfy $X > z\tan\theta$ for that reflector to be illuminated at all. Steeply dipping reflectors near salt flanks, and deep targets generally, therefore demand large-aperture surveys that are often unavailable. Aperture also shapes the focus itself. Because the focused response radiates preferentially in a direction fixed by the array geometry \cite{effects2019aperture}, a virtual source near the edge of the array illuminates the subsurface differently from one at the centre, and amplitudes extracted from the two are not directly comparable.

It is tempting to suppose that because the retrieved wavefields contain the multiples, they must also illuminate what the primaries cannot reach. Under single-sided acquisition they do not. An internal multiple arriving at the focal point from a given angle is recovered by the iterative scheme only if the direct wave reached that point from the same angle, so a zone in shadow for primaries stays in shadow for multiples too \cite{van2014interferometric}. RTM faces the same geometric constraint on dip, so the limit is not specific to the Marchenko method, but neither does the method lift it. It inherits whatever the survey provides.

\subsection{Computational Cost for Large 3D Datasets}

As discussed in Section~\ref{sec:gpu}, the computational cost of 3D Marchenko imaging scales as $O(N_s^4 N_t)$ for $N_s$ surface positions per dimension, which is one reason industry-led 3D applications remain scarce. For a modern 3D survey with $10^3 \times 10^3$ surface positions and $10^4$ time samples, this is $O(10^{16})$ operations per focal point. Plane-wave compression \cite{brackenhoff2022three}, machine-learning acceleration \cite{wang2025accelerating}, and sparse representations \cite{zhang2020marchenko} all reduce the effective cost. The binding constraint, though, is data movement rather than arithmetic. Koehne et al.\ \cite{koehne2021multi} measured disk reads and host-to-device transfers at 37\% to 87\% of total runtime on 2D datasets between 3 and 250 GB, with the fraction worst when few focal points are imaged; a full 3D reflection volume makes that ratio worse, not better. Faster kernels will not close the gap on their own without compression or a change in acquisition.

\subsection{Anisotropy and Anelasticity}

Most Marchenko theory assumes isotropic, lossless acoustic media. Anisotropy introduces angular dependence in the phase velocity, requiring more elastic parameters, up to 21 in the most general case. Anelasticity introduces frequency-dependent phase velocity and amplitude decay. Both complicate wavefield decomposition and the definition of the focusing function. Generalisations to anisotropic media are under investigation, but no complete implementation has been demonstrated on field data.

\section{Future Directions}
\label{sec:future}          

The acoustic theory is established and robust, several open-source implementations are maintained, and field data results are growing more compelling year by year. The open problems below concern extending the method to wavefields it does not yet handle, and making the 3D case affordable.

\subsection{Full Elastic and Multi-Component Extensions}

An elastic implementation that handles P-to-S mode conversion, multi-component sources and receivers, and anisotropy is the extension with the largest gap between theory and practice. The framework has existed since 2014 \cite{da2014elastodynamic, da2016elastic}. What it demands is multi-component data on both source and receiver sides, a decomposition into four one-way constituents rather than two, and correspondingly larger coupled systems. Land surveys with 3-component geophones and multi-component ocean-bottom nodes supply the data; the missing piece is a field result that shows the elastic scheme outperforming the acoustic one on the same line.

\subsection{Joint Marchenko Inversion and FWI}

A natural next step from target-oriented FWI (Section~\ref{sec:targeting}) is joint inversion of the velocity model and reflectivity within the Marchenko framework. The Marchenko equations would act as a constraint on the internal consistency of the wavefields through the focusing condition, while the FWI objective drives the velocity model toward the true medium, combining the amplitude fidelity of FWI with the multiple suppression of the Marchenko scheme. Whether that combination is stable in practice is untested, and so is the prior question of whether the focusing constraint carries information about the velocity model or acts only as regularisation. There is a circularity to resolve as well: the focusing functions are computed from a traveltime estimate derived from the velocity model that the inversion is trying to update.

\subsection{Machine Learning Acceleration and Neural Operators}

The self-supervised approach of Wang et al.\ \cite{wang2025accelerating, wang2026deep} is the first step toward routine machine-learning-accelerated Marchenko imaging, and it sidesteps the labelling problem by training on data self-consistency. Generalisation is the open issue. A network trained on one geological setting has no guarantee of transferring to another, and the failure mode is quiet: a plausible-looking focusing function that produces a plausible-looking image with events in the wrong place. Neural operators \cite{li2020neural}, which learn the mapping from reflection data to focusing functions as a function-to-function map rather than a function-to-parameter map, are a candidate for better transfer and admit clearer uncertainty estimates. Enforcing the Marchenko equations as a soft constraint during training should improve reliability, at a training cost that may cancel part of the inference-time gain.

\subsection{Marchenko Methods for Ambient Noise and Passive Data}

All existing Marchenko implementations use active-source data. An interesting possibility is applying the Marchenko framework to passive (ambient noise) data, in which the Green's function is retrieved by cross-correlating long records of background seismic noise \cite{lobkis2001emergence}. The challenge is that ambient noise sources are distributed over a finite region rather than a complete enclosing surface, introducing biases in the retrieved Green's functions. The Marchenko approach could potentially correct for single-sided illumination in ambient noise interferometry. Whether the required travel time information can be extracted without active sources remains unclear.

\subsection{Marchenko Methods for Borehole and OBN Geometry}

Ocean-bottom node surveys and borehole VSP geometry offer denser subsurface sampling and higher data quality than towed streamers, and they place receivers where the water column above them is accurately known. Ravasi \cite{ravasi2017rayleigh} exploited this in the Rayleigh--Marchenko scheme, and the upside-down and reciprocal variants \cite{wang2023upside, wang2024upside, ravasi2023reciprocal} adapt it to the sparse, irregular receiver geometries that ocean-bottom surveys actually have. Dedicated formulations for downhole-to-surface and downhole-to-downhole configurations are less developed, and the sparse-source case, where nodes are dense but shots are not, is the one that most ocean-bottom acquisition presents.

\subsection{Marchenko Monitoring of Dynamic Subsurface Processes}

The time-lapse applications reviewed in Section~\ref{sec:timelapse} treat the subsurface as quasi-static between surveys. Dynamic processes such as earthquake rupture, slow-slip events, and induced seismicity from injection operations require imaging the subsurface at much higher time resolution, possibly from continuous recordings. Adapting the Marchenko framework to time-varying media, using a time-windowed or sliding-window formulation, is an open research problem with direct applications to reservoir monitoring, geothermal energy, and seismic hazard assessment.

\section{Conclusions}
\label{sec:conclusions}

The Marchenko method retrieves subsurface Green's functions, internal multiples included, from single-sided surface recordings, using only a smooth background velocity as input. A decade of work has confirmed that experimentally, extended it to elastic and electromagnetic media, and produced field results on marine data.

The mathematical structure is as follows. In one dimension the inverse scattering problem is solvable through the GLM equation, which relates the scattering potential to the measured reflection response without approximation. In higher dimensions the Schr\"{o}dinger reduction has no counterpart, because no single traveltime coordinate parameterises all wave paths, but the same kind of result, a coupled pair of integral equations for the focusing functions, follows instead from the acoustic reciprocity theorems. The focusing functions act as the inverse of the transmission response of the overburden. Once retrieved from the reflection data by iterating the coupled equations, or by solving them in least-squares form where the iteration does not converge, they yield the upgoing and downgoing Green's functions that imaging, redatuming, and inversion require, free of internal multiple contamination. Elastic, electromagnetic, and dissipative extensions follow from the corresponding reciprocity theorems, with a larger wavefield decomposition and larger coupled systems.

The costs are specific. The method needs an accurate initial traveltime estimate, and errors in it shift the time window and displace events in the image. The standard formulation ignores elastic mode conversion. The 3D problem remains expensive, and the binding constraint there is data movement rather than arithmetic, which GPU and machine-learning acceleration do not remove. Applying the method to land data with a complex near-surface is harder than applying it offshore, and the land results published so far have leaned on VSP constraints rather than on surface data alone.

That divides the ground fairly clearly. Marine imaging under complex overburdens, internal multiple elimination in deep water, and time-lapse monitoring where overburden noise is the limiting factor are settings where the method works and has been shown to work, on data from the Santos Basin, the North Sea, and the Gulf of Mexico. Routine land application, elastic implementation on multi-component data, and large-scale 3D industrial processing are settings where it has not yet been shown to work.

Three problems stand between the second list and the first. Joint inversion of velocity and reflectivity through the focusing constraint is theoretically natural and practically untested, and carries an unresolved circularity in that the focusing functions depend on the model being updated. The elastic extension has existed since 2014 without a convincing field result. Machine-learning acceleration works on the data it was trained near, and its generalisation across geological settings has not been demonstrated. None of the three is blocked on new theory.

\appendices
\setcounter{equation}{0}
\renewcommand{\theequation}{A-\arabic{equation}}

\section{Detailed Derivation of the 1D Gelfand--Levitan--Marchenko Equation}
\label{app:1D}

\subsection{Wronskian Relations and the Fundamental Solutions}

Let $\hat{\psi}_+(\tau, \omega)$ and $\hat{\psi}_-(\tau, \omega)$ denote the two fundamental (Jost) solutions to the 1D Schr\"{o}dinger equation \eqref{eq:schrodinger} in the frequency domain, with boundary conditions:
\begin{align}
\hat{\psi}_+(\tau, \omega) &\sim e^{i\omega\tau} \text{ as } \tau \to -\infty \\  &\text{(incident wave from the left)} \nonumber
\end{align}
\begin{align}
\hat{\psi}_-(\tau, \omega) &\sim e^{-i\omega\tau} \text{ as } \tau \to +\infty \\ &\text{(incident wave from the right)} \nonumber
\end{align}

\noindent The Wronskian $W[\hat{\psi}_+, \hat{\psi}_-] = \hat{\psi}_+ \partial_\tau \hat{\psi}_- - \hat{\psi}_- \partial_\tau \hat{\psi}_+$ is constant (independent of $\tau$) by Abel's identity, because the Schr\"{o}dinger equation has no first-derivative term. Evaluating it in the two free regions, where the solutions are known plane waves, is what relates $\hat{R}$ and $\hat{T}$ below.

The physical scattering solution normalised to unit incident amplitude from the left is denoted $\hat{\psi}$. For $\tau < 0$, the free region to the left of the medium, it is the sum of the incident and reflected waves:

\begin{equation}
\hat{\psi}(\tau, \omega) = e^{i\omega\tau} + \hat{R}(\omega)\, e^{-i\omega\tau}, \quad \tau < 0
\end{equation}

\noindent where $\hat{R}(\omega)$ is the Fourier transform of the reflection response. For $\tau > \tau_{\max}$, the free region to the right, only the transmitted wave remains:
\begin{equation}
\hat{\psi}(\tau, \omega) = \hat{T}(\omega)\, e^{i\omega\tau}, \quad \tau > \tau_{\max}
\end{equation}

\noindent where $\hat{T}(\omega)$ is the transmission coefficient. The Wronskian condition at $\tau \to \pm\infty$ gives the following:
\begin{align}
|\hat{R}(\omega)|^2 + |\hat{T}(\omega)|^2 = 1 \\
\text{(energy conservation for real, lossless media)} \nonumber
\end{align}

\subsection{Construction of the GLM Kernel}

The construction rests on a transformation operator. The claim is that the solution $\psi_a$ for the medium with potential $q$ can be written as the corresponding free-medium solution plus an integral operator acting on it,
\begin{equation}
\psi_a(\tau, t) = \psi_0(\tau, t) + \int_{-\tau}^{\tau} A(\tau, s)\, \psi_0(s, t)\, ds \label{eq:transformop}
\end{equation}

\noindent where $\psi_0$ is the solution in the absence of scattering and $A$ is a kernel independent of frequency. Both arguments of $A$ are measured in traveltime, which after the transformation of Section~\ref{sec:1D} carries the same units as $\tau$; this is what allows a single kernel to encode the entire scattering history. The integration variable $s$ occupies the \emph{first} slot of $\psi_0$, so that for an impulsive free solution $\psi_0(s,t) = \delta(t-s)$ the integral sifts out $A(\tau,t)$ and equation \eqref{eq:transformop} reduces to the annihilator \eqref{eq:annihilator}, as it must.

Requiring that \eqref{eq:transformop} satisfy the Schr\"{o}dinger equation \eqref{eq:schrodinger} for every $t$, while $\psi_0$ satisfies the free equation, forces $A$ to obey the Goursat problem
\begin{equation}
\frac{\partial^2 A}{\partial \tau^2} - \frac{\partial^2 A}{\partial t^2} = q(\tau)\,A \label{eq:goursat}
\end{equation}

\noindent with $A$ vanishing on $t = -\tau$ and, on the diagonal $t = \tau$, satisfying
\begin{equation}
\frac{d}{d\tau}A(\tau, \tau) = -\tfrac{1}{2}\,q(\tau)
\end{equation}

\noindent which is equation \eqref{eq:qrecovery} and identifies $A(\tau,\tau)$ with the amplitude coefficient $a_1(\tau)$ found from the ray expansion in Section~\ref{sec:1D}. The potential appears in \eqref{eq:goursat} only through its first argument, which is what makes the diagonal condition a complete recipe for recovering $q$.

Both signs appear in the literature, and the difference is entirely in how the kernel is defined. The standard statement of the Marchenko equation for $-u'' + qu = k^2u$ recovers the potential as $q = -2\,\mathrm{d}K(x,x)/\mathrm{d}x$ \cite{chadan1997introduction}, matching the convention used here. Burridge \cite{burridge1980gelfand} writes $q = +2\,\mathrm{d}K_1(x,x)/\mathrm{d}x$, but his kernel is the negative of the one above, so the two agree. What fixes the sign unambiguously in the present treatment is the ray expansion, which gives $A(\tau,\tau) = a_1(\tau) = -\tfrac{1}{2}\int_0^{\tau} q\,\mathrm{d}\tau'$ with no assumption about the kernel at all. A reader comparing against either source should check the kernel's sign before comparing the recovery formula.

\subsection{Derivation of the GLM Equation}

The integral equation follows from the annihilation condition at the surface, using
\begin{align}
\psi_+(0, t) &= \delta(t) + R(t), \quad R(t) = 0 \text{ for } t < 0
\end{align}
\begin{align}
\psi_a(0, t) &= \delta(t) + A(0, t)
\end{align}

\noindent The reflection response is causal, so $R$ vanishes for $t < 0$, which fixes the lower limit of the integral below.

The equation for $A$ now follows from the defining property of the annihilator, which is that it produces no upgoing field at the surface within the window $|t| < \tau$. Propagating the transformation-operator representation \eqref{eq:transformop} to $\tau = 0$ and collecting the terms that reach the surface inside that window gives three contributions: the kernel itself, the reflection response convolved with the incident delta, and the reflection response acting on the kernel. Setting their sum to zero,
\begin{equation}
A(\tau, t) + R(t + \tau) + \int_{-\tau}^{\tau} A(\tau, s)\, R(s + t)\, ds = 0, \quad |t| < \tau
\end{equation}

\noindent which is the GLM equation \eqref{eq:GLM}. It has the structure of the classical Marchenko equation $K(x,y) + F(x+y) + \int K(x,z)F(z+y)\,\mathrm{d}z = 0$ term for term: the kernel, the data evaluated at the sum of both arguments, and an integral in which the variable of integration occupies the second slot of the kernel and enters the data through the same summed argument \cite{chadan1997introduction, burridge1980gelfand}. What differs is the range of integration. The classical half-line problem integrates over $z > x$, whereas the focusing formulation used here integrates over the finite window $|s| < \tau$ set by the direct-arrival traveltime, which is the same truncation that reappears as the time-window operator $\Theta$ in the multidimensional case.

Two remarks on the status of this derivation. The step from the annihilation condition to the integral equation uses the completeness of the Jost solutions, which is what guarantees that vanishing on the window implies the vanishing of the bracketed expression pointwise; the classical treatments \cite{gelfand1951determination, marchenko1955reconstruction, burridge1980gelfand, chadan1997introduction} establish this and we take it as given. The Wronskian relations of Appendix~\ref{app:1D}-A are used only to relate $\hat{R}$ and $\hat{T}$ and to establish that the two Jost solutions are independent; they do not enter the derivation of the integral equation itself.

\subsection{Neumann Series Solution}

The GLM equation is a Fredholm equation of the second kind:
\begin{equation}
(I + K_\tau) A = -R_\tau
\end{equation}

where $(K_\tau A)(\tau, t) = \int_{-\tau}^{\tau} A(\tau, s) R(s + t)\, ds$ and $R_\tau(t) = R(t + \tau)$. The Neumann series solution is:
\begin{align}
A^{(0)} &= -R_\tau\\
A^{(k+1)} &= -R_\tau - K_\tau A^{(k)}
\end{align}

\noindent Convergence of the series requires $\|K_\tau\| < 1$ in the operator norm on $L^2(-\tau, \tau)$. Since $K_\tau$ is built from $R$ restricted to a finite window, its norm grows with both the strength of the reflectors above $\tau$ and the length of the window, so the condition is a statement about accumulated scattering in the overburden rather than about $|\hat{R}(\omega)|$ alone. Energy conservation in a lossless medium bounds $|\hat{R}(\omega)| \leq 1$ and rules out the pathological cases, but it does not by itself guarantee contraction. Where the series fails to contract, the equation itself remains well posed and can be solved directly, as a Fredholm equation of the second kind, by the matrix inversion described in Section~\ref{sec:1D}.

The $k$-th iterate accounts for internal multiples up to order $k$: $A^{(0)}$ carries the primaries only, $A^{(1)}$ adds the first-order internal multiples, and so on. The converged solution accounts for all orders, which is why Marchenko-based methods suppress internal multiples without modelling them explicitly.

\section{Derivation of the Multidimensional Marchenko Equations}
\label{app:2D3D}

\setcounter{equation}{0}
\renewcommand{\theequation}{B-\arabic{equation}}

\subsection{Configuration and States}

We derive the coupled Marchenko equations \eqref{eq:marchenko1}--\eqref{eq:marchenko2} from first principles using the convolution- and correlation-type reciprocity theorems.

\textbf{Domain:} The domain $\mathbb{D}^+$ is the slab between the acquisition surface $\dD_0$ (at $x_3 = 0$) and the virtual depth level $\dD_i$ (at $x_3 = x_{3,i}$). The boundary $\partial\mathbb{D} = \dD_0 \cup \dD_i$ with outward normal $\mathbf{n}$.

\textbf{State A -- Focusing function:} We choose state A to be the focusing function state, with:
\begin{align}
P_A(\xv, \xA, t) &= f_1(\xv, \xA, t) = f_1^+(\xv, \xA, t) + f_1^-(\xv, \xA, t)\\
q_A(\xv, \xA, t) &= 0 \quad \text{(no sources in } \mathbb{D}^+\text{)}
\end{align}

The focusing function $f_1$ is defined in the truncated medium, which coincides with the physical medium inside $\mathbb{D}^+$ and is homogeneous below $\dD_i$. The two states therefore share the same medium parameters throughout $\mathbb{D}^+$, which is what the reciprocity theorems require; they differ only below $\dD_i$, outside the integration domain.

\textbf{State B -- Physical Green's function:} State B is the actual physical medium:
\begin{align}
P_B(\xv, \xA, t) &= G(\xv, \xA, t) = G^+(\xv, \xA, t) + G^-(\xv, \xA, t)\\
q_B(\xv, \xA, t) &= \delta(\xv - \xA)\,\delta(t) \quad \text{(point source at } \xA \in \dD_i\text{)}
\end{align}

\noindent with $q$ the volume injection rate of equation \eqref{eq:continuity}, so that state B satisfies equation \eqref{eq:greendef}.

\subsection{Applying the Convolution-Type Reciprocity Theorem}

Substituting states A and B into the convolution-type reciprocity theorem \eqref{eq:recipconv} and noting that:

\begin{itemize}
    \item The source $q_A = 0$ everywhere in $\mathbb{D}^+$, so state A contributes nothing to the volume integral.
    \item On $\dD_i$, the focusing condition \eqref{eq:focusdef} makes the downgoing constituent of $f_1$ collapse to $\delta(\mathbf{x}_H - \mathbf{x}_{A,H})\,\delta(t)$, and the upgoing constituent vanishes there because the truncated medium is reflection-free below $\dD_i$.
    \item Both fields obey the radiation condition on the lateral closing surface, which is taken to infinity so that its contribution vanishes.
\end{itemize}

The boundary integral over $\partial\mathbb{D} = \dD_0 \cup \dD_i$ then involves $f_1$ and $G$ on the two boundaries. On $\dD_i$, the focusing condition collapses the integral to a contribution at the focal point $\xA$; on $\dD_0$, decomposing both fields into pressure-normalised up- and downgoing constituents (Appendix~\ref{app:decomp}) and expressing the upgoing field through the reflection operator ($G^- = R \conv G^+$ at the surface) reorganises the boundary terms into:
\begin{align}
(\mathcal{R}f_1^+)(\xH, \xA, t) = f_1^-(\xH, \xA, t) + G^{+}(\xH, \xA, t) \tag{\ref{eq:rep1}}
\end{align}

\subsection{Applying the Correlation-Type Reciprocity Theorem}

Substituting the time-reversal of state A (replacing $f_1(\xv, \xA, t)$ with $f_1(\xv, \xA, -t)$) and state B into the correlation-type reciprocity theorem \eqref{eq:recipcorr}, with similar boundary conditions:
\begin{align}
(\mathcal{R}^{\star}f_1^-)(\xH, \xA, t) = f_1^+(\xH, \xA, t) + G^{-}(\xH, \xA, -t) \tag{\ref{eq:rep2}}
\end{align}

\subsection{Applying the Time-Window Operator}

\noindent Equations \eqref{eq:rep1} and \eqref{eq:rep2} each contain two unknowns: $(f_1^+, f_1^-)$ and $(G^+, G^-)$. To separate them, observe that:

\begin{itemize}
    \item $f_1^-(\xH, \xA, t) = 0$ for $|t| > t_d(\xH, \xA)$ (equation \eqref{eq:f1minus_support}), so $\Theta f_1^- = f_1^-$.
    \item $G^{+}(\xH, \xA, t) = 0$ for $t < t_d(\xH, \xA)$, by causality: the response of a virtual source at $\xA$ cannot reach $\xH$ before the direct wave. Hence $\Theta G^{+} = 0$, and by the same argument $\Theta\{G^{-}(-t)\} = 0$.
    \item $f_1^+(\xH, \xA, t) = f_{1,d}^+(\xH, \xA, t) + M^+(\xH, \xA, t)$, with the direct term a delta at $t = -t_d$ and the coda $M^+$ vanishing for $t \leq -t_d$ and $t \geq t_d$ \cite{slob2014seismic}. Therefore $\Theta f_1^+ = M^+$: the window keeps the coda and discards the direct arrival.
\end{itemize}

Applying $\Theta$ to equation \eqref{eq:rep1} eliminates $G^{+}$ and leaves
\begin{equation}
f_1^-(\xH, \xA, t) = \Theta\!\left\{ (\mathcal{R}f_1^+)(\xH, \xA, t) \right\} \tag{\ref{eq:marchenko1}}
\end{equation}

\noindent Applying $\Theta$ to equation \eqref{eq:rep2} eliminates $G^{-}(-t)$ and leaves $M^+$; adding $f_{1,d}^+$ back to both sides reconstitutes $f_1^+$ and yields equation \eqref{eq:marchenko2}. This completes the derivation.

\noindent Sign conventions differ across the literature and are easy to mix. In the pressure-normalised convention used here, following Thorbecke et al.\ \cite{thorbecke2017implementation} and Brackenhoff et al.\ \cite{brackenhoff2022three}, both coupled equations carry a positive $\Theta$ term and the minus signs appear in equations \eqref{eq:Gm}--\eqref{eq:Gp}. The internal check on any implementation is that subtracting $f_1^-$ from $\mathcal{R}f_1^+$ must leave precisely the part of the synthesis that the window rejected; if it does not, the signs or the normalisation are inconsistent.

\subsection{Reduction to the 1D GLM Equation}

\noindent In one spatial dimension the surface integral collapses, the convolution over $\xH'$ disappears, and the coupled equations reduce to a pair of scalar time-domain relations:
\begin{align}
f_1^-(t) &= \Theta\!\left\{ (R \conv f_1^+)(t) \right\}\\
f_1^+(t) &= f_{1,d}^+(t) + \Theta\!\left\{ (R \star f_1^-)(t) \right\}
\end{align}

\noindent with $\conv$ the convolution of equation \eqref{eq:synthesis} and $\star$ the correlation of equation \eqref{eq:synthesiscorr}, both reduced to one dimension.

\noindent where $\star$ denotes correlation and $\Theta$ passes $|t| < t_d$. Substituting the first relation into the second gives a single equation in $f_1^+$ alone. Writing $\tau = t_d$ for the one-way traveltime to the focal depth and identifying the coda of $f_1^+$ with the GLM kernel through $A(\tau, t) = f_1^-(\tau, -t)$, that equation has the kernel structure of equation \eqref{eq:GLM}: an unknown, a term linear in the measured reflection response, and an integral of the unknown against $R$ over the symmetric window $(-\tau, \tau)$. The correspondence is exact for a scalar medium in the impedance-normalised variables of Section~\ref{sec:1D}, which is the sense in which the multidimensional equations contain the GLM equation as their 1D limit \cite{wapenaar2014marchenko, broggini2012connection}. It is a structural equivalence between the two integral equations, not a term-by-term identity of the wavefields, which carry different normalisations in the two formulations.

\section{Wavefield Decomposition into One-Way Components}
\label{app:decomp}

\setcounter{equation}{0}
\renewcommand{\theequation}{C-\arabic{equation}}

\subsection{One-Way Decomposition of the Two-Way System}

For the acoustic wave equation with varying density, the pressure-velocity system \eqref{eq:euler}--\eqref{eq:continuity} in the frequency domain takes the form
\begin{equation}
\partial_{x_3}\begin{pmatrix} \hat{P} \\ \hat{V}_3 \end{pmatrix} = \begin{pmatrix} 0 & i\omega\rho \\ \frac{i\omega}{\kappa} - \frac{\nabla_H^2}{i\omega\rho} & 0 \end{pmatrix} \begin{pmatrix} \hat{P} \\ \hat{V}_3 \end{pmatrix}
\end{equation}

where $\nabla_H^2 = \partial_{x_1}^2 + \partial_{x_2}^2$. This is a first-order system in $x_3$ with a $2 \times 2$ matrix. The eigenvalue decomposition yields two eigenvalues $\pm ik_{x_3}$ corresponding to downgoing ($+$) and upgoing ($-$) modes.

The pressure-normalised one-way fields are:
\begin{align}
\hat{P}^+ &= \frac{1}{2}\hat{P} + \frac{\omega\rho}{2k_{x_3}}\hat{V}_3\\
\hat{P}^- &= \frac{1}{2}\hat{P} - \frac{\omega\rho}{2k_{x_3}}\hat{V}_3
\end{align}

These satisfy:
\begin{align}
\partial_{x_3}\hat{P}^+ &= ik_{x_3}\hat{P}^+ + \nonumber \\ &\text{(coupling terms due to medium heterogeneity)}
\end{align}
\begin{align}
\partial_{x_3}\hat{P}^- &= -ik_{x_3}\hat{P}^- + \text{(coupling terms)}
\end{align}

In a homogeneous layer, the coupling terms vanish and $\hat{P}^{\pm}$ propagate independently with vertical wavenumber $\pm k_{x_3}$.

The alternative \emph{flux-normalised} fields follow by rescaling,
\begin{equation}
\hat{Q}^{\pm} = \sqrt{\frac{k_{x_3}}{\omega\rho}}\;\hat{P}^{\pm}
\end{equation}

\noindent for which $|\hat{Q}^+|^2 - |\hat{Q}^-|^2$ equals the vertical acoustic power flux, since the flux $\propto \Re\{\hat{P}^*\hat{V}_3\}$ evaluates to $\tfrac{k_{x_3}}{\omega\rho}(|\hat{P}^+|^2 - |\hat{P}^-|^2)$, the cross terms being purely imaginary. Which normalisation is adopted changes where the minus signs fall in the coupled Marchenko equations, so the two conventions cannot be mixed. This review uses pressure normalisation throughout, following the implementation literature \cite{thorbecke2017implementation, brackenhoff2022three}.

\subsection{Wavenumber Domain Decomposition}

In the wavenumber domain $(\kH, \omega)$:

\begin{equation}
k_{x_3}(\kH, x_3, \omega) = \sqrt{\frac{\omega^2}{c^2(x_3)} - |\kH|^2}
\end{equation}

For $|\kH| < |\omega|/c$ (propagating modes), $k_{x_3}$ is real; for $|\kH| > |\omega|/c$ (evanescent modes), $k_{x_3}$ is imaginary. The standard derivation retains propagating modes only, and the evanescent contribution is dropped on the grounds that it decays exponentially with depth. That approximation is safe at exploration depths and offsets, but it is an approximation rather than a limitation of the theory: Wapenaar \cite{wapenaar2020marchenko} and Brackenhoff and Wapenaar \cite{brackenhoff2023evanescent} formulated Marchenko schemes that account for evanescent waves explicitly, which matters for shallow, high-frequency, or large-aperture configurations where the evanescent part of the spectrum is not negligible.

\subsection{Decomposition at the Acquisition Surface}

At $x_3 = 0$, the medium is assumed known (e.g.\ water with $c_0 = 1500$~m/s, $\rho_0 = 1000$~kg/m$^3$). The decomposition is:

\begin{align}
\hat{P}^+(\kH, 0, \omega) &= \frac{1}{2}\hat{P} + \frac{\rho_0 c_0}{2\cos\theta}\hat{V}_3 \\
\hat{P}^-(\kH, 0, \omega) &= \frac{1}{2}\hat{P} - \frac{\rho_0 c_0}{2\cos\theta}\hat{V}_3
\end{align}

where $\cos\theta = k_{x_3}/(\omega/c_0) = \sqrt{1 - c_0^2|\kH|^2/\omega^2}$

In practice, this decomposition requires both pressure $P$ and vertical particle velocity $V_3$, hence the importance of dual-sensor streamers or ocean-bottom sensors for Marchenko applications. Single-sensor recordings (hydrophone only) require additional processing to estimate $V_3$ from $P$ using the water-column velocity.

\section{Iterative Algorithm for Solving the Marchenko Equations}
\label{app:algorithm}

\setcounter{equation}{0}
\renewcommand{\theequation}{D-\arabic{equation}}

\subsection{Algorithm Summary}

The complete iterative algorithm for solving the Marchenko equations and retrieving subsurface Green's functions is given below.

\noindent \textbf{Input:}
\begin{itemize}
    \item Reflection response $R(\mathbf{x}_{H,s}, \mathbf{x}_{H,r}, t)$: $N_s \times N_r \times N_t$ array.
    \item Background velocity model $c_0(\xv)$.
    \item Focal point $\xA = (\mathbf{x}_{A,H}, x_{3,A})$.
    \item Convergence tolerance $\epsilon$ and maximum iterations $K$.
\end{itemize}

\noindent \textbf{Step 1 -- Compute direct travel time:}
\begin{align*}
    t_d(\xH, \xA) = \text{solve eikonal from subsurface} \\ \text{focusing point,} \ \xA, \text{through} \, c_0(\xv) \, \text{to surface} \, \dD_0
\end{align*}

\noindent \textbf{Step 2 -- Construct initial estimate:}
\begin{align*}
    f_1^{+,0}(\xH, \xA, t) = G_d(\xH, \xA, -t) = A_d(\xH, \xA)\,\delta\!\left(t + t_d\right)
\end{align*}

\noindent the time-reversed direct arrival modelled in $c_0$, with $A_d \equiv 1$ for a kinematic-only initialisation. This approximates the exact $T_d^{\,\mathrm{inv}}$ up to the amplitude error described in Section~\ref{sec:initestimate}.

\noindent \textbf{Step 3 -- Iterative update: for $k = 0, 1, \ldots, K-1$:}
\begin{align*}
M_k(\xH, t) &= (\mathcal{R}f_1^{+,k})(\xH, \xA, t)
\end{align*}
\begin{align*}
f_1^{-,k+1}(\xH, \xA, t) &= \Theta_{t_d}\!\left\{ M_k(\xH, t) \right\}
\end{align*}
\begin{align*}
N_k(\xH, t) &= (\mathcal{R}^{\star}f_1^{-,k+1})(\xH, \xA, t)
\end{align*}
\begin{align*}
f_1^{+,k+1}(\xH, \xA, t) &= f_1^{+,0}(\xH, \xA, t) + \Theta_{t_d}\!\left\{ N_k(\xH, t) \right\}
\end{align*}

Check convergence: if $\|f_1^{+,k+1} - f_1^{+,k}\|_2 / \|f_1^{+,k}\|_2 < \epsilon$, stop. Each application of $\mathcal{R}$ or $\mathcal{R}^{\star}$ is one synthesis, so an iteration costs two.

\noindent \textbf{Step 4 -- Retrieve Green's functions:}
\begin{align*}
G^{+}(\xH, \xA, t) &= M_K(\xH, t) - f_1^{-,K}(\xH, \xA, t) \\
G^{-}(\xH, \xA, -t) &= N_K(\xH, t) - f_1^{+,K}(\xH, \xA, t)
\end{align*}%

\noindent Each Green's function is the part of the corresponding synthesis that the time window rejected, so no extra synthesis is needed once the iteration has converged.
\[
G_{\text{total}} = G^{+}  + G^{-}
\]
\noindent \textbf{Step 5 -- Form image using imaging conditions:}
\[
I(\xA) = \int_{\dD_0} \int_{-\infty}^{\infty} G^+(\xH, \xA, t)\, G^-(\xH, \xA, t)\, dt\, d\xH
\]

This is the cross-correlation imaging condition applied to the Marchenko-retrieved wavefields; the alternative imaging conditions of Section~\ref{sec:imaging} can be substituted here.

\subsection{Computational Complexity}

The dominant cost in each iteration is the multi-dimensional convolution in Step 3. Using the FFT-based overlap-add method \cite{pfister2017discrete}:

\begin{itemize}
    \item \textbf{2D case:} $O(N_s N_r N_t \log N_t)$ per iteration per focal point. For $N_s = N_r = 256$ and $N_t = 2048$: $\sim 10^9$ operations per focal point per iteration.
    \item \textbf{3D case:} $O(N_s^2 N_r^2 N_t \log N_t)$ per iteration per focal point. For $N_s = N_r = 64$ per dimension and $N_t = 2048$: $\sim 10^{11}$ operations.
\end{itemize}

\noindent Both figures count the $\log N_t$ factor of the transform. In a practical implementation the reflection data are transformed once and the iteration proceeds in the frequency domain \cite{thorbecke2017implementation}, so the recurring per-iteration cost is the $O(N_s N_r N_f)$ cross-multiplication and the transforms are amortised.

\subsection{Stability and Convergence}

The Neumann iteration converges when the operator $\mathbf{R}\Theta\mathbf{R}^{*}\Theta$ composed over one full cycle of equations \eqref{eq:iter1}--\eqref{eq:iter2} is a contraction. A practical necessary condition is that the windowed reflection operator does not amplify energy,
\begin{equation}
\left\| \Theta\, \mathbf{R} \right\|_2 < 1
\end{equation}

which is the multi-dimensional counterpart of the condition on $\|K_\tau\|$ in Appendix~\ref{app:1D}. Contraction can fail when the reflection response has not been properly deghosted or wavelet-deconvolved, when the overburden is strongly scattering, and when the formulation retains free-surface multiples, since the free surface adds reflection legs at every bounce \cite{dukalski2017marchenko}. Damping the update stabilises the iteration in practice,
\begin{equation}
f_1^{+,k+1} = \left(1 + \lambda\right)^{-1}\left[f_1^{+,0} + \Theta N_k\right]
\end{equation}

with $\lambda \sim 10^{-3}$--$10^{-1}$ depending on the noise level in the data. This is a relaxation of the fixed-point iteration, not Tikhonov regularisation in the variational sense; the least-squares formulations of Section~\ref{sec:lsmi}, which minimise a damped $\ell^2$ misfit, are the properly regularised alternative and do not require the series to contract at all \cite{de2021marchenko, ravasi2020implementation}.

\section*{Materials and Resources}
The PyMarchenko Python library, developed by the Deep Imaging Group at KAUST, implements several of the algorithms discussed in this review: Marchenko redatuming by iterative substitution and by inversion, Marchenko redatuming with irregular sources, Marchenko-based data reconstruction, and Rayleigh--Marchenko redatuming (\url{https://github.com/DIG-Kaust/pymarchenko}).

Lomas and Curtis \cite{lomas2019introduction} released MATLAB codes with their introductory paper on Marchenko methods and imaging (\url{https://wiki.seg.org/wiki/Software:Marchenko_for_imaging}). A GPU-optimised CuPy port of that code, intended for teaching and able to run on a single entry-level GPU, accompanies this review and is available at (\url{https://drive.google.com/drive/u/1/folders/1SOQUvALC9upk8hrCzLSN_JkUvkq5ezh_}).

Jan Thorbecke maintains an open-source repository of tools for geophysical modelling, including Marchenko algorithms written in C (\url{https://github.com/JanThorbecke/OpenSource}).


\ifCLASSOPTIONcaptionsoff
  \newpage
\fi

\bibliographystyle{IEEEtran}
\bibliography{bibtex/bib/IEEEabrv, bibtex/bib/references}




\end{document}